%% file: main.tex
\def\FULLVERSION{}\def\FINALVERSION{}
\documentclass[conference]{IEEEtran}

\usepackage[T1]{fontenc}
\usepackage{cite}
\usepackage{amsmath,amssymb,amsthm,mathtools}
\usepackage{graphicx}
\usepackage{booktabs}
\usepackage{multirow}
\usepackage{algorithm}
\usepackage[noend]{algpseudocode}
\usepackage{xcolor}
\usepackage{tikz}
\usetikzlibrary{calc,arrows.meta}
\usepackage{url}
\usepackage{xr-hyper}
\usepackage[hidelinks]{hyperref}
\newcommand{\authororcid}[2]{\href{https://orcid.org/#2}{#1\,\raisebox{-0.15ex}{\includegraphics[height=1.65ex]{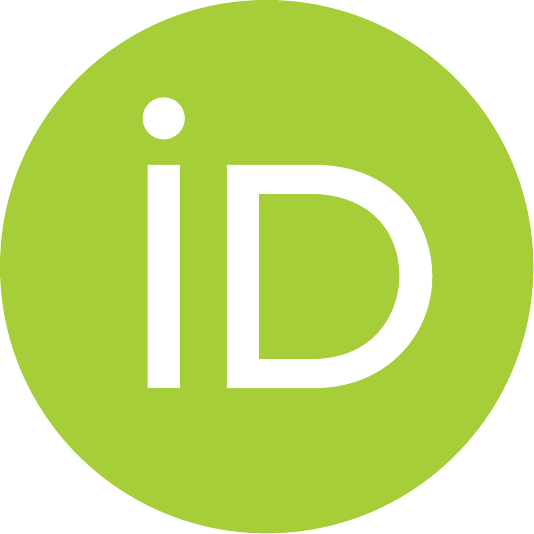}}}}
\usepackage[capitalise,noabbrev]{cleveref}

\newtheorem{theorem}{Theorem}
\newtheorem{lemma}{Lemma}
\newtheorem{proposition}{Proposition}
\theoremstyle{definition}
\newtheorem{definition}{Definition}
\newtheorem{problem}{Problem}
\crefname{problem}{Problem}{Problems}
\crefname{figure}{Fig.}{Figs.}
\Crefname{figure}{Fig.}{Figs.}
\input{macros}
\newif\iffull
\ifdefined\FULLVERSION\fulltrue\else\fullfalse\fi
\iffull\else\IfFileExists{main_full.aux}{\externaldocument[TR-]{main_full}}{}\fi
\newcommand{\trtab}[1]{\iffull(\cref{#1})\else\cite[Table~\ref*{TR-#1}]{techreport}\fi}
\newcommand{\trapp}[1]{\iffull(\cref{#1})\else\cite[App.~\ref*{TR-#1}]{techreport}\fi}
\newcommand{\trsec}[1]{\iffull(\cref{#1})\else\cite[Section~\ref*{TR-#1}]{techreport}\fi}

\begin{document}

\title{Community Search under Member-Specific Tolerances}

\author{
\IEEEauthorblockN{\authororcid{Yanghao Liu}{0009-0001-7979-2885}\IEEEauthorrefmark{1},
\authororcid{Fangda Guo}{0000-0003-2401-6499}\IEEEauthorrefmark{2},
\authororcid{Yuxuan Zhang}{0000-0002-9141-7447}\IEEEauthorrefmark{1},
\authororcid{Xueqi Cheng}{0000-0002-5201-8195}\IEEEauthorrefmark{1}}
\IEEEauthorblockA{\IEEEauthorrefmark{1}State Key Laboratory of AI Safety\\
Institute of Computing Technology, Chinese Academy of Sciences\\
\IEEEauthorrefmark{2}State Key Laboratory of Robotics and Intelligent Systems\\
Shenyang Institute of Automation, Chinese Academy of Sciences\\
liuyanghao@ict.ac.cn; guofangda@sia.cn; zhangyuxuan@ict.ac.cn; cxq@ict.ac.cn}
}

\maketitle

\input{sections/00_abstract}

\input{sections/01_intro}
\input{sections/02_related}
\input{sections/03_problem}
\input{sections/04_theory}
\input{sections/05_algorithm}
\input{sections/06_experiments}
\input{sections/07_conclusion}
\iffull\input{sections/09_supplement}\fi
\input{sections/08_statements}

\ifdefined\NOREFS\else
\bibliographystyle{IEEEtran}
\bibliography{refs}
\fi

\end{document}

%% file: macros.tex
\newcommand{\R}{\mathbb{R}}
\newcommand{\X}{\mathcal{X}}
\newcommand{\diam}{\operatorname{diam}}
\newcommand{\tmin}{\theta_{\min}}
\newcommand{\tacc}{\theta^{\mathrm{acc}}}
\newcommand{\trej}{\theta^{\mathrm{rej}}}
\newcommand{\Ck}{C_k}
\newcommand{\OPT}{\mathrm{OPT}}
\newcommand{\UB}{\mathrm{UB}}
\newcommand{\conv}{\operatorname{conv}}
\newcommand{\MaxLCC}{\textup{\textsc{MaxLCC}}}
\newcommand{\LCCDec}{\textup{\textsc{LCC-Dec}}}
\newcommand{\Gc}{G_{\theta}}
\newcommand{\Tq}{T_q}
\newcommand{\norm}[1]{\lVert #1 \rVert}
\newcommand{\Clique}{\textup{\textsc{Clique}}}
\newcommand{\MCClique}{\textup{\textsc{Multicolored Clique}}}
\newcommand{\LCCacc}{LCC\textsuperscript{acc}}
\newcommand{\LCCrej}{LCC\textsuperscript{rej}}
\newcommand{\pp}{\,pp}

\ifdefined\FINALVERSION

\else

\fi

%% file: sections/00_abstract.tex
\begin{abstract}
We study community search when each member has its own tolerance for differences in position, as when members of a group interact repeatedly and influence one another.
A latitude-consistent community (LCC) contains the query vertex, is connected, has minimum degree at least~$k$, and bounds its diameter by the smallest member tolerance.
Social judgment theory motivates the bound. The bound is exactly the condition under which every interaction in a connected closed group stays accepted under bounded-confidence dynamics for every choice of positive weights. It also prevents rejection under social-judgment dynamics, whereas the pairwise condition and other natural relaxations preserve neither guarantee.
Finding a maximum LCC is NP-hard even in the plane, and with unbounded dimension, deciding whether an LCC exists is NP-complete for every fixed $k\ge2$.
LatSweep fixes the answer's smallest tolerance and turns the coupled bound into a best-first sweep over candidate thresholds, yielding an exact branch-and-bound algorithm, a $(1,2)$-bicriteria approximation and two feasible heuristics.
Across eight graphs with up to 1.8 billion edges, six with synthetic positions and tolerances, the exact algorithm proves optimality on all but 9 of 3{,}700 efficiency queries and usually certifies absence within microseconds when no LCC exists; in a shared implementation on two large graphs, its branch-and-bound solver proves more queries than a reimplemented maximum-$(k,r)$-core algorithm, a mixed-integer program and CP-SAT.
In simulations of closed groups, LCC groups keep every accepted interaction under uneven influence, whereas 3--40\% of pairwise-compatible groups lose some.
On three location-based networks, tolerance-agnostic groups often place another member's home beyond a member's estimated travel radius.
Enforcing every member's tolerance causes no measurable loss of co-visits relative to spatial communities of the same size. Relative to the larger pairwise-compatible groups, it can cause a size-adjusted loss.
\end{abstract}


%% file: sections/01_intro.tex
\section{Introduction}\label{sec:intro}

For a query vertex $q$, community search identifies a cohesive subgraph containing $q$~\cite{sozio2010community,fang2020survey}.
When vertices have positions, such as opinions, interest profiles or homes, existing models enforce compactness using a shared threshold~\cite{zhang2017engagement,wang2018radius,chen2018colocated} or by minimizing spread~\cite{fang2017spatial,liu2020vac}.
However, a shared threshold cannot express how much variation each individual accepts: social judgment theory (SJT)~\cite{sherif1961social} describes a personal latitude of acceptance whose width varies across individuals, just as users of location-based social networks differ in how far they are willing to travel.

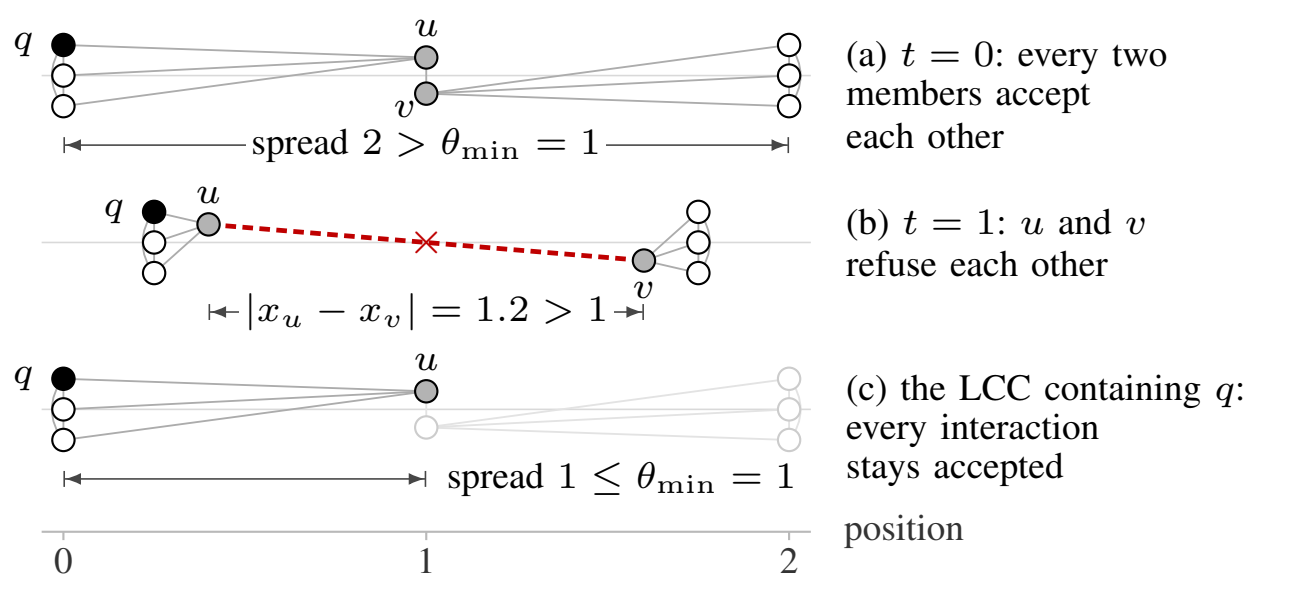
\begin{figure}[t]
\centering
\input{figures/fig_problem}
\caption{Pairwise acceptance versus the smallest-tolerance bound ($k=3$; grey $u$ and $v$ have tolerance~1, all others~2; in (c), non-members are faded; vertical offsets only aid legibility).}\label{fig:problem}
\end{figure}

We search for \emph{latitude-consistent communities} (LCCs): connected subgraphs with minimum degree at least $k$, containing $q$, whose diameter in the position space is bounded by the smallest member tolerance.
In deliberative settings such as a platform assembling a discussion panel on a contested topic, members adjust their positions as they influence one another over repeated rounds, so a group selected only for initial pairwise compatibility can lose acceptable exchanges.
In \cref{fig:problem}(a), every pair initially accepts each other, so the group meets the pairwise condition, yet a single round of equal-weight averaging (\cref{eq:hk} in \cref{sec:theory:dynamics}) separates $u$ and $v$~(b), whereas bounding the spread by the smallest tolerance preserves acceptance in the LCC of~(c).
With a common tolerance $r$, a largest LCC is a maximum $(k,r)$-core~\cite{zhang2017engagement} containing $q$. Member-specific tolerances complicate the search because adding a less tolerant member tightens the bound for every pair; whether two vertices can coexist depends on the rest of the group, preventing direct pruning under a fixed threshold.
We take member tolerances as inputs, as community search takes attributes or query radii, and study what the bound guarantees and how to search under it.

\begin{itemize}
\item \emph{Model and guarantees} (\cref{sec:problem,sec:theory:dynamics}).
If tolerances are the radii of the latitudes of acceptance, every interaction in a closed LCC stays accepted under bounded-confidence dynamics with personal bounds~\cite{hegselmann2002opinion,lorenz2010heterogeneous} and any positive weights, and the group converges to consensus. If they are the distances beyond which members reject, no member of an LCC ever rejects another under SJT dynamics.
Conversely, in every connected group that violates the bound, some positive weighting of the averaging makes a member refuse an interaction. Neither a slack of~$\varepsilon$, nor exempting two members, nor the pairwise condition $d(x_u,x_v)\le\min(\theta_u,\theta_v)$ preserves the guarantees.
For members who also listen to outside friends, we give a sufficient stability condition.
\item \emph{Complexity} (\cref{sec:theory:complexity}).
Finding a largest LCC is NP-hard even with equal tolerances in the plane and on the sphere. With the dimension as part of the input, it is not approximable within $n^{1-\varepsilon}$ unless $\mathrm{P}=\mathrm{NP}$, and for each fixed $k\ge2$, even deciding whether an LCC exists is NP-complete. On a line, the problem is polynomial.
\item \emph{Algorithms} (\cref{sec:algorithm}).
LatSweep fixes the smallest tolerance $\tau$ of the answer so that every member must satisfy $d(x_v,x_q)\le\tau\le\theta_v$, turning the coupled constraint into a sweep over survival intervals with a common threshold at each step (\cref{fig:latsweep}).
The sweep yields BallCore-F and a greedy variant, which return only LCCs; BallCore-R, a polynomial-time $(1,2)$-bicriteria approximation whose diameter factor cannot be improved in general metrics unless $\mathrm{P}=\mathrm{NP}$; and an exact branch-and-bound algorithm that refines upper bounds best-first and, if stopped early, returns the best LCC found, if any, with a proven bound on the optimum.
A query-independent index shrinks the search region of all but BallCore-R.
\item \emph{Experiments} (\cref{sec:experiments}).
We ask how far exact search scales, how groups with and without the bound behave under repeated interaction, and what personal tolerances change on real positions.
On eight graphs with up to 1.8 billion edges, six with synthetic positions and tolerances, the exact algorithm proves within microseconds that most queries without an LCC have none. It proves optimality on all but 9 of our 3{,}700 efficiency queries and, on those, returns an LCC of at least 0.689 times a proven upper bound. On two large graphs, with only the subproblem solver exchanged, it proves more queries than a reimplemented maximum-$(k,r)$-core algorithm, a mixed-integer program and CP-SAT.
In simulations of closed groups, LCC groups keep every accepted interaction, whereas groups that satisfy only the pairwise condition lose some when influence is uneven.
On three location-based datasets, tolerance-agnostic groups often place another member's home beyond a member's radius; enforcing every member's tolerance causes no measurable loss of co-visits relative to spatial communities of the same size, and it can cause a size-adjusted loss relative to the larger pairwise groups.
\end{itemize}
Our code and processing scripts are available at \url{https://github.com/YangH-L/LACS}\iffull; compared with the conference version, this report provides full proofs and an appendix of supplementary results\else, and full proofs and supplementary tables appear in our technical report~\cite{techreport}\fi.

%% file: figures/fig_problem.tex
\begin{tikzpicture}[x=2.4cm,y=0.92cm,
    every node/.style={font=\scriptsize},
    m/.style={circle,draw=black,fill=white,inner sep=0pt,minimum size=4.2pt},
    t/.style={circle,draw=black,fill=gray!60,inner sep=0pt,minimum size=4.2pt},
    qq/.style={circle,draw=black,fill=black,inner sep=0pt,minimum size=4.2pt},
    ghost/.style={circle,draw=gray!40,fill=white,inner sep=0pt,minimum size=4.2pt},
    e/.style={draw=gray!65,line width=0.35pt},
    eg/.style={draw=gray!22,line width=0.35pt},
    bad/.style={draw=red!75!black,line width=0.9pt,dash pattern=on 2pt off 1.2pt},
    dim/.style={draw=gray!55!black,line width=0.3pt,{Bar[width=3pt]Latex[length=3pt]}-{Latex[length=3pt]Bar[width=3pt]}},
    base/.style={draw=gray!30,line width=0.3pt},
    lbl/.style={anchor=west,align=left,font=\scriptsize,inner sep=1pt},
  ]
  \def\rowA{0}\def\rowB{-1.2}\def\rowC{-2.4}
  \foreach \y in {\rowA,\rowB,\rowC} {\draw[base] (-0.06,\y) -- (2.06,\y);}

  \newcommand{\grp}[5]{%
    \coordinate (h1) at (#1,#5+0.22); \coordinate (h2) at (#1,#5); \coordinate (h3) at (#1,#5-0.22);
    \coordinate (g1) at (#4,#5+0.22); \coordinate (g2) at (#4,#5); \coordinate (g3) at (#4,#5-0.22);
    \coordinate (u) at (#2,#5+0.13); \coordinate (v) at (#3,#5-0.13);}

  \grp{0}{1}{1}{2}{\rowA}
  \foreach \x in {h1,h2,h3} {\draw[e] (u) -- (\x);}
  \foreach \x in {g1,g2,g3} {\draw[e] (v) -- (\x);}
  \draw[e] (h1) -- (h2) -- (h3); \draw[e] (h1) to[bend right=40] (h3);
  \draw[e] (g1) -- (g2) -- (g3); \draw[e] (g1) to[bend left=40] (g3);
  \draw[e] (u) -- (v);
  \node[qq] at (h1) {}; \node[m] at (h2) {}; \node[m] at (h3) {};
  \node[m] at (g1) {}; \node[m] at (g2) {}; \node[m] at (g3) {};
  \node[t] at (u) {}; \node[t] at (v) {};
  \node[anchor=east] at ($(h1)+(-0.03,0)$) {$q$};
  \node[anchor=south] at ($(u)+(0,0.02)$) {$u$};
  \node[anchor=north east,inner sep=1pt] at ($(v)+(-0.01,-0.01)$) {$v$};
  \draw[dim] (0,\rowA-0.5) -- node[fill=white,inner sep=1pt] {spread $2>\theta_{\min}=1$} (2,\rowA-0.5);
  \node[lbl] at (2.14,\rowA+0.13) {(a) $t=0$: every two};
  \node[lbl] at (2.14,\rowA-0.15) {members accept};
  \node[lbl] at (2.14,\rowA-0.43) {each other};

  \grp{0.25}{0.4}{1.6}{1.75}{\rowB}
  \foreach \x in {h1,h2,h3} {\draw[e] (u) -- (\x);}
  \foreach \x in {g1,g2,g3} {\draw[e] (v) -- (\x);}
  \draw[e] (h1) -- (h2) -- (h3); \draw[e] (h1) to[bend right=40] (h3);
  \draw[e] (g1) -- (g2) -- (g3); \draw[e] (g1) to[bend left=40] (g3);
  \draw[bad] (u) -- (v);
  \node[red!75!black,font=\footnotesize] at (1,\rowB) {$\times$};
  \node[qq] at (h1) {}; \node[m] at (h2) {}; \node[m] at (h3) {};
  \node[m] at (g1) {}; \node[m] at (g2) {}; \node[m] at (g3) {};
  \node[t] at (u) {}; \node[t] at (v) {};
  \node[anchor=east] at ($(h1)+(-0.03,0)$) {$q$};
  \node[anchor=south] at ($(u)+(0,0.02)$) {$u$};
  \node[anchor=north] at ($(v)+(0,-0.02)$) {$v$};
  \draw[dim] (0.4,\rowB-0.5) -- node[fill=white,inner sep=1pt] {$|x_u-x_v|=1.2>1$} (1.6,\rowB-0.5);
  \node[lbl] at (2.14,\rowB+0.13) {(b) $t=1$: $u$ and $v$};
  \node[lbl] at (2.14,\rowB-0.15) {refuse each other};

  \grp{0}{1}{1}{2}{\rowC}
  \foreach \x in {g1,g2,g3} {\draw[eg] (v) -- (\x);}
  \draw[eg] (g1) -- (g2) -- (g3); \draw[eg] (g1) to[bend left=40] (g3);
  \draw[eg] (u) -- (v);
  \node[ghost] at (v) {}; \node[ghost] at (g1) {}; \node[ghost] at (g2) {}; \node[ghost] at (g3) {};
  \foreach \x in {h1,h2,h3} {\draw[e] (u) -- (\x);}
  \draw[e] (h1) -- (h2) -- (h3); \draw[e] (h1) to[bend right=40] (h3);
  \node[qq] at (h1) {}; \node[m] at (h2) {}; \node[m] at (h3) {};
  \node[t] at (u) {};
  \node[anchor=east] at ($(h1)+(-0.03,0)$) {$q$};
  \node[anchor=south] at ($(u)+(0,0.02)$) {$u$};
  \draw[dim] (0,\rowC-0.5) -- (1,\rowC-0.5);
  \node[anchor=west,inner sep=1pt] at (1.04,\rowC-0.5) {spread $1\le\theta_{\min}=1$};
  \node[lbl] at (2.14,\rowC+0.13) {(c) the LCC containing $q$:};
  \node[lbl] at (2.14,\rowC-0.15) {every interaction};
  \node[lbl] at (2.14,\rowC-0.43) {stays accepted};

  \draw[gray!55] (-0.06,\rowC-0.88) -- (2.06,\rowC-0.88);
  \foreach \x in {0,1,2} {\draw[gray!55] (\x,\rowC-0.88) -- (\x,\rowC-0.94); \node[anchor=north,gray!40!black,inner sep=1.5pt] at (\x,\rowC-0.93) {\x};}
  \node[anchor=west,gray!40!black] at (2.1,\rowC-0.88) {position};
\end{tikzpicture}

%% file: sections/02_related.tex
\section{Related Work}\label{sec:related}

\paragraph{Community search}
Community search identifies a cohesive subgraph containing a query vertex, usually a connected $k$-core~\cite{sozio2010community,cui2014local} or $k$-truss~\cite{huang2014querying}; a survey is given in~\cite{fang2020survey}.
Attributed variants add conditions on attributes. ACQ~\cite{fang2016effective} maximizes the keywords shared by all members; VAC~\cite{liu2020vac} minimizes the largest pairwise attribute distance in a $k$-truss, which is NP-hard and admits a 2-approximation; and skyline community search~\cite{guo2026skyline} returns the cohesive communities of a bipartite graph that no other community dominates on multi-dimensional edge attributes.
$(k,r)$-cores~\cite{zhang2017engagement} require all pairs of members to be similar within one threshold, and $(k,\mathcal{P})$-cores~\cite{fang2020heterogeneous} extend $k$-cores to heterogeneous information networks, where significant-attributed community search~\cite{liu2024sach} maximizes the smallest significance of a member and FCS-HGNN~\cite{chen2024fcs} learns to return communities that span several vertex types.
In density personalized group queries~\cite{shen2023density}, members individually bound their internal degree; in an LCC, group composition sets the distance bound for every pair.
Signed-network models take positive and negative edges as observed~\cite{bonchi2019discovering,ordozgoiti2020finding}, and Liu et al.~\cite{liu2026rethinking} cluster cognition-aware embeddings; an LCC keeps an explicit cohesion constraint and derives acceptance from positions and personal tolerances.

\paragraph{Spatial and geo-social groups}
Spatial-aware community search (SAC)~\cite{fang2017spatial} returns the $k$-core containing $q$ whose minimum covering circle is smallest.
Radius-bounded $k$-cores~\cite{wang2018radius} must fit in a circle of given radius, and maximum co-located communities~\cite{chen2018colocated} keep all pairwise distances within a given threshold.
Geo-social group queries~\cite{yang2012socio,zhu2017geosocial} select a group, near a rally point or within a query region, in which every member knows a given number of others, and skyline cohesive groups~\cite{li2020skyline} trade social cohesion against the members' travel cost to a meeting point within everyone's reach.
On road-social networks, cohesive group nearest neighbour queries~\cite{guo2019cohesive,guo2021cohesive} pair a cohesive group with the meeting point that minimizes its travel distance, under one or several criteria, and multi-attributed community search~\cite{guo2021multi} ranks cohesive groups by attributes with uncertain weights.
Rather than imposing a shared spatial radius or objective, LCCs bound group spread by the least tolerant member through a hard metric constraint, analogous to how least-misery group recommendation scores groups by their least satisfied member~\cite{ameryahia2009group,basuroy2015group}.

\paragraph{Opinion dynamics}
Averaging models repeatedly update agents to weighted averages of their neighbours' opinions~\cite{degroot1974reaching}, possibly with an anchor at their initial opinions~\cite{friedkin1990social}.
Bounded-confidence models~\cite{deffuant2000mixing,hegselmann2002opinion} average only opinions within a confidence bound; with heterogeneous bounds, whether consensus arises depends on the bounds and the network~\cite{lorenz2010heterogeneous,mirtabatabaei2012opinion}, and, for scalar opinions, the dynamics converge in finite time if the initial opinions span at most twice the smallest bound~\cite{su2017partial}.
Opinions fixed outside the group act as constant signals, for example from radical groups or charismatic leaders~\cite{hegselmann2015opinion}.
Models based on SJT add noncommitment and rejection~\cite{jager2005uniformity}; bounded-confidence models have been fitted to social traces~\cite{monti2020learning}, and individual users' confidence intervals have been measured as the span of their neighbours' opinions~\cite{luan2026homophily}.
We use these dynamics as a group-selection criterion: the search returns groups with guarantees about subsequent interactions (\cref{sec:theory}), whereas the cited studies follow an initially given population.

%% file: sections/03_problem.tex
\section{Problem Definition}\label{sec:problem}

\paragraph{Member-specific tolerances}
Social judgment theory~\cite{sherif1961social} describes how people assess positions against a personal anchor. Nearby positions are assimilated within the \emph{latitude of acceptance}; distant positions are contrasted within the \emph{latitude of rejection}, with noncommitment between these regions. Their widths differ across individuals.
We give each vertex $v$ a position $x_v$ in a metric space $(\X,d)$ and two radii $0<\tacc_v\le\trej_v$: vertex $v$ accepts a position $y$ if $d(x_v,y)\le\tacc_v$, rejects it if $d(x_v,y)>\trej_v$, and is noncommittal otherwise.
The same search model applies to geographic positions.
In meetups (\cref{sec:experiments}), $x_v$ is a user's home and $\theta_v$ estimates, from past visits, how far the user is willing to travel. Each member need only reach every other member's home.

\begin{table}[t]
\caption{Notation.}\label{tab:notation}
\centering\footnotesize\setlength{\tabcolsep}{4pt}
\input{tables/tab_notation}
\end{table}

\paragraph{Notation}
Let $G=(V,E)$ be an undirected simple graph with $n=|V|$.\footnote{For a heterogeneous information network, $G$ can be the homogeneous graph induced by a meta-path, as for $(k,\mathcal{P})$-cores~\cite{fang2020heterogeneous}.}
For $H\subseteq V$, $G[H]$ is the induced subgraph and $N_H(v)$ the set of neighbours of $v$ in $G[H]$.
Each vertex has a position $x_v\in\X$ and a tolerance $\theta_v>0$; we write $\diam(H)=\max_{u,v\in H}d(x_u,x_v)$ and $\tmin(H)=\min_{w\in H}\theta_w$; \cref{tab:notation} lists the main notation.

\begin{definition}[Latitude-consistent community]\label{def:lcc}
For a query vertex $q$ and an integer $k\ge 1$, we call a set $H\subseteq V$ a \emph{latitude-consistent community} (LCC) when
(i)~$q\in H$;
(ii)~$G[H]$ is connected and has minimum degree at least $k$; and
(iii)~$\diam(H)\le\tmin(H)$.
\end{definition}

Choosing $\theta=\tacc$ yields an \emph{\LCCacc}, in which every member accepts every other member's position; $\theta=\trej$ yields the weaker \emph{\LCCrej}, where no member rejects another.

\begin{problem}[\MaxLCC]\label{prob:maxlcc}
For inputs $G$, $x$, $\theta$, $q$ and $k$, return an LCC with the largest possible number of vertices.
The decision problem \LCCDec{} asks whether there exists an LCC containing at least $s$ vertices.
\end{problem}

\paragraph{Why the minimum}
\Cref{sec:theory} shows that condition~(iii) is exactly what keeps every interaction in a connected group accepted at all times, for all positive averaging weights (\cref{prop:char}). In the worst case, it cannot be weakened even for equal weights: neither a slack of $\varepsilon$, nor exempting two members, nor the pairwise condition $d(x_u,x_v)\le\min(\theta_u,\theta_v)$ preserves the guarantee (\cref{prop:tight}, \cref{fig:problem}).
In a normed space, pairwise acceptance is equivalent to every member accepting every weighted average of the members' positions, which suffices while those positions stay fixed.
Removing members can only decrease $\diam(H)$ and increase $\tmin(H)$, so condition~(iii) is hereditary. Yet the condition couples three members at a time: $d(x_u,x_v)\le\theta_w$ for all $u,v,w\in H$. This dependence on other members prevents a fixed pairwise decomposition.
For instance, $u$ and $v$ at distance $1$ with tolerance $1$ form a feasible pair, and so does each with a vertex $w$ midway between them with tolerance $1/2$, yet $\{u,v,w\}$ violates~(iii).
With uniform tolerances $\theta_v\equiv r$, condition~(iii) becomes the pairwise constraint of $(k,r)$-cores~\cite{zhang2017engagement}, with distance in place of similarity. \MaxLCC{} then asks for a maximum $(k,r)$-core containing $q$. Pruning for $(k,r)$-cores relies on this fixed pairwise structure, which LatSweep (\cref{sec:algorithm}) recovers by fixing the threshold.

\paragraph{When to require~(iii)}
The guarantees in \cref{sec:theory} apply to closed groups interacting repeatedly under unknown, possibly uneven influence weights, such as platform-formed study groups, deliberative-poll panels, project teams~\cite{lappas2009finding} or online deliberation groups.
For a one-off meetup with fixed positions, the pairwise condition already suffices and permits larger groups (\cref{sec:exp:dynamics,sec:exp:meetup}).

\paragraph{$k$-cores}
For $S\subseteq V$, the $k$-core~\cite{seidman1983network} of $G[S]$ is its unique maximal subgraph with minimum degree at least $k$, and $\Ck(S,q)$ denotes the component of this $k$-core that contains $q$, or the empty set if $q$ is not in the $k$-core; peeling computes it in $O(|S|+|E(S)|)$ time~\cite{batagelj2003cores}, and we use the same notation for peeling in a spanning subgraph of $G$.
Throughout, we use the \emph{$k$-core fact}: every connected $H\subseteq S$ with $q\in H$ and minimum degree at least $k$ satisfies $H\subseteq\Ck(S,q)$.

%% file: tables/tab_notation.tex
\begin{tabular}{@{}ll@{}}
\toprule
Symbol & Meaning \\
\midrule
$G=(V,E)$, $n$ & social graph, $n=|V|$ \\
$(\X,d)$ & metric space of positions \\
$q$, $k$ & query vertex, minimum degree \\
$x_v$, $\theta_v$ & position and tolerance of $v$ \\
$\tacc_v\le\trej_v$ & radii of acceptance and of rejection \\
$H$, $h$ & group, $h=|H|$ \\
$G[H]$, $N_H(v)$ & induced subgraph; neighbours of $v$ in it \\
$\diam(H)$, $\tmin(H)$ & largest distance and smallest tolerance in $H$ \\
$\OPT$ & size of a largest LCC (0 if none) \\
$x_v(t)$, $A_v(t)$ & position at step $t$; neighbours accepted at $t$ \\
$w_{vu}(t)\ge\eta$ & averaging weights, bounded below by $\eta>0$ \\
$\Ck(S,q)$ & component with $q$ of the $k$-core of $G[S]$ \\
$\Gc$ & subgraph of $G$ on mutually accepted edges \\
$\rho_v$, $I_v$ & $d(x_v,x_q)$; survival interval of $v$ \\
$R$ & query region around $q$ \\
$\tau$, $\Tq$ & threshold; candidate thresholds \\
$S(\tau)$, $C(\tau)$ & vertices of $R$ with $\tau\in I_v$; $\Ck(S(\tau),q)$ \\
$\OPT_\tau$, $\beta(\tau)$ & optimum at threshold $\tau$ and its upper bound \\
$B$, $\UB$ & best set found; proven upper bound on $\OPT$ \\
\bottomrule
\end{tabular}

%% file: sections/04_theory.tex
\section{Theoretical Properties}\label{sec:theory}

The theory addresses two questions: which groups preserve acceptance under opinion dynamics (\cref{sec:theory:dynamics}), and how difficult it is to find a largest such group or approximate its size (\cref{sec:theory:complexity}).

\subsection{Guarantees under Opinion Dynamics}\label{sec:theory:dynamics}

Here, positions are opinions in $\R^m$ under any norm $\norm{\cdot}$.
A group $H$ starts from $x_v(0)=x_v$; the graph and the tolerances are fixed, interactions follow edges of $G[H]$, and $\diam(t)=\max_{u,v\in H}\norm{x_u(t)-x_v(t)}$.

\paragraph{Bounded-confidence dynamics}
In the network version of the Hegselmann--Krause (HK) model~\cite{hegselmann2002opinion} with agent-specific bounds~\cite{lorenz2010heterogeneous}, vertex $v$ averages the opinions of the neighbours it accepts:
\begin{equation}\label{eq:hk}
x_v(t+1)=\sum_{u\in A_v(t)\cup\{v\}}w_{vu}(t)\,x_u(t),
\end{equation}
where $A_v(t)=\{u\in N_H(v):\norm{x_u(t)-x_v(t)}\le\theta_v\}$, the weights are non-negative with unit sum, and $w_{vu}(t)\ge\eta>0$ for $u\in A_v(t)\cup\{v\}$; equal weights\iffull{} $w_{vu}(t)=1/(|A_v(t)|+1)$\fi{} give $\eta=1/|H|$.
An interaction along $(u,v)$ is \emph{accepted} at time $t$ if $u\in A_v(t)$.

\begin{lemma}\label{lem:hull}
If every $x_v(t+1)$ is a convex combination of $\{x_u(t):u\in H\}$, then $\diam(t+1)\le\diam(t)$, because, by convexity of the norm, a point set and its convex hull have the same diameter.
\end{lemma}

\begin{theorem}[Consensus]\label{thm:consensus}
Let $H$ be an \LCCacc{} with $h=|H|\ge2$ that evolves by~\eqref{eq:hk} with $\theta=\tacc$.
Then for every $t\ge0$,
(a)~every interaction along every edge of $G[H]$ is accepted, and
(b)~$\diam(t)$ is non-increasing and\iffull{} $\diam(t)\le(1-h\eta^{h-1})^{\lfloor t/(h-1)\rfloor}\diam(0)$\else{} converges to $0$ geometrically\fi, so all opinions converge to a common point.
\iffull
(c)~With equal weights, the limit is the $(\deg_H(v)+1)$-weighted average of the initial opinions.
\fi
\end{theorem}
\begin{proof}
(a)~Each update~\eqref{eq:hk} is a convex combination, so by \cref{lem:hull}, $\diam(t)\le\diam(0)\le\tmin(H)\le\theta_v$ for all $t$ and $v$; hence $A_v(t)=N_H(v)$.

(b)~By~(a), the update matrix $W(t)$ is row-stochastic with entries at least $\eta$ on the diagonal and on the edges of $G[H]$.
Since $G[H]$ is connected, any two vertices are joined by a walk of exactly $h-1$ steps when self-loops are allowed, so every entry of $\Pi=W(t+h-2)\cdots W(t)$ is at least $\gamma=\eta^{h-1}$.
\iffull
For two rows $a,b$ of $\Pi$, let $\delta=\frac12\sum_j|a_j-b_j|=1-\sum_j\min(a_j,b_j)\le1-h\gamma$; the vector $\sum_j(a_j-b_j)x_j$ is $0$ if $\delta=0$ and otherwise $\delta$ times a difference of two convex combinations of the $x_j$, so its norm is at most $\delta\cdot\diam$.
Hence
\else
By the Dobrushin coefficient of $\Pi$~\cite{seneta2006nonnegative},
\fi
the diameter is multiplied by at most $1-h\gamma$ every $h-1$ steps.
Because every later opinion remains in the convex hull of the opinions at time $t$, the trajectories are Cauchy and share a common limit.
\iffull

(c)~With equal weights, $W(t)\equiv W$ is the random walk on $G[H]$ with self-loops, whose stationary distribution $\pi_v\propto\deg_H(v)+1$ satisfies $\pi^{\top}x(t)=\pi^{\top}x(0)$.
\fi
\end{proof}

Condition~(iii) fixes the accepted interactions, so the dynamics reduce to linear averaging on $G[H]$~\cite{degroot1974reaching}. Connectivity then ensures convergence; the role of $k$ is structural cohesion.
Consensus does not require condition~(iii), although it suffices: heterogeneous bounded-confidence dynamics can reach consensus from much more dispersed configurations and remain only partly understood~\cite{lorenz2010heterogeneous,mirtabatabaei2012opinion}. Natural relaxations of~(iii) do not guarantee consensus (\cref{prop:tight}).
\iffull
Under asynchronous updates, in which only some vertices apply~\eqref{eq:hk} at each step, \cref{thm:consensus}(a) and the monotonicity of $\diam(t)$ still hold, and if every vertex updates at least once in every $\kappa$ consecutive steps, applying~(b) to blocks of $\kappa$ steps gives convergence (cf.~\cite{moreau2005stability}).
The loose rate bound in~(b) serves only to establish geometric convergence.
\else
\Cref{thm:consensus}(a) and the monotonicity of $\diam(t)$ also hold under asynchronous updates, as does convergence if every vertex updates at bounded intervals~\cite{techreport}.
\fi

\paragraph{Social-judgment dynamics}
SJT adds noncommitment and rejection.
Following Jager and Amblard~\cite{jager2005uniformity}, vertex $v$ moves towards accepted neighbours, ignores noncommittal ones, and moves away from rejected ones:
\begin{multline}\label{eq:sjt}
x_v(t+1)=x_v(t)+\sum_{u\in A_v(t)}\alpha_{vu}\big(x_u(t)-x_v(t)\big)\\
-\sum_{u\in R_v(t)}\beta_{vu}\big(x_u(t)-x_v(t)\big),
\end{multline}
where $A_v(t)$ uses $\tacc_v$, $R_v(t)=\{u\in N_H(v):\norm{x_u(t)-x_v(t)}>\trej_v\}$, and $\alpha_{vu},\beta_{vu}\ge0$ may vary with $t$, with $\sum_u\alpha_{vu}\le1$; updates may be asynchronous.

\begin{theorem}[No rejection]\label{thm:norejection}
If an \LCCrej{} $H$ follows~\eqref{eq:sjt}, then $R_v(t)=\emptyset$ for all $v\in H$ and $t\ge0$, and $\diam(t)$ never increases; consensus need not occur.
\end{theorem}
\begin{proof}
Suppose $\diam(t)\le\tmin(H)$, which holds at $t=0$.
Then every edge satisfies $\norm{x_u(t)-x_v(t)}\le\trej_v$, so $R_v(t)=\emptyset$, and the update\iffull{} $x_v(t+1)=(1-\sum_u\alpha_{vu})x_v(t)+\sum_u\alpha_{vu}x_u(t)$\fi{} is a convex combination; \cref{lem:hull} gives $\diam(t+1)\le\diam(t)$, and induction completes the proof.
Finally, split $K_{2k+2}$ equally between positions $0$ and $1$ with $\tacc=1/2$ and $\trej=1$: this \LCCrej{} has only noncommittal cross pairs and never moves.
\end{proof}

\begin{proposition}[Tightness]\label{prop:tight}
Fix $k\ge1$, $\theta>0$ and $\varepsilon>0$.
(a)~There is a connected group with minimum degree at least $k$, diameter $\theta+\varepsilon$ and uniform tolerance $\theta$ in which acceptance and consensus fail under~\eqref{eq:hk}; with $\tacc=\theta/2$, $\trej=\theta$ and $\beta>0$, rejection occurs under~\eqref{eq:sjt} and the diameter grows.
(b)~There is a connected group with minimum degree at least $k$ in which every pair satisfies $d(x_u,x_v)\le\min(\theta_u,\theta_v)$ and every member $w$ but two satisfies $\diam(H)\le\theta_w$, but in which acceptance and consensus fail under~\eqref{eq:hk}; with $\tacc=\trej=\theta$ and $\beta>0$, rejection occurs under~\eqref{eq:sjt}.
\end{proposition}
\begin{proof}
(a)~Split $K_{2k+2}$ equally between positions $0$ and $\theta+\varepsilon$ on a line.
Under~\eqref{eq:hk}, no cross pair is accepted, so both halves stay where they are; under~\eqref{eq:sjt}, every cross pair is rejected, and the first update pushes the halves apart.

(b)~Let $\ell=\max(k,3)$.
Place cliques $H_0$ and $H_2$ of $\ell$ vertices each at positions $0$ and $2$ with tolerance $2$, and two adjacent vertices $u$ and $v$ at position $1$ with tolerance $1$, where $u$ is adjacent to all of $H_0$ and $v$ to all of $H_2$.
\iffull
Every pair is within the smaller of its two tolerances, every vertex has degree at least $\ell\ge k$, and only $u$ and $v$ have tolerance below $\diam(H)=2$.
With equal weights, one update of~\eqref{eq:hk} moves $H_0$ to $1/(\ell+1)$, $u$ to $2/(\ell+2)$, $v$ to $2-2/(\ell+2)$ and $H_2$ to $2-1/(\ell+1)$.
Then $\norm{x_u-x_v}=2-4/(\ell+2)>1$, so $u$ and $v$ no longer accept each other, and the vertices of $H_0\cup\{u\}$, like those of $H_2\cup\{v\}$, accept only one another.
Each part therefore stays in the convex hull of its own positions at $t=1$, and the two hulls are more than $1$ apart; by the argument of \cref{thm:consensus}(b), each part converges to its own limit.
With $\tacc=\trej$ and $\alpha_{vu}=1/(|A_v(t)|+1)$, the first update of~\eqref{eq:sjt} coincides with that of~\eqref{eq:hk}, after which $u$ and $v$ reject each other.
\else
With equal weights, one update separates $u$ and $v$ by $2-4/(\ell+2)>1$; then $H_0\cup\{u\}$ and $H_2\cup\{v\}$ converge to distinct points, and under~\eqref{eq:sjt} with $\tacc=\trej$, $u$ and $v$ reject each other.
\fi
\end{proof}

Part~(b), illustrated in \cref{fig:problem} for $\ell=3$, explains the minimum in \cref{def:lcc}: even with pairwise acceptance, exempting two members breaks the guarantees.
\iffull
A single member below the bound already breaks acceptance.
Give $H_2$ one more vertex, and replace $u$ and $v$ by one vertex $w$ at position $1$ with tolerance $1$, adjacent to all of $H_0$ and to one vertex $z\in H_2$; one update moves $w$ to $3/(\ell+2)$ and $z$ to $(2\ell+3)/(\ell+2)$, which are $2\ell/(\ell+2)>1$ apart.
However, consensus survives a single exemption: every other member accepts all its neighbours at all times, so $w$ is a root of every interaction graph, and consensus follows~\cite{moreau2005stability}.
\else
A single member $w$ below the bound already breaks acceptance, by a variant of this construction~\cite{techreport}, but not consensus: every other member accepts all its neighbours at all times, so $w$ is a root of every interaction graph, and consensus follows~\cite{moreau2005stability}.
\fi

The counterexamples in \cref{prop:tight} motivate an exact characterization across static and dynamic regimes: whereas the pairwise condition suffices for every member to accept static compromises when positions remain fixed, condition~(iii) is necessary and sufficient to protect a connected group against every positive weighting of~\eqref{eq:hk}.

\begin{proposition}[Characterization]\label{prop:char}
Let $H\subseteq V$ with $|H|\ge2$ and $\theta=\tacc$.
(a)~Every pair $u,v\in H$ satisfies $\norm{x_u-x_v}\le\min(\theta_u,\theta_v)$ if and only if every member of $H$ accepts every point of $\conv\{x_u:u\in H\}$.
(b)~If $G[H]$ is connected, then $\diam(H)\le\tmin(H)$ if and only if, for every $\eta>0$ and all weights admitted by~\eqref{eq:hk}, every interaction along every edge of $G[H]$ is accepted at all times.
Moreover, if $\diam(H)>\tmin(H)$ and all these interactions are accepted at $t=0$, then for every sufficiently small $\eta>0$ some admitted weights make a member refuse, within $2|H|$ steps, an interaction that it accepted at $t=0$.
\end{proposition}
\iffull\begin{proof}\else\begin{proof}[Proof sketch]\fi
\iffull
(a)~For fixed $v$, the function $p\mapsto\norm{p-x_v}$ is convex, so its maximum over the convex hull is attained at one of the points $x_u$.
Hence $v$ accepts the whole hull if and only if $\norm{x_u-x_v}\le\theta_v$ for all $u\in H$; requiring this for every $v$ gives the pairwise condition.

(b)~The proof of \cref{thm:consensus}(a) requires only that each update be a convex combination, so it applies to all admitted weights and establishes the forward direction.
For the converse, assume $\diam(H)>\tmin(H)$ and that every interaction along an edge of $G[H]$ is accepted at $t=0$, since otherwise there is nothing to prove; let $D=\diam(H)$ and $0<\delta\le\frac12$.
In a \emph{copy step} along an edge $(z,y)$, member $y$ puts weight $1-\delta$ on $z$ and spreads $\delta$ evenly over its other accepted neighbours and itself, while every other member puts weight $1-\delta$ on itself and spreads $\delta$ evenly over its accepted neighbours, or keeps weight $1$ if it accepts none.
Every weight on an accepted neighbour or on the member itself is at least $\delta/|H|$, so these weights are admitted for every $\eta\le\delta/|H|$, provided that $y$ still accepts $z$; if it does not, an interaction accepted at $t=0$ has been refused.
The same holds if any member refuses any neighbour during the steps below, so assume that this never happens.
By \cref{lem:hull}, the diameter stays at most $D$; hence in a copy step $y$ moves to within $\delta D$ of the position of $z$, and every other member moves by at most $\delta D$.
\emph{Copying along a path} $p_0,\dots,p_L$, that is, performing copy steps along $(p_0,p_1),\dots,(p_{L-1},p_L)$ in this order, therefore brings $p_L$ to within $L\delta D$ of where $p_0$ started, and moves every member outside $\{p_1,\dots,p_L\}$ by at most $L\delta D$.

If some pair $a,b\in H$ has $\norm{x_a-x_b}>\theta_a$, then $a$ and $b$ are not adjacent.
A shortest path from $b$ to $a$, without its last vertex $a$, ends at a neighbour $y$ of $a$ and has length $1\le L<|H|$.
Copying along it brings $y$ to within $L\delta D$ of $x_b$, while $a$ moves by at most $L\delta D$, so $a$ refuses $y$ at $t=L$ once $2L\delta D<\norm{x_a-x_b}-\theta_a$.

Otherwise the pairwise condition holds.
Let $\theta_w=\tmin(H)$ and $\norm{x_u-x_v}=D$; then $w\notin\{u,v\}$, because $\norm{x_w-x_y}\le\theta_w<D$ for all $y\in H$.
There is a path from $u$ to $w$ that avoids $v$, or one from $v$ to $w$ that avoids $u$: if $v$ lies on every path from $u$ to $w$, then every path from $v$ to $w$ continues inside the component of $G[H]-v$ that contains $w$, and $u$ is not in this component.
Swapping $u$ and $v$ if necessary, copy along a path from $u$ to $w$ that avoids $v$, and then along a shortest path from $v$ to $w$ without its last vertex $w$; the latter ends at a neighbour $y$ of $w$, possibly $y=v$.
After these $T<2|H|$ steps, $w$ is within $T\delta D$ of $x_u$ and $y$ within $T\delta D$ of $x_v$, so $\norm{x_w(T)-x_y(T)}\ge(1-2T\delta)D>\theta_w$ once $2T\delta D<D-\theta_w$: member $w$ refuses $y$, which it accepted at $t=0$.
\else
(a)~The distance to $x_v$ is convex, so its maximum over the hull is attained at some $x_u$.
(b)~The forward direction follows from the proof of \cref{thm:consensus}(a), which holds for all admitted weights.
For the converse, suppose that every interaction is accepted at $t=0$, and let one member at a time put weight $1-\delta$ on a neighbour, so that it nearly copies that neighbour's opinion while all others nearly keep theirs; these weights are admitted for every $\eta\le\delta/|H|$.
If the pairwise condition fails for a pair $a,b$, they are not adjacent; copy $x_b$ along a path to a neighbour of $a$.
Otherwise let $\norm{x_u-x_v}=\diam(H)>\theta_w=\tmin(H)$; copy $x_u$ to $w$ along a path that avoids $v$ (swapping $u$ and $v$ if needed), and then $x_v$ to a neighbour of $w$ along a path that avoids $w$.
In both cases, for small $\delta$, a member refuses a neighbour that it accepted at $t=0$ within fewer than $2|H|$ steps.
The technical report~\cite{techreport} gives the full proof.
\fi
\end{proof}

Part~(b) is a worst-case statement over all weightings. Its converse requires weights that may be arbitrarily uneven: in the proof, each moving member puts almost all its weight on one neighbour. Under equal weights, a group that violates~(iii) may or may not lose an accepted interaction (\cref{prop:tight}(b)). For instance, let $u,w,v$ form a triangle in $G$, at positions $0$, $1/2$ and $1$ on a line, with tolerances $1$, $0.6$ and $1$: every pair satisfies the pairwise condition and (iii) fails, but with $\eta=\frac13$ every update is the equal-weight one, so the group meets at its centroid after one step and no interaction is ever refused.
The proof requires only a connected interaction graph, so~(b) also holds when all members hear one another.
With $\tacc=\trej$, the same steps are admitted by~\eqref{eq:sjt}, where a refusal is a rejection, so (iii) is also exactly what rules out rejection under~\eqref{eq:sjt} for all coefficients\iffull; this needs $\tacc=\trej$, since, for instance, with $\tacc\equiv0.4$ in the triangle above every pair is noncommittal and nobody ever moves\fi.
\Cref{sec:exp:dynamics} measures how often other methods' groups lose an accepted interaction under equal and uneven weights.

\paragraph{Open groups}
When members average with accepted neighbours outside the group, condition~(iii) alone no longer suffices; \cref{prop:open} supplies an additional sufficient condition.

\begin{proposition}[Open groups]\label{prop:open}
Let each $v\in H$ also interact with its outside neighbours $O_v=N_G(v)\setminus H$, whose positions $x_o$ stay fixed, and let every update move $x_v$ to a convex combination, with non-negative weights that may vary with time and state, of $x_v(t)$ and the in-group and outside neighbours that $v$ accepts at time~$t$.
Let $Y_0$ be the set of outside positions accepted by some member at $t=0$, and $K_0=\conv(\{x_v(0):v\in H\}\cup Y_0)$.
If (a)~$\diam(K_0)\le\tmin(H)$ and (b)~$\norm{x_o-p}>\theta_v$ for all $p\in K_0$, $v\in H$ and $o\in O_v$ with $x_o\notin Y_0$, then $x_v(t)\in K_0$ for all $v$ and~$t$, and at every time each member accepts exactly the neighbours it accepted at $t=0$.
\end{proposition}
\iffull\begin{proof}\else\begin{proof}[Proof sketch]\fi
\iffull
We prove by induction on $t$ that $x_u(t)\in K_0$ for all $u\in H$; the claim holds at $t=0$.
Assume it holds at time $t$ and fix $v$.
If $o\in O_v$ and $x_o\notin Y_0$, then $\norm{x_v(t)-x_o}>\theta_v$ by~(b), so $v$ does not accept $o$; every other point that $v$ can average, namely $x_v(t)$, the positions $x_u(t)$ of its in-group neighbours and the outside positions in $Y_0$, lies in the convex set $K_0$, so $x_v(t+1)\in K_0$.
Any two points of $K_0$ are within $\diam(K_0)\le\tmin(H)\le\theta_v$, so at every time $v$ accepts all its in-group neighbours and all outside neighbours with positions in $Y_0$, and no others; by the definition of $Y_0$, these are exactly the neighbours it accepts at $t=0$.
As a point set and its convex hull have the same diameter (\cref{lem:hull}), (a) can be checked on the finite set $\{x_v(0)\}_{v\in H}\cup Y_0$.
The argument requires neither connectivity nor positive weights nor synchronous updates, but it does require fixed outside positions.
\else
By induction, all members stay in $K_0$: by~(b), no member accepts an outside neighbour whose position is not in $Y_0$, so every update is a convex combination of points of $K_0$, and any two points of $K_0$ are within $\tmin(H)$.
By \cref{lem:hull}, (a) can be checked on the finite set $\{x_v(0)\}_{v\in H}\cup Y_0$.
\fi
\end{proof}

\iffull
The fixed outside neighbours are the constant signals of Hegselmann and Krause~\cite{hegselmann2015opinion}; what matters is the certificate.
The condition of Mirtabatabaei and Bullo~\cite[Prop.~4.11]{mirtabatabaei2012opinion}, under which an isolated all-to-all group with scalar opinions reaches consensus if all members but one have bounds larger than its opinion range, permits one exemption but no outside neighbours, the reverse of \cref{prop:open}; neither condition therefore implies the other. In our network model, two exemptions are too many by \cref{prop:tight}(b).
\else
Mirtabatabaei and Bullo's consensus condition~\cite[Prop.~4.11]{mirtabatabaei2012opinion} permits one exemption but no outside neighbours, the reverse of \cref{prop:open}, so neither implies the other.
\fi
\iffull
Condition~(a) of \cref{prop:open} alone does not suffice, even with equal weights, because members can begin to accept outside neighbours that none accepted at $t=0$ and drift apart, an instance of the indirect influence described in~\cite{hegselmann2015opinion}.
For example, let adjacent members $a$ and $b$ start at $0$ and $\frac12$ with $\theta\equiv1$, let $a$ have 20 outside neighbours at $0$, and $b$ have 10 at $\frac{19}{20}$ and 10 at $\frac85$.
Then $\diam(\{x_a,x_b\}\cup Y_0)=\frac{19}{20}\le1$, but (b) fails; under equal-weight HK, $b$ starts to accept the neighbours at $\frac85$ after one step, and at $t=2$ the two members are $\frac{51}{44}>1$ apart and refuse each other.
The argument needs outside positions to stay fixed, and open groups need not reach consensus: if $a$ has 20 outside neighbours at $0$ and $b$ has 10 at $0.95$, with $\theta\equiv1$ and members at $0$ and $\frac12$, the two members converge to $\frac{19}{460}$ and $\frac{399}{460}$ under equal weights, so the members' diameter grows from $\frac12$ to $\frac{19}{23}$ while $K_0$ stays fixed.
\else
Examples in~\cite{techreport} show that open groups need not reach consensus and that condition~(a) of \cref{prop:open} alone does not suffice, even with equal weights: members can begin to accept outside neighbours that none accepted at $t=0$ and drift apart, an instance of the indirect influence described in~\cite{hegselmann2015opinion}.
\fi
\Cref{sec:exp:dynamics} evaluates how often conditions~(a) and~(b) hold in real networks and how groups behave when external weights vary.

\subsection{Complexity}\label{sec:theory:complexity}

We study \LCCDec{} in $\R^m$ with the Euclidean norm and rational coordinates and tolerances; in parts (a), (b) and~(f), $m$ is part of the input and we use the following embedding.

\begin{lemma}\label{lem:embed}
Given a graph $F=(U,E_F)$ with $|U|=n$, points $y_u\in\{-1,0,1\}^m$ with $m=O(n^2)$ and an integer $r$ can be computed in polynomial time such that, for distinct $u,w\in U$, $\norm{y_u-y_w}\le r$ if and only if $\{u,w\}\in E_F$.
\end{lemma}
\iffull
\begin{proof}
Let $\bar F$ denote the complement of $F$, and let $\Delta$ denote its maximum degree.
Take the least positive integer $j$ satisfying $2j^2\ge\Delta$, and put $D=2j^2$.
For every edge $\{u,w\}$ in $\bar F$ ordered by $u<w$, introduce a coordinate assigning $y_u=+1$, $y_w=-1$, and every other point the value $0$.
For each $u$, introduce another $D-\deg_{\bar F}(u)$ private coordinates with value $1$.
This gives $\norm{y_u}^2=D$ for every $u$; distinct points can share a non-zero coordinate only when they are adjacent in $\bar F$.
Thus $\langle y_u,y_w\rangle$ is $-1$ when $\{u,w\}\notin E_F$ and $0$ when $\{u,w\}\in E_F$.
Thus $\norm{y_u-y_w}^2$ takes value $2D+2$ in the former case and $2D$ in the latter, so the threshold $r=2j$ distinguishes them.
\end{proof}
\else
\begin{proof}[Proof sketch]
Give each edge of $\bar F$ a coordinate that is $+1$ at one end, $-1$ at the other and $0$ elsewhere, and pad with private unit coordinates to $\norm{y_u}^2=D=2j^2$, for the least $j$ with $2j^2\ge\Delta(\bar F)$; squared distances are then $2D$ on edges of $F$ and $2D+2$ otherwise, separated by $r=2j$.
\end{proof}
\fi

\begin{theorem}[Hardness]\label{thm:hardness}
\begin{itemize}
\item[(a)] For each fixed $k\ge1$, \LCCDec{} is NP-complete.
This holds even for complete $G$, equal tolerances, and coordinates restricted to $\{-1,0,1\}$.
\item[(b)] For every fixed $\varepsilon\in(0,1)$, unless $\mathrm{P}=\mathrm{NP}$, no polynomial-time algorithm can approximate \MaxLCC{} to within a factor of $n^{1-\varepsilon}$.
\item[(c)] The conclusions of (a) and (b) extend to general metrics supplied as distance matrices.
\item[(d)] When $m=1$, \MaxLCC{} can be solved in $O(n^2(n+|E|))$ time.
\item[(e)] For each fixed $k\ge1$ and $m\ge2$, \LCCDec{} remains NP-complete, also on the sphere with chord distance, even with equal tolerances.
\item[(f)] For each fixed $k\ge2$, deciding whether an LCC exists is NP-complete, even with equal tolerances, both for distances in $\{1,2\}$ and in $\R^m$ with $m$ part of the input.
Hence, for each fixed $k\ge2$, unless $\mathrm{P}=\mathrm{NP}$, no polynomial-time algorithm returns an LCC for every input that has one; in particular, \MaxLCC{} cannot be approximated within any factor.
\end{itemize}
\end{theorem}
\iffull\begin{proof}\else\begin{proof}[Proof sketch]\fi
\iffull
Feasibility can be checked by comparing squared distances and squared tolerances using rational arithmetic, so membership in NP follows directly.

(a)~The reduction is from \Clique{}, which is NP-complete~\cite{karp1972reducibility}.
For an instance $(F,c)$, add a clique $K$ of $k+1$ new vertices, including $q$, and make every vertex of $K$ adjacent to every vertex of $U$.
Apply the embedding of \cref{lem:embed} to this augmented graph, take $G$ to be complete, assign tolerance $r$ to every vertex, and set $s=c+k+1$.
Any $c$-clique $Q$ in $F$ gives the set $Q\cup K$, whose pairwise distances are at most $r$ and whose minimum degree in $G$ is $c+k\ge k$.
Conversely, every pair of vertices in an LCC has distance at most $r$; by \cref{lem:embed}, these vertices form a clique in the augmented graph, and deleting $K$ leaves a clique in $F$ with at least $c$ vertices.

(b)~Embed $F$ itself by \cref{lem:embed}, with complete $G$ and tolerance $r$.
Checking all sets of at most $k+1$ vertices in $n^{O(k)}$ time either finds a maximum clique, if its size $\omega$ is at most $k$, or shows $\omega>k$.
In the latter case, apply a $\rho$-approximation algorithm to each query $q\in U$ that lies in a $(k+1)$-clique.
If $q$ belongs to a maximum clique, the corresponding optimum equals $\omega$.
Every LCC returned by the algorithm is a clique in $F$, and the largest returned set therefore contains at least $\omega/\rho$ vertices.
This gives a polynomial-time approximation for \Clique{} within $\rho$; taking $\rho=n^{1-\varepsilon}$ would imply $\mathrm{P}=\mathrm{NP}$~\cite{zuckerman2007linear}.

(c)~The embedding would give irrational distances, so we bypass it: set $d(u,w)=1$ if $\{u,w\}$ is an edge of the graph being embedded and $d(u,w)=2$ for every other pair $u\ne w$, and give every vertex tolerance $1$.
Any symmetric matrix with zero diagonal and off-diagonal entries in $\{1,2\}$ satisfies the triangle inequality, and the arguments of (a) and (b) apply unchanged.

(d)~Take an optimal solution $H^*$ and define $\tau^*=\tmin(H^*)$ and $a^*=\min_{v\in H^*}x_v$.
The inequality $\diam(H^*)\le\tau^*$ places $H^*$ inside the window $[a^*,a^*+\tau^*]$.
For each pair $(a,\tau)\in\{x_u\}\times\{\theta_u\}$, evaluate $C(a,\tau)=\Ck(S_{a,\tau},q)$, where $S_{a,\tau}=\{v:\theta_v\ge\tau,\ a\le x_v\le a+\tau\}$.
Every non-empty $C(a,\tau)$ is an LCC: its positions occupy a window of length $\tau$, and each tolerance is at least $\tau$.
The $k$-core fact gives $H^*\subseteq C(a^*,\tau^*)$, establishing optimality of the largest $C(a,\tau)$.
The enumeration considers $n^2$ pairs and takes $O(n+|E|)$ time for each.

(e)~We reduce from 3-SAT~\cite{karp1972reducibility} with variables $x_1,\dots,x_\nu$, $\nu\ge2$, and clauses $c_1,\dots,c_\mu$.
Let $t_i=i/(\nu+1)$ and $p_i=\big((1-t_i^2)/(1+t_i^2),\,2t_i/(1+t_i^2)\big)$, a rational point of the unit circle at angle $\phi_i=2\arctan t_i\in(0,\pi/2)$.
Place a clique $K$ of $k+1$ vertices, including $q$, at the origin; for each $i$, place $k$ \emph{copies} of the literal $x_i$ at $p_i$ and $k$ copies of $\neg x_i$ at $-p_i$; and place one vertex per clause at the origin.
Every copy is adjacent to all of $K$, and each clause vertex is adjacent to all copies of its literals.
Copies of $x_i$ and $\neg x_i$ are at distance $2$; every other pair is closer than $r=2-1/(16(\nu+1)^2)$.
Indeed, pairs involving the origin and pairs of copies of the same literal are at distance at most $1$.
Copies of literals of distinct variables $x_i$ and $x_j$ are at distance $2\sin(\alpha/2)$ or $2\cos(\alpha/2)$, where $\alpha=|\phi_i-\phi_j|<\pi/2$; both are at most $2\cos(\delta/2)$, where $\delta\le\alpha$ is the smallest difference between distinct angles $\phi_i$.
Since the derivative of $2\arctan t$ is at least $1$ on $[0,1]$ and consecutive values $t_i$ differ by $1/(\nu+1)$, we have $\delta\ge1/(\nu+1)$; with $y=\delta/2<1$ and $\cos y\le1-y^2/2+y^4/24\le1-11y^2/24$, this gives $2\cos(\delta/2)\le2-11\delta^2/48<r$.
With tolerance $r$ for every vertex, an LCC contains copies of at most one literal per variable, and therefore at most $s=k+1+k\nu+\mu$ vertices.
An LCC with $s$ vertices contains $K$, all copies of one literal per variable, and every clause vertex; each clause vertex then needs $k$ neighbours, that is, a chosen literal, so the chosen literals satisfy the formula.
Conversely, the true literals of a satisfying assignment, together with $K$ and all clause vertices, form an LCC with $s$ vertices.
On the sphere with chord distance, put the copies on the equator and the other vertices at the north pole, at distance $\sqrt2<r$ from every copy; the argument is unchanged.
Padding the coordinates with zeros extends the planar construction to every fixed $m\ge2$.

(f)~Membership in NP is as above.
We reduce from \MCClique: given a graph $F=(U,E_F)$ whose vertex set is partitioned into classes $U_1,\dots,U_m$ with $m\ge2$, decide whether some $u_1\in U_1,\dots,u_m\in U_m$ are pairwise adjacent.
It is NP-complete, since an instance $(F',c)$ of \Clique{} with $c\ge2$ reduces to it by taking $U_i=V(F')\times\{i\}$ for $i\le c$ and joining $(a,i)$ and $(b,j)$ if and only if $i\ne j$ and $\{a,b\}\in E(F')$.
Add an original vertex $\hat q$ and arrange the classes $\{\hat q\},U_1,\dots,U_m$ on a ring, so that the \emph{ring neighbours} of $U_i$ are $U_{i-1}$ and $U_{i+1}$, with $U_0=U_{m+1}=\{\hat q\}$.
Replace every original $a\in U\cup\{\hat q\}$ by $k-1$ copies, let $q$ be a copy of $\hat q$, and join two copies in $G$ if and only if their originals lie in ring-neighbouring classes.
Call two originals \emph{compatible} if they are equal, if one of them is $\hat q$, or if they lie in different classes and are adjacent in $F$.
Put copies of compatible originals at distance $1$ and all other pairs at distance~$2$, with tolerance $1$ for every vertex; for Euclidean distances, embed the compatibility graph on $U\cup\{\hat q\}$ by \cref{lem:embed}, put every copy at the point of its original, and give every vertex tolerance $r$.
In both cases, a set satisfies~(iii) if and only if its members are copies of pairwise compatible originals.
If $u_1,\dots,u_m$ form a multicolored clique, all copies of $\hat q,u_1,\dots,u_m$ form an LCC: the subgraph they induce is connected, each of its vertices has $2(k-1)\ge k$ neighbours in it, and their originals are pairwise compatible.
Conversely, let $H$ be an LCC.
Distinct originals in the same class are incompatible, so the members of $H$ among the copies of a class are copies of a single original.
The neighbours in $G[H]$ of a member therefore lie in its two ring-neighbouring classes, at most $k-1$ in each, and minimum degree $k$ forces both of these classes to contribute members.
Starting from $q$, whose ring neighbours are $U_1$ and $U_m$, and following the ring, every class $U_i$ contributes copies of some $u_i$; these originals are pairwise compatible and lie in different classes, so they form a multicolored clique.
Finally, if a polynomial-time algorithm returned an LCC for every input that has one, checking whether its output is an LCC would decide existence in polynomial time.
\else
(a)~Join a $(k+1)$-clique containing $q$ to a \Clique{} instance $F$~\cite{karp1972reducibility}, embed the result by \cref{lem:embed}, and take $G$ complete with all tolerances $r$: LCCs with $c+k+1$ vertices correspond to $c$-cliques of $F$.
(b)~Embedding $F$ itself, a $\rho$-approximation run from every $q$ approximates maximum clique within $\rho$~\cite{zuckerman2007linear}.
(c)~Use distance $1$ on edges and $2$ elsewhere.
(d)~Peel the vertices of tolerance at least $\tau$ in each window $[a,a+\tau]$.
(e)~Reduce from 3-SAT~\cite{karp1972reducibility} with $k$ copies of each literal at antipodal rational points of the unit circle and tolerance just below $2$, so that only complementary literals conflict.
(f)~Reduce from \MCClique: place $q$ and the colour classes on a ring, join copies of vertices in consecutive classes, call two vertices compatible exactly when they are equal, one of them is $q$, or they lie in different classes and are adjacent in $F$, and replace every vertex by $k-1$ pairwise non-adjacent copies.
Each class can then contribute copies of only one vertex, so a member has at most $k-1$ neighbours on either side, and minimum degree $k$ forces an LCC around the whole ring that picks one vertex per class.
Full proofs, including that of \cref{lem:embed}, appear in~\cite{techreport}.
\fi
\end{proof}

Parts (a) and~(b) transfer the hardness of \Clique{} through \cref{lem:embed}, a simple instance of representing graphs by distances as in the study of sphericity~\cite{maehara1984space}, and (c) through distances in $\{1,2\}$; (e) and~(f) require new constructions.
With uniform tolerances, (a) and~(e) also show that the query version of the $(k,r)$-core problem~\cite{zhang2017engagement} is NP-hard under Euclidean distance, even in the plane, and by~(f), for each fixed $k\ge2$, even deciding whether $q$ lies in any $(k,r)$-core is NP-complete when the dimension is part of the input.
\iffull
The assumption that $k$ is fixed matters: if $k$ is part of the input, the construction of~(a) with $K=\{q\}$ and $k=c$ already makes existence NP-complete.
\fi
Parts~(b) and~(f) motivate relaxing the diameter, but not the size, in \cref{sec:algorithm:ballcore}: by~(f), for each fixed $k\ge2$ and unless $\mathrm{P}=\mathrm{NP}$, every polynomial-time algorithm that returns only LCCs, such as BallCore-F and the greedy variant, returns nothing on some inputs that have an LCC, whereas BallCore-R is empty only if no LCC exists.
\iffull
Part~(b) is not subsumed by~(f): it also holds for $k=1$, and its proof gives the same bound even on inputs with an LCC that is easy to find, a $(k+1)$-clique of $F$ containing $q$.
\else
Part~(b), unlike~(f), also holds for $k=1$.
\fi
\paragraph{Open cases}
It remains open whether \MaxLCC{} is hard to approximate in dimensions~2 and~3 and whether, for $k\ge2$, deciding the existence of an LCC is hard in fixed dimension; on the sphere with chord distance, as in our meetup data, only~(e) applies.
\iffull
If $G$ is complete, \cref{lem:sweep} reduces \MaxLCC{} to finding, for each candidate threshold, a maximum clique containing $q$ in a unit ball graph.
This is polynomial in the plane~\cite{clark1990unit}, even with member-specific tolerances, so the hardness in~(e) stems from the graph; in dimension~4 and above, it is NP-hard even for complete $G$~\cite{bonamy2021eptas}, while these results leave dimension~3 unresolved.
\else
With complete $G$, \MaxLCC{} is polynomial in the plane, even with member-specific tolerances~\cite{clark1990unit}; the hardness in~(e) stems from the graph.
\fi

%% file: sections/05_algorithm.tex
\section{The LatSweep Framework}\label{sec:algorithm}

\begin{figure}[t]
\centering
\input{figures/fig_latsweep}
\caption{The LatSweep framework: (a)~survival intervals and the candidates $S(\tau)$ (dots) at two thresholds; (b)~index, query pipeline and the bounds $\beta(\tau)$ that the sweep refines.}\label{fig:latsweep}
\end{figure}
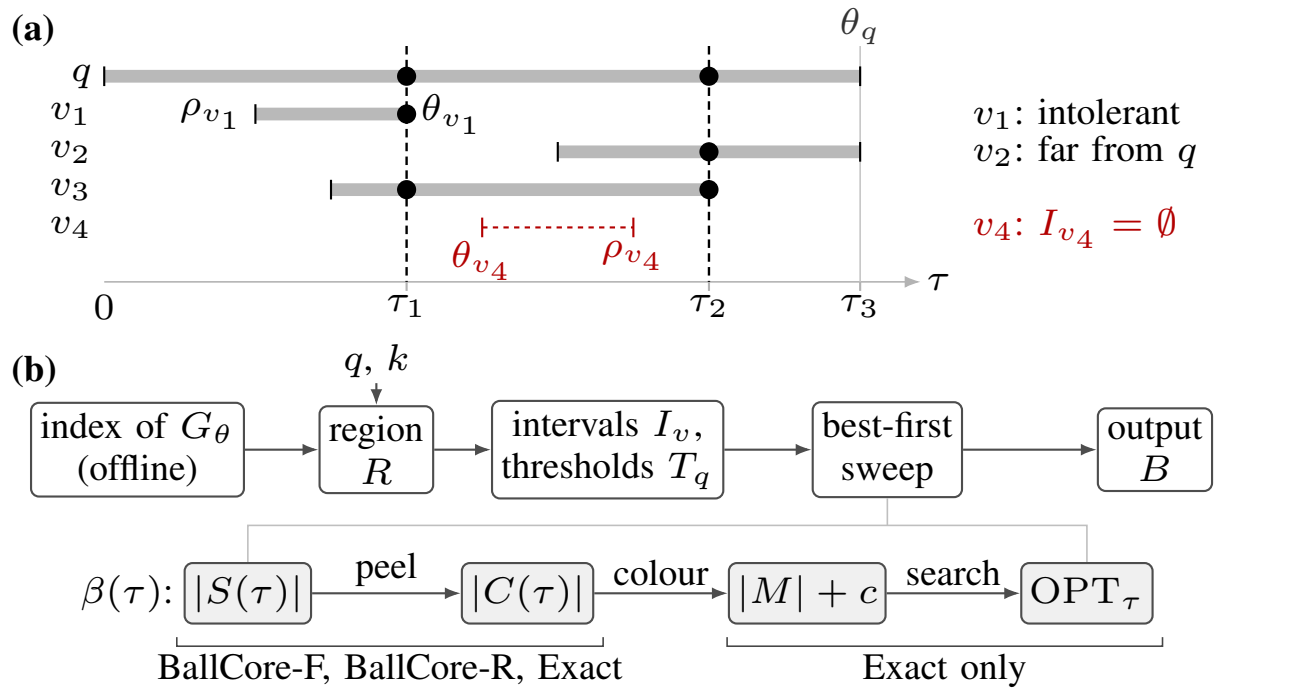

LatSweep makes the coupled constraint~(iii) searchable by fixing the smallest tolerance $\tau$. Each value yields a fixed-threshold subproblem in a reduced region around $q$ (\cref{fig:latsweep}).
The sweep yields four algorithms: BallCore-F and a greedy variant return only LCCs, BallCore-R is a polynomial-time $(1,2)$-bicriteria approximation, and LatSweep-Exact (\emph{Exact}) is exact and, if stopped early, returns the best LCC found, if any, with an upper bound on the optimum.
For a fixed threshold, the subproblem is a search for a maximum $(k,r)$-core containing $q$~\cite{zhang2017engagement}, which we solve with known tools: peeling, filtering of conflicts (pairs farther apart than the threshold) and colouring bounds~\cite{tomita2003efficient}.
Survival intervals give each threshold its candidate set; these sets are not nested. Best-first refinement then allocates search across thresholds and supports early stopping with a certificate; a compatibility index reduces the search region.
We write $\rho_v=d(x_v,x_q)$ and let $\OPT$ be the size of a largest LCC, or $0$ if none exists; all results hold in arbitrary metric spaces.

\subsection{Survival Intervals and Candidate Reduction}\label{sec:algorithm:sweep}

\begin{lemma}[Sound reductions]\label{lem:reduce}
Let $H$ be an LCC and $\tau=\tmin(H)$.
(a)~Every $v\in H$ satisfies $\rho_v\le\tau\le\theta_v$, and hence $\rho_v\le\min(\theta_v,\theta_q)$.
(b)~Let $\Gc=(V,E_\theta)$ with $E_\theta=\{(u,v)\in E:d(x_u,x_v)\le\min(\theta_u,\theta_v)\}$.
Then $G[H]=\Gc[H]$; hence $H$ lies in the component of the $k$-core of $\Gc$ that contains $q$.
\end{lemma}
\begin{proof}
(a)~As $q\in H$, we have $\rho_v\le\diam(H)\le\tmin(H)=\tau\le\theta_v$, and $\tau\le\theta_q$.
(b)~Every edge $(u,v)$ inside $H$ satisfies $d(x_u,x_v)\le\diam(H)\le\tmin(H)\le\min(\theta_u,\theta_v)$; the $k$-core fact applied to $\Gc$ completes the proof.
\end{proof}

Lemma~\ref{lem:reduce}(a) assigns each vertex $v$ a \emph{survival interval} $I_v=[\rho_v,\min(\theta_v,\theta_q)]$ for the smallest tolerance of an LCC containing it. A vertex with an empty interval, such as $v_4$ in \cref{fig:latsweep}(a), cannot belong to an LCC.
The core numbers of $\Gc$ form a \emph{compatibility index}; it is independent of the query and is built once per tolerance assignment (say, for $\tacc$ and for $\trej$) in $O(t_d|E|+n)$ time, where $t_d$ is the cost of one distance evaluation.
For a query and a graph $G'\in\{G,\Gc\}$, the \emph{query region} $R$ is obtained by a traversal from $q$ along edges of $G'$ that visits only vertices with a non-empty survival interval and core number at least $k$ in $G'$.
By \cref{lem:reduce}, every LCC lies in $R$ for either choice of $G'$, and building $R$ inspects only $R$ and its neighbours; BallCore-R uses $G'=G$, and hence the core numbers of $G$ rather than the index, all other variants $G'=\Gc$.

\paragraph{Candidate thresholds}
For $\tau>0$, let $S(\tau)=\{v\in R:\tau\in I_v\}$ and let $C(\tau)=\Ck(S(\tau),q)$, peeled in $G'$.
Let $\Tq=\{\min(\theta_v,\theta_q):v\in R\}$, and let $\OPT_\tau$ be the largest size of a connected set $H\subseteq S(\tau)$ with $q\in H$, minimum degree at least $k$ and $\diam(H)\le\tau$.

\begin{lemma}\label{lem:sweep}
$\OPT=\max_{\tau\in\Tq}\OPT_\tau$, and $\OPT_\tau\le|C(\tau)|\le|S(\tau)|$ for all $\tau$.
\end{lemma}
\begin{proof}
Every set counted in $\OPT_\tau$ has all tolerances at least $\tau\ge\diam(H)$ and is therefore an LCC.
Conversely, an optimal $H^*$ has $\tau^*=\tmin(H^*)\le\theta_q$, so $\tau^*\in\Tq$, and $H^*\subseteq S(\tau^*)$ by \cref{lem:reduce}(a); thus $H^*$ is counted in $\OPT_{\tau^*}$.
The bound $\OPT_\tau\le|C(\tau)|$ is the $k$-core fact.
\end{proof}

\paragraph{Best-first sweep}
\Cref{alg:latsweep} keeps an upper bound $\beta(\tau)$ on $\OPT_\tau$ for every threshold and repeatedly refines the bound of the threshold with the largest one, in stages of increasing cost (\cref{fig:latsweep}(b)): the interval count $|S(\tau)|$, obtained for all thresholds by one sort of the interval end points; the size $|C(\tau)|$ of the peeled set; for Exact, the bound that branch and bound computes at its root (\cref{sec:algorithm:exact}); and finally the search itself, which settles $\OPT_\tau$.
The root bound holds only for solutions larger than the incumbent $B$, which suffices because $B$ never shrinks.
By \cref{lem:sweep}, no threshold can improve the incumbent once every remaining bound is at most $|B|$.
Ordering by $|S(\tau)|$ alone can prioritize expensive subproblems that have small optima; refining $\beta(\tau)$ instead focuses the search on promising thresholds (\cref{sec:exp:efficiency}).
Unlike with a uniform radius, $S(\tau)$ is not monotone in $\tau$: a larger $\tau$ admits vertices farther from $q$ but drops those with $\theta_v<\tau$; in \cref{fig:latsweep}(a), from $\tau_1$ to $\tau_2$, $S(\tau)$ loses $v_1$ and gains $v_2$.
The sort takes $O(n_q\log n_q)$ time and one peeling $O(n_q+e_q)$, where $n_q=|R|$ and $e_q=|E(R)|$.

\begin{algorithm}[t]
\caption{LatSweep}\label{alg:latsweep}
\begin{algorithmic}[1]
\Require $G$, $x$, $\theta$, query $q$, integer $k$, variant $\in\{\textsf{F},\textsf{R},\textsf{Exact}\}$
\State $\lambda\gets2$ for \textsf{F}, else $1$;\ \ $G'\gets G$ for \textsf{R}, else $\Gc$
\State $R\gets$ query region in $G'$, with $\lambda\rho_v$ for $\rho_v$ \Comment{\cref{lem:reduce}}
\State $I_v\gets[\lambda\rho_v,\min(\theta_v,\theta_q)]$ for $v\in R$
\State $T\gets\{\min(\theta_v,\theta_q):v\in R\}$;\ \ $\beta(\tau)\gets|S(\tau)|$ for $\tau\in T$
\State $B\gets\emptyset$, or for \textsf{Exact} the larger of the BallCore-F and greedy results
\While{$T\ne\emptyset$ and $\max_{\tau\in T}\beta(\tau)>|B|$}
  \State $\tau\gets$ a threshold in $T$ with the largest $\beta(\tau)$
  \If{$\tau$ has not been peeled}
    \State $C(\tau)\gets\Ck(S(\tau),q)$ in $G'$;\ \ $\beta(\tau)\gets|C(\tau)|$
    \If{the variant is \textsf{F} or \textsf{R}}
      \State \textbf{if} $|C(\tau)|>|B|$ \textbf{then} $B\gets C(\tau)$
      \State remove $\tau$ from $T$
    \EndIf
  \ElsIf{$\tau$ has no root bound yet}
    \State $\beta(\tau)\gets$ root bound of $C(\tau)$ \Comment{\cref{sec:algorithm:exact}}
  \Else
    \State $H\gets\textsc{BranchAndBound}(C(\tau),\tau,|B|)$
    \State \textbf{if} $|H|>|B|$ \textbf{then} $B\gets H$
    \State remove $\tau$ from $T$ \Comment{the threshold is settled}
  \EndIf
\EndWhile
\State \Return $B$
\end{algorithmic}
\end{algorithm}

\subsection{BallCore: Feasible and Bicriteria Solutions}\label{sec:algorithm:ballcore}

\paragraph{Feasible solutions}
Variant \textsf{F} doubles the left end points of the survival intervals ($\lambda=2$ in \cref{alg:latsweep}); it is fast but conservative, since it admits only vertices within $\tau/2$ of $q$.
The greedy variant avoids this restriction. It starts from $C(\tau)$, peeled in $\Gc$, at each of the 20 thresholds with the largest $|S(\tau)|$.
It repeatedly deletes the vertex other than $q$ with the most \emph{conflicts}, that is, partners at distance greater than $\tau$, and re-peels after each deletion until no conflict remains; if peeling removes $q$, the threshold yields nothing.

\begin{proposition}[Feasibility]\label{prop:feasible}
Every non-empty output of BallCore-F or of the greedy variant is an LCC.
\end{proposition}
\begin{proof}
By peeling, each output contains $q$, is connected, and has minimum degree at least $k$ and tolerances at least $\tau$.
In BallCore-F, any two members satisfy $d(x_u,x_v)\le\rho_u+\rho_v\le\tau/2+\tau/2=\tau$; in the greedy output, no pair conflicts.
\end{proof}

\paragraph{BallCore-R}
Variant~\textsf{R}, for \emph{relaxed}, omits the doubling and uses the largest peeled set as a bicriteria approximation.
It peels in $G'=G$, so that $C(\tau)$ is the largest connected $k$-core containing $q$ whose members lie within distance $\tau$ of $q$ and have tolerance at least $\tau$.
Between consecutive values of $\Tq$, $S(\tau)$ can only grow. Maximizing over $\tau\in\Tq$ therefore solves exactly the relaxation of \MaxLCC{} that replaces~(iii) by the requirement that $\rho_v\le\tau\le\theta_v$ hold for all members and some $\tau$. Every LCC satisfies this relaxation by \cref{lem:reduce}(a).

\begin{theorem}[Bicriteria approximation]\label{thm:bicriteria}
Let $\hat\tau$ maximize $|C(\tau)|$ over $\tau\in\Tq$, and let $B=C(\hat\tau)$.
Then $|B|\ge\OPT$; in particular, $B$ is empty only if no LCC exists.
If $B$ is non-empty, it contains $q$, is connected, has minimum degree at least $k$, and satisfies $\diam(B)\le2\,\tmin(B)$.
Given the region, BallCore-R finds $B$ in $O(n_q\log n_q+|\Tq|(n_q+e_q))$ time.
\end{theorem}
\begin{proof}
By the proof of \cref{lem:sweep} and the $k$-core fact, an optimal $H^*$ satisfies $H^*\subseteq C(\tau^*)$ for some $\tau^*\in\Tq$, so $|B|\ge|C(\tau^*)|\ge\OPT$.
For any $u,v\in C(\tau)$, we have $d(x_u,x_v)\le\rho_u+\rho_v\le2\tau$, and every tolerance in $C(\tau)$ is at least $\tau$; hence $\diam(C(\tau))\le2\tau\le2\,\tmin(C(\tau))$.
\Cref{alg:latsweep} stops only when every remaining threshold has $|S(\tau)|\le|B|$ and therefore cannot improve $B$.
\end{proof}

The factor $2$ is attained: place $q$ at $0$ and $k$ vertices at each of $-\tau$ and $+\tau$ on a line, take $G$ complete, and give every vertex tolerance $\tau$.
Then $C(\tau)$ contains all $2k+1$ vertices and has diameter $2\tau$, whereas an optimum consists of $q$ and the $k$ vertices on one side.
It is also best possible in general metrics: with distances in $\{1,2\}$ as in \cref{thm:hardness}(c), every set of diameter below $2$ is a clique, so unless $\mathrm{P}=\mathrm{NP}$, no polynomial-time algorithm achieves size ratio $n^{1-\varepsilon}$ with diameter factor $2-\delta$.
BallCore-R relaxes condition~(iii), so its output has none of the dynamical guarantees of \cref{sec:theory}; a completed sweep returns at least $\OPT$ vertices within twice their smallest tolerance, and thus bounds $\OPT$ and certifies absence whenever $B=\emptyset$.

\subsection{LatSweep-Exact}\label{sec:algorithm:exact}

By \cref{lem:sweep}, an optimal LCC is found by solving, at every $\tau\in\Tq$ whose bound exceeds the incumbent, the fixed-threshold subproblem on $C(\tau)$: two vertices \emph{conflict} if their distance exceeds $\tau$, and the goal is a largest conflict-free subset containing $q$ that is connected and has minimum degree at least $k$.
The exact variant peels in $\Gc$, which is sparser and by \cref{lem:reduce}(b) loses no solution, and initializes the incumbent with the BallCore-F and greedy solutions, which are feasible by \cref{prop:feasible}.

\paragraph{Branch and bound}
$\textsc{BranchAndBound}(C,\tau,b)$ returns a largest feasible subset of $C$ if it has more than $b$ vertices, and the empty set otherwise.
A search node $(M,P)$ consists of selected vertices $M\ni q$ and undecided vertices $P$, with no conflicts within $M$ or between $M$ and $P$; the root has $M=\{q\}$ and $P=C(\tau)\setminus\{q\}$, which satisfies this invariant because $\rho_v\le\tau$ for every $v\in C(\tau)$.
Let $b$ denote the size of the incumbent.
Each node first applies rules~(1) and~(2) of \cref{lem:prune} to reduce $M\cup P$.
If no conflict remains in $P$, then $M\cup P$ is feasible and optimal for the subtree; otherwise, the node tests rule~(3) and branches on a vertex $p\in P$ with the most conflicts in $P$, either deleting $p$ or moving it to $M$ and deleting its conflicting partners from $P$.
Together, the two branches retain every solution in the subtree, and both maintain the invariant.
The \emph{root bound} used by \cref{alg:latsweep} is $|M|+c$ at the root after rules~(1) and~(2), where $c\le|P|$ is the colour count of rule~(3); by \cref{lem:prune}, no feasible subset of $C(\tau)$ with more than $b$ vertices exceeds it.
Rules~(2) and~(3) adapt the degree and colouring bounds of maximum-clique search~\cite{tomita2003efficient}.

\begin{lemma}[Sound pruning]\label{lem:prune}
The rules below preserve every feasible solution with more than $b$ vertices in the current node's subtree.
(1)~Delete edges joining conflicting vertices, peel $G[M\cup P]$ down to the component containing $q$ in its $k$-core, and prune the node whenever this removes a vertex of $M$.
(2)~Prune the node when $|M\cup P|\le b$.
If this test fails, examine any vertex having fewer than $b$ conflict-free partners in $M\cup P$: delete it if it belongs to $P$, or prune the node if it belongs to $M$.
Apply rules (1) and (2) alternately until both leave $M\cup P$ unchanged.
(3)~Greedily colour the vertices of $P$, in increasing order of their number of conflicts in $P$, so that every colour class is a set of pairwise conflicting vertices; prune when $|M|$ plus the colour count $c$ does not exceed $b$.
\end{lemma}
\begin{proof}
Take any feasible $H$ satisfying $|H|>b$ and $M\subseteq H\subseteq M\cup P$.
\iffull
(1)~Since $H$ contains no conflicting pair, none of the deleted edges belongs to $G[H]$; the $k$-core fact therefore ensures that $H$ remains after peeling.
(2)~Every $v\in H$ has the remaining $|H|-1\ge b$ vertices of $H$ as conflict-free partners, and all are contained in $M\cup P$.
(3)~$H\setminus M$ is pairwise conflict-free, so it can contain at most one vertex from each colour class.
\else
It uses no deleted edge, so it survives peeling by the $k$-core fact; each of its vertices has its other $|H|-1\ge b$ members as conflict-free partners; and it contains at most one vertex of each colour class.
\fi
\end{proof}

Colouring vertices with few conflicts first tends to use fewer colours than an arbitrary order~\cite{tomita2003efficient} and gives a smaller additional speed-up in our experiments.
\iffull
Conflicts are stored as bitsets over $C(\tau)$, which takes $O(|C(\tau)|^2)$ time and space per threshold.
\fi
The reductions shrink the region by orders of magnitude for most evaluated queries (\cref{sec:experiments}). The remaining search is still exponential in the worst case, as expected from \cref{thm:hardness}.

\paragraph{Time limits and certificates}
Our implementation checks a time limit at every search node and does not search sets $C(\tau)$ with more than 3{,}000 vertices; such thresholds keep their bound, and root bounds are computed only for sets of up to 20{,}000 vertices, which Greedy also skips if larger.
If Exact stops before every threshold is settled, one more pass outside the time limit, lasting at most 5\,s or 10\% of the limit, whichever is larger, tightens the bounds of the unsettled thresholds $T$. Exact then returns the incumbent $B$ with $\UB=\max(|B|,\max_{\tau\in T}\beta(\tau))\ge\OPT$.
Every answer is thus proven optimal ($\UB=|B|$, which includes proven empty answers), feasible with a gap $|B|/\UB$ ($0<|B|<\UB$), or undecided ($B=\emptyset$ and $\UB>0$).

%% file: figures/fig_latsweep.tex
\begin{tikzpicture}[x=1cm,y=1cm,
    every node/.style={font=\scriptsize},
    bar/.style={draw=gray!55,line width=2.4pt},
    tick/.style={draw=black,line width=0.4pt},
    sweep/.style={draw=black,line width=0.45pt,dash pattern=on 1.6pt off 1pt},
    hit/.style={circle,fill=black,inner sep=0pt,minimum size=3.6pt},
    box/.style={draw=black!70,rounded corners=1.5pt,align=center,inner sep=1.8pt,line width=0.4pt,minimum height=0.5cm},
    sbox/.style={draw=black!70,fill=gray!12,rounded corners=1.5pt,align=center,inner sep=1.5pt,line width=0.4pt,minimum height=0.4cm},
    arr/.style={-{Latex[length=3pt]},draw=black!70,line width=0.4pt},
    brace/.style={draw=gray!55!black,line width=0.3pt},
    lbl/.style={anchor=west,font=\scriptsize\bfseries,inner sep=0pt},
  ]
  \def\xo{0.5}\def\s{0.5}   %
  \def\ya{0}\def\yb{-0.25}\def\yc{-0.5}\def\yd{-0.75}\def\ye{-1.0}\def\yx{-1.36}
  \newcommand{\iv}[3]{%
    \draw[bar] (\xo+\s*#2,#1) -- (\xo+\s*#3,#1);
    \draw[tick] (\xo+\s*#2,#1-0.065) -- (\xo+\s*#2,#1+0.065);
    \draw[tick] (\xo+\s*#3,#1-0.065) -- (\xo+\s*#3,#1+0.065);}
  \node[lbl] at (-0.12,0.3) {(a)};
  \draw[gray!60,line width=0.3pt,-{Latex[length=3pt]}] (\xo,\yx) -- (\xo+\s*10.8,\yx);
  \node[anchor=west,inner sep=1pt] at (\xo+\s*10.8,\yx) {$\tau$};
  \foreach \t/\n in {4/1,8/2,10/3} {
    \draw[gray!60] (\xo+\s*\t,\yx) -- (\xo+\s*\t,\yx-0.06);
    \node[anchor=north,inner sep=1.2pt] at (\xo+\s*\t,\yx-0.04) {$\tau_\n$};}
  \node[anchor=north,inner sep=1.2pt] at (\xo,\yx-0.04) {$0$};
  \draw[gray!45,line width=0.3pt] (\xo+\s*10,0.2) -- (\xo+\s*10,\yx);
  \node[anchor=south,inner sep=1pt,gray!40!black] at (\xo+\s*10,0.17) {$\theta_q$};
  \draw[sweep] (\xo+\s*4,0.2) -- (\xo+\s*4,\yx);
  \draw[sweep] (\xo+\s*8,0.2) -- (\xo+\s*8,\yx);
  \foreach \y/\v in {\ya/q,\yb/v_1,\yc/v_2,\yd/v_3,\ye/v_4} {\node[anchor=east,inner sep=1pt] at (\xo-0.04,\y) {$\v$};}
  \iv{\ya}{0}{10}
  \iv{\yb}{2}{4}
  \iv{\yc}{6}{10}
  \iv{\yd}{3}{8}
  \node[anchor=east,inner sep=0.8pt] at (\xo+\s*2-0.04,\yb) {$\rho_{v_1}$};
  \node[anchor=west,inner sep=0.8pt] at (\xo+\s*4+0.06,\yb) {$\theta_{v_1}$};
  \draw[red!70!black,line width=0.4pt,dash pattern=on 1pt off 1pt] (\xo+\s*5,\ye) -- (\xo+\s*7,\ye);
  \draw[tick,red!70!black] (\xo+\s*5,\ye-0.065) -- (\xo+\s*5,\ye+0.065);
  \draw[tick,red!70!black] (\xo+\s*7,\ye-0.065) -- (\xo+\s*7,\ye+0.065);
  \node[anchor=north,inner sep=0.6pt,red!70!black] at (\xo+\s*5,\ye-0.06) {$\theta_{v_4}$};
  \node[anchor=north,inner sep=0.6pt,red!70!black] at (\xo+\s*7,\ye-0.06) {$\rho_{v_4}$};
  \foreach \y in {\ya,\yb,\yd} {\node[hit] at (\xo+\s*4,\y) {};}
  \foreach \y in {\ya,\yc,\yd} {\node[hit] at (\xo+\s*8,\y) {};}
  \node[anchor=west,inner sep=1pt] at (\xo+\s*11.4,\yb) {$v_1$: intolerant};
  \node[anchor=west,inner sep=1pt] at (\xo+\s*11.4,\yc) {$v_2$: far from $q$};
  \node[anchor=west,inner sep=1pt,red!70!black] at (\xo+\s*11.4,\ye) {$v_4$: $I_{v_4}=\emptyset$};

  \begin{scope}[yshift=-2.47cm]
  \node[lbl] at (-0.12,0.52) {(b)};
  \node[anchor=south,inner sep=1pt] (qk) at (2.3,0.44) {$q$, $k$};
  \node[box] (idx) at (0.72,0) {index of $\Gc$\\(offline)};
  \node[box] (reg) at (2.3,0) {region\\$R$};
  \node[box] (int) at (3.83,0) {intervals $I_v$,\\thresholds $\Tq$};
  \node[box] (bf) at (5.68,0) {best-first\\sweep};
  \node[box] (out) at (7.45,0) {output\\$B$};
  \draw[arr] (idx) -- (reg);
  \draw[arr] (qk) -- (reg);
  \draw[arr] (reg) -- (int);
  \draw[arr] (int) -- (bf);
  \draw[arr] (bf) -- (out);
  \node[anchor=east,inner sep=1pt] at (1.0,-0.95) {$\beta(\tau)$:};
  \node[sbox] (s1) at (1.45,-0.95) {$|S(\tau)|$};
  \node[sbox] (s2) at (3.3,-0.95) {$|C(\tau)|$};
  \node[sbox] (s3) at (5.15,-0.95) {$|M|+c$};
  \node[sbox] (s4) at (7.0,-0.95) {$\OPT_\tau$};
  \draw[arr] (s1) -- node[above,inner sep=1.2pt] {peel} (s2);
  \draw[arr] (s2) -- node[above,inner sep=1.2pt] {colour} (s3);
  \draw[arr] (s3) -- node[above,inner sep=1.2pt] {search} (s4);
  \draw[gray!50,line width=0.3pt] (bf.south) -- ++(0,-0.18) coordinate (j);
  \draw[gray!50,line width=0.3pt] (s1.north) |- (j) -| (s4.north);
  \draw[brace] (0.98,-1.24) -- ++(0,-0.06) -- (3.8,-1.30) -- ++(0,0.06);
  \node[anchor=north,inner sep=1.2pt] at (2.39,-1.30) {BallCore-F, BallCore-R, Exact};
  \draw[brace] (4.62,-1.24) -- ++(0,-0.06) -- (7.5,-1.30) -- ++(0,0.06);
  \node[anchor=north,inner sep=1.2pt] at (6.06,-1.30) {Exact only};
  \end{scope}
\end{tikzpicture}

%% file: sections/06_experiments.tex
\section{Experiments}\label{sec:experiments}

The experiments ask how far exact search scales and where it stops (\cref{sec:exp:efficiency}), how groups with and without the bound behave under uneven influence and outside ties (\cref{sec:exp:dynamics}), and what personal tolerances change on real positions and what enforcing them costs (\cref{sec:exp:meetup}\iffull{}; two parliaments in \cref{sec:exp:congress}\fi).
Shares are compared in percentage points (pp); brackets denote 95\% bootstrap confidence intervals (CIs). Between-method CIs resample queries 2{,}000 times with the two groups paired within each query. Groups may overlap across queries; the CIs therefore describe the sampled queries within each graph.
\iffull
Appendix~\ref{app:supp} gives supplementary tables.
\else
The technical report~\cite{techreport} gives supplementary tables.
\fi

\subsection{Setup}\label{sec:exp:setup}

\paragraph{Graphs and positions}
Gowalla and Brightkite~\cite{cho2011friendship} provide friendships and time-stamped check-ins, which we split by time into a training (70\%), a validation (10\%) and a test period (20\%).
We retain users with at least 10 training check-ins and the friendships between them.
The home $x_v$ of a user is the median check-in position in the user's most frequent cell of about 25\,km, and $\theta_v$ is the median distance of the user's training check-ins from home, at least 1\,km.
Distances are chord lengths between Earth-centred coordinates\iffull, within 0.1\% of great-circle distances below 300\,km\fi.
DBLP, YouTube, LiveJournal, Orkut and Friendster~\cite{yang2012defining}, together with Pokec~\cite{takac2012pokec}, lack positions. For these SNAP graphs~\cite{leskovec2014snap}, we generate community-aligned positions.
Label propagation~\cite{raghavan2007near} with a size cap forms groups of at most 1{,}000 vertices, whose members scatter with unit Gaussian spread in $\R^m$ around centres about eight times the within-group scale $\sqrt{2m}$ apart.
Tolerances are $\theta_v=s\sqrt{2m}\exp(\sigma Z_v-\sigma^2/2)$ with standard normal $Z_v$ and $\sigma=0.4$.
The scale $s$ is calibrated on 30 queries: it is the smallest value, on a grid refined by bisection, at which Exact (with a 5\,s limit per query) finds LCCs with a median of at least 10 members. Where no value of the grid achieves this (YouTube, and Friendster on a coarser grid), we take the largest one. The calibrated $s$ ranges from 1.25 on Orkut to 3 on YouTube.

\paragraph{Queries and algorithms}
Efficiency queries are drawn at random from the $k$-core of $G$ (on synthetic positions, from groups of at least 20 members), 50 per setting.
The default setting ($k=3$, all vertices and, on synthetic positions, $m=16$ and the calibrated scale) occurs in every parameter sweep on Gowalla, Brightkite, LiveJournal and Orkut. We pool its disjoint samples into 200 queries per graph, or 150 on Gowalla and Brightkite, which have no dimension sweep. The other four graphs use only the default setting, with 200 queries each.
We run Exact (with a time limit of 60\,s), the greedy variant (\emph{Greedy}), BallCore-F and BallCore-R.
The connected $k$-core component of $q$ and the global $(k,r)$-core, found by a heuristic search with the median tolerance as radius, serve as references.
All algorithms we time are written in C++ (g++~11.4, \texttt{-O3 -march=native}) and run single-threaded on an Intel Xeon Gold 6530 with 64\,GB of memory\iffull; a two-pass loader converts Friendster's edge list in 13 minutes\fi.
Query times exclude loading the graph, building the index and the bound pass after a time limit (\cref{sec:algorithm:exact}), which took at most 4\,\textmu{}s on our efficiency queries.

\paragraph{Baselines}
The baselines separate the effects of the LCC constraint, personal tolerances and structural cohesion; each dynamics or application analysis uses the relevant subset.
The largest group satisfying the pairwise condition (\emph{pairwise}) is also NP-hard to find: under equal tolerances, the pairwise condition coincides with~(iii), so the proof of \cref{thm:hardness}(a) applies. Our branch and bound solves it on a fixed conflict graph, in which $u$ and $v$ conflict if $d(x_u,x_v)>\min(\theta_u,\theta_v)$, with a limit of 60\,s (10\,s in the heterogeneity sweep, where Exact keeps 60\,s).
\emph{Edge-accept} is the $k$-core on the edges accepted by both endpoints.
Several baselines ignore tolerances. The $(k,r)$-core at the LCC's own threshold $\tau$ is an oracle baseline, since $\tau$ is known only after the search; the \emph{global} $(k,r)$-core uses the median tolerance instead. SAC~\cite{fang2017spatial}, in its AppInc approximation, returns the connected $k$-core containing $q$ in the smallest ball centred at $q$ that holds one, and \emph{size-matched SAC} returns the $k$-core in the smallest such ball whose $k$-core has at least the LCC's size, so its groups may be larger than the LCC. Finally, we use a random connected group of the LCC's size and the $k$-core component.

\subsection{Efficiency}\label{sec:exp:efficiency}

At the default setting, exact search is fast and proves its answers; once tolerances span several communities, it often cannot finish within the time limit.

\begin{table*}[t]
\caption{Exact and Greedy at the default setting ($k=3$).}\label{tab:eff}
\centering\small
\input{tables/tab_eff}
\end{table*}

\paragraph{Queries with and without an LCC}
At the default setting, most queries without an LCC are settled before any search: the reductions in \cref{lem:reduce} eliminate every candidate for 74--100\% of them, and Exact certifies absence in a median of at most 4.7\,\textmu{}s and a maximum of 0.81\,s.
Between 13\% (Brightkite) and 57\% (DBLP) of the queries have an LCC, and \cref{tab:eff} reports these queries separately. On them, Exact takes a median of 0.07\,ms (Brightkite) to 861\,ms (Pokec), and it proves optimality on all but 4 queries on Pokec and 3 on Friendster, which reach the 60\,s limit (column Unproven).
As a reference, we extract $q$'s $k$-core component, which ignores positions and tolerances (median times in \cref{tab:eff}). On the seven graphs where it ran alongside Exact, Exact is faster on 70\% (Pokec) to 99\% (Brightkite) of the default queries, and on 41--97\% of those with an LCC. On Friendster, the extraction was timed in a separate run of the same code on the same queries (median 70.5\,s).
For queries with an LCC, Greedy finds one on 77--100\%. On queries with a proven optimum, Greedy's mean size relative to the optimum is 0.62--0.98, where a query counts as 0 when Greedy finds no LCC.
On LiveJournal, Orkut and Pokec, BallCore-F finds an LCC for only 4--6\% of queries that have one.
BallCore-R is evaluated as an upper bound because its output need not be an LCC. On the four largest graphs, at most 3\% of its non-empty outputs satisfy~(iii), and its sets have a median size of 1.3--2.0 times the optimum; the global $(k,r)$-core is never an LCC there.

\paragraph{Unproven queries}
Of the 3{,}700 queries across all graphs and settings, 1{,}352 have an LCC, and 9 of these remain unproven by Exact: 2 among the 2{,}900 parameter-sweep queries and 7 among the 800 queries on the other four graphs.
For each of them, Exact still returns an LCC of at least 0.689 times a proven upper bound on $\OPT$ (\cref{sec:algorithm:exact}): the ratio is 0.955 and 0.982 on LiveJournal and Orkut, and 0.689--0.939 on Pokec and Friendster.
The unproven queries have 560--918 candidate vertices, and no efficiency query has more than 1{,}502, so the size caps of \cref{sec:algorithm:exact} never apply in this setting~\trtab{tab:tr:unproven}.

\paragraph{Scalability}
On Friendster, with 65.6 million vertices and 1.8 billion edges, Exact proves 98.5\% of the queries. It takes 1.9\,s on average, with a median of 0.9\,ms and a 90th percentile of 2.0\,s. On the queries with an LCC, its median is 352\,ms and its 90th percentile (p90 in \cref{tab:eff}) is 7.9\,s, whereas extracting $q$'s $k$-core component takes a median of 70.5\,s.
The hardest graph is Pokec (median 861\,ms, p90 21.6\,s for queries with an LCC).
The graph $\Gc$ keeps 10--61\% of the edges. The index is built once per tolerance assignment, in 5\,ms (Brightkite) to 9.7\,s (Orkut), and in 4.7 minutes on Friendster.

\paragraph{Ablation and parameters}
On the 50 queries of the $k=3$ sample on LiveJournal and Orkut, Exact solves every query within the limit. Best-first order, conflict colouring and rule~(3) each contribute, and rule~(3) the most. Sweeping in non-increasing order of $|S(\tau)|$ instead of best-first (\emph{count order}) leaves 6\% and 8\% of these queries unproven; colouring in index order instead of by conflicts leaves one query (2\%) each; and dropping rule~(3) of \cref{lem:prune} leaves 12\% each.
Without the index, the median query region on LiveJournal grows from 7 to 29 vertices and the median time from 0.07 to 0.48\,ms. Without any reductions or pruning, every query processes $q$'s entire $k$-core component (2.7 and 3.0 million vertices), and 12\% and 14\% of the queries reach the limit.
Across $k\in\{2,\dots,5\}$, vertex samples of 25--100\% and $m\in\{2,4,8,16,32\}$ dimensions, the median time over all queries stays below 0.5\,ms on LiveJournal and 19\,ms on Orkut (below 96 and 525\,ms over those with an LCC), and no query reaches the limit.
The tolerance scale $s$ matters most. At half the calibrated $s$, almost no LCC exists (0\% and 4\% of the queries). At twice and three times the calibrated $s$, most queries have one (68--88\%), and at twice the calibrated $s$, Exact leaves one query unproven on each graph.

\begin{table}[t]
\caption{Solvers in the count-order sweep: queries proven optimal (\%) and median slowdown against ours on hard queries.}\label{tab:solvers}
\centering\footnotesize\setlength{\tabcolsep}{3.5pt}
\input{tables/tab_solvers}
\end{table}

\paragraph{Solver comparison}
We compare solvers for the fixed-threshold subproblem in a common Python framework that uses the count-order sweep and exchanges only this solver.
Our branch and bound proves 96.5\% of the 200 default queries per graph on LiveJournal and Orkut, against 72.5--90.5\% for the other solvers (\cref{tab:solvers}, which also gives the shares among the queries with an LCC found by some solver). It also proves every query that any other solver proves, and all solvers agree on every proven optimum.
Besides the sweep, the framework fixes the reductions, the cap on searched sets, the time limit of 60\,s and the initial incumbent. The alternative solvers are our C++ reimplementation of AdvMax, the maximum $(k,r)$-core algorithm of Zhang et al.~\cite{zhang2017engagement}, adapted to contain $q$ and run with its own and with a simple branching order; a mixed-integer program with flow-based connectivity, solved by HiGHS~\cite{huangfu2018parallelizing}; and the same model in CP-SAT~\cite{ortools}.
All solvers, including a separate C++ copy of our branch and bound, run single-threaded inside this Python framework, so only relative times are meaningful; 150 of the queries also appear in \cref{tab:eff}.
On hard queries, which take our solver at least 100\,ms and are proven by both solvers of a pair (12--28 per pair), the other solvers are slower by median factors of 5.7 to 175. On queries of 1--100\,ms, the AdvMax reimplementations match our branch and bound (median ratios 0.93--1.05), and below 1\,ms the others are about twice as fast, a difference of microseconds.

\paragraph{Against a single threshold}
Once the index is built, sweeping personal tolerances costs no more per query than a single-threshold search in our engine.
On the 200 queries of \cref{tab:solvers}, we compute the maximum $(k,r)$-core containing $q$. At the median tolerance, it times out on 71 queries on Orkut and 19 on LiveJournal, where best-first Exact proves all 200. At the LCC's own threshold $\tau^*$, run on the 115 and 95 queries with an LCC, it times out on 37 and 16.
For queries with an LCC, the sweep is faster on 92--94\% of the queries on Orkut (median time ratios 0.065 and 0.135). On LiveJournal, where the sweep answers most of the 200 queries in under 1\,ms, the two searches take similar times (median ratios 0.86--0.89, with a 1\,ms floor).
Only the sweep needs the index, and on LiveJournal its construction would dominate if amortized over 200 queries~\trapp{app:single}.
A single threshold also ignores the personal tolerances in the candidate region, and the single-threshold answers satisfy~(iii) on at most 4.2\% of the queries.

\paragraph{Heterogeneity}
Heterogeneous tolerances are what separate LCCs from $(k,r)$-cores.
With equal tolerances ($\sigma=0$), the LCC coincides with the $(k,r)$-core at the same threshold on all 220 queries of LiveJournal and Orkut on which Exact and the exact $(k,r)$-core and pairwise searches all finish. Exact then has a single threshold to sweep; in count order, it reaches its limit on 0.5\% (LiveJournal) and 45\% (Orkut) of the 200 queries per graph, against 14\% and 26.5\% at $\sigma=0.2$ and 4\% and 7\% at the default $\sigma=0.4$~\trtab{tab:tr:e2}; the copy of our solver in \cref{tab:solvers} leaves 3.5\% unproven there.
From $\sigma=0.2$ on, $(k,r)$-cores at the LCC's threshold satisfy~(iii) on at most 9\% of the queries, and 12--65\% of their members face another member beyond their own tolerance.

\paragraph{Tolerances that span communities enlarge the search}
Once tolerances let LCCs span several communities, the candidate regions grow by up to two orders of magnitude.
The calibrated scales of \cref{sec:exp:setup} were chosen to yield LCCs with a median of at least 10 members.
Here we consider tolerances of $c$ times the median distance between friends; since most friends lie in different communities, $c\ge0.75$ (Orkut) or $c\ge1$ (LiveJournal) lets LCCs span several communities.
On Orkut, the queries with an LCC have a median of 8{,}360 candidate vertices at $c=0.75$ and 210{,}146 at $c=1$, whereas no efficiency query has more than 1{,}502; on LiveJournal at $c=1$, the median is 1{,}218~\trapp{app:caps}.
In these settings, Exact reaches the time limit on 35--99\% of the queries with an LCC, and 0--1.7\% stop at a cap. The bounds are then loose: the median ratio of the returned solution to the bound is 0.03--7.6\%. At $c=1$, Exact proves only 64\% (LiveJournal) and 20\% (Orkut) of these queries\iffull, and for $c=0.5$, 0.75, 1, 1.5 and 2, 100\%, 100\%, 64\%, 15\% and 3\% on LiveJournal and 100\%, 51\%, 20\%, 3\% and 0.6\% on Orkut\fi~\trtab{tab:tr:e4}.
At the calibrated scale, drawing the queries from the whole $k$-core rather than from groups of at least 20 members still leaves no query unproven.
Larger caps and time limits do not close the gap (\cref{sec:algorithm:exact}). We test them at $c=1$ on both graphs and at $c=0.75$ on Orkut: raising the cap on searched sets to 20{,}000 or 50{,}000, or the root-bound cap to 100{,}000, changes the proven share by at most $+1.1$\pp; letting Greedy use all thresholds lowers it by up to 3.3\pp; and a limit of 300\,s proves only 0--9\% of the remaining queries~\trtab{tab:tr:caps}.

\subsection{Dynamics on Returned Groups}\label{sec:exp:dynamics}

\begin{table}[t]
\caption{Groups (\%) that lose an interaction accepted at the start, on $G[H]$ (G, B, L, O: Gowalla, Brightkite, LiveJournal, Orkut).}\label{tab:dyn}
\centering\footnotesize\setlength{\tabcolsep}{2.6pt}
\input{tables/tab_dyn}
\end{table}

In closed groups, the bound keeps every accepted interaction, whereas groups without it, including pairwise groups, lose some under uneven influence. Outside ties make groups of every method lose interactions, LCC groups less often on Gowalla and Brightkite.
We run the dynamics of \cref{sec:theory} on the groups that each method returns for queries with an LCC. These are a random 1{,}000 of the meetup queries of \cref{sec:exp:meetup} on Gowalla, all 462 on Brightkite, and the 95 and 115 default queries of \cref{tab:solvers} on LiveJournal and Orkut; for pairwise, we also include the 3, 2, 12 and 14 groups of queries without an LCC. Interactions follow the edges of $G[H]$ for at most 300 steps. We test whether any initially accepted interaction is later refused.
In addition to equal-weight HK~\eqref{eq:hk}, the variants in \cref{tab:dyn} draw each member's weights from a Dirichlet distribution with $\alpha=0.1$, which concentrates them on a few neighbours, or let random adjacent pairs move towards each other (Deffuant--Weisbuch gossip~\cite{deffuant2000mixing}). Randomized variants run 10 times, and a group counts as losing an interaction if any run does.
These variants serve as stress tests between equal weights and the weights in the proof of \cref{prop:char}(b), which put almost all of a member's weight on one neighbour.

\paragraph{Closed groups}
No group satisfying~(iii) loses an accepted interaction in 83{,}594 group--variant pairs across 11 variants, including complete interaction graphs, as \cref{thm:consensus} predicts. Under SJT dynamics~\eqref{eq:sjt}, none of 15{,}392 shows a new rejection. The zero counts check the implementation against the theorem; the other rows of \cref{tab:dyn} compare groups without its guarantee.
Even with equal weights, groups found without tolerances lose accepted interactions: 4--17\% of the $(k,r)$-cores at $\tau$ do. With uneven influence, this happens much more often (34--98\%).
With equal weights, pairwise groups never lose an accepted interaction, although 62--98\% violate~(iii) and fall outside \cref{thm:consensus}. Under gossip, 3--17\% lose one, as do 32--40\% under concentrated weights.\iffull\else{} Per run, these shares are 1--5\% and 5--13\%~\trtab{tab:tr:pairwise}.\fi{}
When a pairwise group loses an interaction, usually a single pair is affected: the median share of refused interactions is at most 0.3\%. On Gowalla and Brightkite with concentrated weights, refusal occurs mostly in the first step, and acceptance resumes in about 99\% of cases. Refusals start mostly at the least tolerant members, whom the LCC tends to leave out. In 73\% and 94\% of the refusals, the refuser is among the three least tolerant members. The LCC of the same query excludes the refuser, the refused member or both in 84\% and 82\% of them, while these pairwise groups have a median of 22.5 and 5 members outside the LCC~\trapp{app:p4}.
Consensus is about as frequent in these groups as in LCC groups.
\iffull Per run rather than per group, pairwise groups lose an interaction in 1--5\% of the runs under gossip and 5--13\% under concentrated weights, against 23--67\% and 32--90\% for $(k,r)$-cores at $\tau$ and size-matched SAC groups. \fi
If the tolerances that drive the dynamics deviate from those used in the search by a factor $\exp(0.3Z)$ with standard normal $Z$, 0--3.5\% of the LCC groups lose an interaction, against 0.9--4.1\% of the pairwise groups and 25--38\% of the $(k,r)$-cores at $\tau$.

\paragraph{LCC or pairwise}
The LCC bound trades group size for robustness to uneven influence.
Pairwise groups are at least as large by definition, since every LCC satisfies the pairwise condition. Where both are proven optimal~\trtab{tab:tr:pairwise}, the pairwise group is larger for 60\%, 61\%, 93\% and 97\% of the queries on Gowalla, Brightkite, LiveJournal and Orkut, with a median per-query size ratio of 1.08--1.52.
In exchange, LCC groups lose no accepted interaction in \cref{tab:dyn}, whereas pairwise groups do under uneven influence.
This is the distinction drawn in \cref{sec:problem}: the pairwise condition suffices when positions remain fixed (\cref{prop:char}(a)), as in a one-off meetup, whereas for groups that interact repeatedly with unknown weights, only~(iii) guarantees acceptance for every weighting.

\paragraph{Open groups}
We add outside influence to test the closed-group assumption.
Each accepted outside friend gets weight $w$, against weight $1$ for the member itself and for each accepted group neighbour.
On Gowalla and Brightkite, LCC groups lose an interaction less often than both size-matched SAC groups (by 3.6--16.9\pp) and pairwise groups (by 2--7\pp) at every $w>0$\iffull, but not on Congress (\cref{sec:exp:congress})\fi\iffull; on the synthetic graphs, where accepted outside friends lie in the same community, no LCC group loses one even at $w=1$\fi.
When members average with accepted outside neighbours at fixed positions, \cref{prop:open} preserves acceptance under conditions (a) and~(b), but only 4.6\% and 2.6\% of LCC groups on these datasets meet them, mainly because accepted outside friends extend the hull $K_0$ of \cref{prop:open} beyond $\tmin$.
Already at $w=0.01$, 34\% and 23\% of the LCC groups lose an interaction accepted at the start, and at $w=1$, 65\% and 62\% do; no group that satisfies the conditions ever does.
The loss is small and occurs early: it starts at a median step of 1 and affects a median 0.4--12\% of the initially accepted interactions~\trtab{tab:tr:p22}. Consensus is rare for open groups of every method (at most 5.3\%).

\subsection{Meetups: What Member-Specific Tolerances Change}\label{sec:exp:meetup}

On real positions, we ask what member-specific tolerances change and what enforcing them costs. Groups found without tolerances often leave a member's home beyond another member's travel radius, while enforcing every member's tolerance causes no measurable loss of co-visits relative to spatial communities of the same size.
Here $\theta_v$ is a travel radius. A meetup needs only the pairwise condition of \cref{sec:problem}, which~(iii) strengthens, so these data do not test the dynamics of \cref{sec:theory}.
We replicate the Gowalla and Brightkite application on the Yelp Open Dataset~\cite{yelp2022dataset}.
We keep the users with at least 10 reviews up to 2017 whose home lies in one of 11 metropolitan areas, and the friendships between users of the same area (66{,}412 users, 293{,}785 friendships). As for check-ins, a user's home is the median position of the reviewed businesses in the user's most frequent cell of about 25\,km, and $\theta_v$ is the median distance of the reviewed businesses from home, at least 1\,km.
Two users \emph{co-visit} if they check in at the same venue on the same day (Gowalla, Brightkite) or review the same business in the test year 2019 (Yelp; 2018 serves for validation).
\iffull Homes, tolerances and design choices use only the training and validation periods, except the Yelp time split, chosen by the share of all friend pairs that co-visit in the test period (3.9\%, above a minimum of 0.5\%), which involves no method's groups; the validation-period share (6.8\%) gives the same split, and a second split ending in 2018 gives the same interpretation~\trapp{app:yelpsplit}.\else Homes, tolerances and design choices use only the training and validation periods, except the Yelp time split, chosen by a test-period quantity that involves no method's groups~\trapp{app:yelpsplit}.\fi{}
Every test-active user in the $k$-core is a query, with $k=3$: 15{,}829, 5{,}260 and 10{,}363 queries, of which 2{,}469, 462 and 5{,}010 have an LCC.
Yelp produces much larger LCCs, with a median of 370 members against 16 and 10, and Exact proves 94.8\% of its queries within 60\,s.
\iffull An LCC differs from a $k$-core only where tolerances are narrow relative to the spread of positions, as travel radii are: when we placed Yelp users at the mean category profile of their reviewed businesses, with tolerances defined in the same way, 99.9\% of the friend pairs accepted each other and the LCC was almost the whole $k$-core (median size ratio 0.998).\fi

\paragraph{Tolerance violations}
Groups found without tolerances leave many members with another member's home beyond their estimated travel radius.
In groups of at most 3{,}000 members on Gowalla, Brightkite and Yelp, only 7\%, 2\% and 27\% of the members of SAC groups have every other member's home within their radius; the shares are 29\%, 42\% and 42\% for global $(k,r)$-cores and 65\%, 72\% and 84\% for $(k,r)$-cores at $\tau$. All members of LCC and pairwise groups do, by construction.
\iffull Condition~(iii) is stricter: on the Gowalla queries with an LCC, it holds for 39\% of the pairwise groups, 57\% of the $(k,r)$-cores at $\tau$ and 33\% of the size-matched SAC groups, which are larger (median 22, 27 and 22 members, against 16).\fi

\begin{figure}[t]
\centering
\includegraphics[width=0.8\columnwidth]{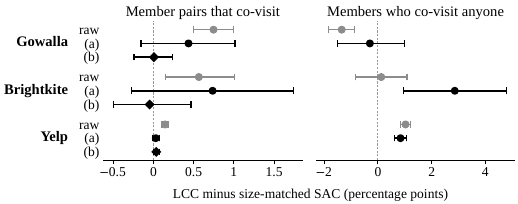}
\caption{LCC minus size-matched SAC in co-visit shares (pp, 95\% CIs): raw, at equal size with other members~(a), and size-adjusted~(b, for pairs only).}\label{fig:meetup}
\end{figure}

\paragraph{Co-visits}
Enforcing every member's tolerance causes no measurable loss of co-visits relative to spatial communities of the same size on Gowalla, Brightkite and Yelp.
Where both groups exist, LCC groups have a higher raw share of co-visiting member pairs than size-matched SAC on all three datasets, mainly owing to size (\cref{fig:meetup}\iffull{} and \cref{tab:meetup}\else; values in~\cite[Table~\ref*{TR-tab:meetup}]{techreport}\fi).
We control for size in two ways, developed on Gowalla and Brightkite and replicated on Yelp: (a)~on the 610, 108 and 2{,}387 queries where SAC and LCC groups have equal size but different members, and (b)~by regressing the difference on the log ratio of the two sizes. Against size-matched SAC, we count a difference only when both CIs exclude 0, and no dataset meets this test: both CIs contain 0 on Gowalla and Brightkite, and on Yelp, the CI of (a) contains 0 while that of (b) lies just above it, with an intercept of $+0.04$\pp.
On Gowalla, Brightkite and Yelp, the CIs of (a) and~(b) exclude losses of more than about 0.25, 0.5 and 0.01\pp, respectively, while 8.4\%, 5.7\% and 1.7\% of the member pairs in LCC groups co-visit.
A second measure, the share of members who co-visit with anyone in the group (\cref{fig:meetup}, right), favours LCC groups at equal size on Brightkite and Yelp, but not on Gowalla.
\iffull Against $(k,r)$-cores at $\tau$, LCC groups are ahead at equal size only on Yelp ($+0.08$\pp\ $[0.04,0.11]$ in pair rate, by~(b)).
Across $k\in\{2,\dots,5\}$ and tolerances at the 70th and 90th percentiles (on Yelp, 2{,}000 queries per setting), the raw differences against size-matched SAC and $(k,r)$-cores at $\tau$ retain their sign on all three datasets (on Gowalla and Brightkite, as unpaired means). \else The raw differences against size-matched SAC keep their sign for $k$ from 2 to 5 and for tolerances at the 70th and 90th percentiles~\trsec{sec:exp:meetup}.\fi
\iffull
Within tolerance-agnostic groups, whether members who can reach everyone co-visit more depends on the dataset and the baseline once eccentricity and activity are controlled for.
\fi

\iffull
\paragraph{Against pairwise groups}
Relative to the pairwise condition, enforcing~(iii) costs co-visits on Gowalla by a size-adjusted estimate that extrapolates, since size cannot be held fixed here.
On Gowalla and Brightkite, raw differences favour LCC groups: the two groups are identical for 38\% of the queries, and on the 60\% and 61\% where the pairwise group is larger, its member pairs co-visit less often (LCC minus pairwise $+1.49$\pp\ $[1.08,1.95]$ and $+1.77$\pp\ $[1.12,2.50]$).
The size-adjusted difference~(b) is nevertheless $-0.85$\pp\ $[-1.35,-0.37]$ on Gowalla, $-0.29$\pp\ $[-0.69,0.11]$ on Brightkite and $-0.05$\pp\ $[-0.12,0.02]$ on Yelp, whose CIs exclude 0 at $k=4$ ($-0.31$\pp) and with 90th-percentile tolerances ($-0.13$\pp); as equal-size groups are almost always identical (Gowalla, Brightkite) or rare (36 of 4{,}951 Yelp queries), (b) extrapolates here~\trapp{app:dyn}.
\else
Against the larger pairwise groups, the size-adjusted difference~(b), LCC minus pairwise, is $-0.85$\pp\ $[-1.35,-0.37]$ on Gowalla, and its CIs contain 0 on Brightkite and Yelp. This estimate extrapolates, since groups of equal size are almost always identical on Gowalla and Brightkite and rare on Yelp~\trapp{app:dyn}.
\fi

\paragraph{Yelp as a weaker test of travel radii}
\iffull On Yelp, whether friends co-visit in the validation year is unrelated to the distance between their homes (Spearman $-0.009$ $[-0.026,0.008]$, against $-0.152$ on Gowalla), so the interpretation of $\theta_v$ as a travel radius is weaker there than for check-ins, although adding the pair's tighter tolerance to a distance-only model of friends' co-visits still lowers the held-out log loss (by 0.0029 $[0.0020,0.0039]$ per pair).\else On Yelp, friends' co-visits in the validation year are unrelated to the distance between their homes (Spearman $-0.009$, against $-0.152$ on Gowalla), although the pair's tighter tolerance still lowers the held-out log loss of a distance-only model of co-visits~\trsec{sec:exp:meetup}.\fi

\paragraph{An empty LCC}
The absence of an LCC usually reflects a constraint already imposed by accepted friendship edges.
An LCC is absent for most queries (84.4\%, 91.2\% and 51.7\% on Gowalla, Brightkite and Yelp). For 95\%, 97\% and 86\% of these queries, even the $k$-core of $q$'s component restricted to friendships whose two ends accept each other is empty, so $q$ has no $k$ connected friends who accept one another. Only 1.8\%, 0.6\% and 4.4\% of them have pairwise groups but no LCC~\trtab{tab:tr:p21}.
An empty answer thus usually tells a user that $q$ lacks such friends. Only a smaller $k$ or wider tolerances could change this, and even tripling every tolerance gives an LCC to only 22\%, 16\% and 50\% of these queries. Conversely, shrinking every tolerance by 20\% keeps an LCC for 80\%, 80\% and 87\% of the queries that have one, at a median 82\%, 89\% and 69\% of its size~\trtab{tab:tr:p24}.
\iffull
On Gowalla and Brightkite, an empty LCC is informative: SAC groups co-visit more when an LCC exists, by $+4.1$\pp\ $[3.4,4.9]$ and $+3.1$\pp\ $[1.9,4.5]$ in a regression on log size, and global $(k,r)$-cores show the same direction ($+0.74$\pp\ $[0.09,1.42]$ and $+1.12$\pp\ $[-0.35,2.51]$ without size adjustment).
On Yelp, the raw difference is a size effect: SAC groups are much larger when the LCC is empty (median 742 against 75 members), and controlling for log size reverses the raw difference ($+2.09$\pp) to $-0.25$\pp\ $[-0.39,-0.11]$.
Descriptively, the Yelp difference is positive for SAC groups of up to 20 members ($+2.3$ and $+0.9$\pp) and close to zero for larger ones, and for global $(k,r)$-cores it remains positive after the control ($+0.08$\pp\ $[0.01,0.16]$); the share of SAC members who co-visit with anyone in the group points in different directions (raw: $+0.8$ and $+5.4$\pp\ on Gowalla and Brightkite, $-3.9$\pp\ on Yelp, where it stays negative after the control).
On two datasets out of three, an empty answer thus marks queries lacking a cohesive group satisfying~(iii), for which groups found without tolerances also tend to meet less.
\fi

\iffull
\subsection{Positions from Votes: Two Parliaments}\label{sec:exp:congress}

Opinion positions are rarely observed alongside a network. We illustrate the method using the 474 members of the 117th U.S. Congress, then repeat the \iffull main analyses\else premise test\fi{} on the European Parliament.
Positions are the first dimension of DW-NOMINATE~\cite{poole2007ideology,lewis2023voteview}, estimated from roll-call votes; $\theta_v$ is the median distance from $v$ to the sponsors of the bills that $v$ co-sponsored from January 2021 to February 8, 2022~\cite{govinfo_billstatus} (median 0.145, floor 0.02, which no member reaches); and $G$ joins members who retweeted, quoted, replied to or mentioned each other on Twitter between February 9 and June 9, 2022~\cite{fink2023congress,fink2023centrality} (10{,}188 edges).
Positions derive from votes and tolerances from co-sponsorship, rather than measured attitudes. With strong party discipline in Congress, the results concern only this setting.
Tolerances do filter: 42.7\% of the adjacent pairs accept each other, an LCC exists for 96.2\% of the 472 queries, and its median size is 12\% of that of $q$'s $k$-core component.

\paragraph{Premise test}
The premise test asks whether co-sponsorship tolerances predict Twitter interactions beyond ideological distance. They do not.
In a logistic model of whether $u$ interacted with $v$, with sender and receiver effects, a spline in the distance, and indicators of the same party, state and chamber, the coefficient of $\mathbf{1}[d(x_u,x_v)\le\theta_u]$ is $-0.044$ (one-sided permutation $p=0.92$, permuting tolerances within party and chamber), and it does not improve the held-out log loss.
\iffull
Members with wider co-sponsorship radii interact over wider ranges (Spearman 0.71 $[0.65,0.76]$; 0.63 $[0.56,0.68]$ controlling for party, chamber and centrality), but the cutoff at $\theta_u$ adds nothing beyond distance and fixed effects, so this tolerance estimate is not validated on Twitter.
\else
The co-sponsorship radius tracks the range of Twitter interactions (Spearman 0.71), but the cutoff at $\theta_u$ adds nothing beyond distance and fixed effects.
\fi
\iffull
A possible explanation, which we do not test, is the adversarial nature of some replies and mentions; Luan et al.~\cite{luan2026homophily} find more cross-ideological exchanges in reply ties than follow ties.
\fi
Of the other tolerance variants, only 90th-percentile tolerances yield a significantly positive coefficient (0.32, $p=0.001$).
\iffull\else
On the European Parliament of 2014--16~\cite{cherepnalkoski2016mep} (roll-call positions, tolerances from earlier retweets), the premise test is again negative ($-0.07$, $p=0.78$).
\fi

\iffull
\begin{table}[t]
\caption{Tolerance violations in Congress groups ($k=3$, 472 queries).}\label{tab:congress}
\centering\footnotesize\setlength{\tabcolsep}{3pt}
\input{tables/tab_congress}
\end{table}
\fi

\paragraph{Static violations and dynamics}
Groups found without tolerances often contain members who face someone beyond their own tolerance (391 of the 454 size-matched SAC groups; 4.9\%, 8.2\% and 56.7\% of the members of $(k,r)$-cores at $\tau$, size-matched SAC groups and global $(k,r)$-cores\iffull; \cref{tab:congress}\fi); pairwise groups contain none by definition, but only 0.7\% of them satisfy~(iii).
\iffull
\Cref{tab:congress} also gives the groups satisfying~(iii) and the member pairs with at least one refusal.
\fi
In closed dynamics, no LCC group loses an interaction accepted at the start, as \cref{thm:consensus} predicts.
Pairwise groups never do under equal weights or under weights proportional to the observed Twitter interactions, but 26--47\% do under concentrated weights, gossip, and updates that follow one random accepted neighbour, as do 33--49\% of the size-matched SAC groups and $(k,r)$-cores at $\tau$\iffull; global $(k,r)$-cores lose an interaction in 33\% of the cases even under equal weights\fi.
No LCC group satisfies the conditions of \cref{prop:open}; in open dynamics, none loses an interaction for $w\le0.1$, but 22\%, 41\% and 81\% do at $w=0.2$, $0.5$ and $1$, and 98\% with observed weights.
Unlike on Gowalla and Brightkite, LCC groups here lose an interaction more often than size-matched SAC groups at every $w\ge0.2$ (22\% against 11\% at $w=0.2$), and more often than pairwise groups at $w=0.2$ and $0.5$.
Because the premise test fails, these results illustrate simulated dynamics on real ideological positions; they do not establish how legislators respond to one another.
\iffull
\Cref{fig:congress} shows a median-sized query: of the 391 of 454 queries (86\%) whose size-matched SAC group has a member facing another beyond its interval, the one whose LCC size is nearest the median (ties: lowest id); it illustrates \cref{tab:congress}.

\begin{figure}[t]
\centering
\includegraphics[width=\columnwidth]{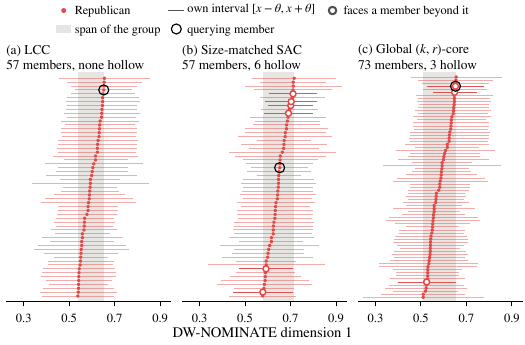}
\caption{Groups for one Congress query ($k=3$).}\label{fig:congress}
\end{figure}
\fi

\paragraph{Co-sponsorship}
\iffull
In the test period (June 10, 2022 to January 3, 2023), the share of member pairs who co-sponsor a bill is not measurably higher in LCC groups than in size-matched SAC groups: $+0.39$\pp\ $[-0.13,0.92]$ raw, $-0.30$\pp\ $[-1.03,0.43]$ at equal size with other members, and $+0.36$\pp\ $[-0.16,0.91]$ size-adjusted; as about two thirds of the member pairs and 99.9\% of the members of LCC groups co-sponsor with each other, this measure discriminates poorly.
\else
From June 10, 2022 to January 3, 2023, the share of member pairs who co-sponsor a bill is not measurably higher in LCC groups than in size-matched SAC groups ($+0.39$\pp\ $[-0.13,0.92]$, nor at equal size), but as about two thirds of the member pairs in LCC groups co-sponsor, this measure discriminates poorly.
\fi

\iffull
\paragraph{A second parliament}
On the European Parliament of 2014--16~\cite{cherepnalkoski2016mep} (roll-call positions, tolerances from earlier retweets), the premise test is again negative ($-0.07$, $p=0.78$); only 90th-percentile tolerances give a positive coefficient (0.42, $p=0.001$, no held-out gain), a hint that acceptance may reach beyond the median distance.
In size-matched SAC groups, 21\% of the members face someone beyond their tolerance, and 21--52\% of the groups lose an initially accepted interaction under unequal influence; no LCC group does.
\fi
\fi

%% file: tables/tab_eff.tex
\begin{tabular}{lrrrrrrrrrr}
\toprule
 & & & & \multicolumn{1}{c}{No LCC} & \multicolumn{3}{c}{LCC exists (ms)} & \multicolumn{2}{c}{Greedy} & \\
\cmidrule(lr){5-5}\cmidrule(lr){6-8}\cmidrule(lr){9-10}
Graph & $|V|$ & $|E|$ & With LCC & Median (\textmu{}s) & Median & p90 & Unproven & Finds & Size & $k$-core (ms) \\
\midrule
Gowalla & 56\,K & 222\,K & 16\% & 0.3 & 0.26 & 27 & 0 & 100\% & 0.98 & 2.2 \\
Brightkite & 24\,K & 108\,K & 13\% & 0.2 & 0.07 & 0.41 & 0 & 95\% & 0.95 & 1.1 \\
\midrule
DBLP & 317\,K & 1.0\,M & 57\% & 0.9 & 0.14 & 5.4 & 0 & 96\% & 0.94 & 22 \\
YouTube & 1.1\,M & 3.0\,M & 23\% & 2.3 & 31 & 349 & 0 & 93\% & 0.90 & 41 \\
Pokec & 1.6\,M & 22.3\,M & 52\% & 3.3 & 861 & 21.6\,s & 4 & 77\% & 0.62 & 373 \\
LiveJournal & 4.0\,M & 34.7\,M & 50\% & 3.4 & 6.0 & 1.9\,s & 0 & 90\% & 0.79 & 630 \\
Orkut & 3.1\,M & 117.2\,M & 55\% & 4.7 & 200 & 3.8\,s & 0 & 84\% & 0.68 & 1.6\,s \\
Friendster & 65.6\,M & 1.81\,B & 55\% & 0.4 & 352 & 7.9\,s & 3 & 89\% & 0.77 & 70.5\,s \\
\bottomrule
\end{tabular}

%% file: tables/tab_solvers.tex
\begin{tabular}{lrrrrrr}
\toprule
 & \multicolumn{2}{c}{Proven} & \multicolumn{2}{c}{Proven, LCC} & \multicolumn{2}{c}{Slowdown} \\
\cmidrule(lr){2-3}\cmidrule(lr){4-5}\cmidrule(lr){6-7}
Solver & LJ & Ork. & LJ & Ork. & LJ & Ork. \\
\midrule
Branch and bound (ours) & 96.5 & 96.5 & 92.6 & 93.9 & 1 & 1 \\
AdvMax, simple branching & 90.5 & 80.5 & 80.0 & 66.1 & 5.7 & 7.2 \\
AdvMax & 89.5 & 76.5 & 77.9 & 59.1 & 14.0 & 14.6 \\
MILP (HiGHS) & 90.5 & 80.0 & 80.0 & 65.2 & 26.7 & 21.2 \\
CP-SAT & 86.0 & 72.5 & 70.5 & 52.2 & 175 & 45.0 \\
\bottomrule
\end{tabular}

%% file: tables/tab_dyn.tex
\begin{tabular}{lrrrrrrrrrrrr}
\toprule
 & \multicolumn{4}{c}{Equal} & \multicolumn{4}{c}{Concentrated} & \multicolumn{4}{c}{Gossip} \\
\cmidrule(lr){2-5}\cmidrule(lr){6-9}\cmidrule(lr){10-13}
Method & G & B & L & O & G & B & L & O & G & B & L & O \\
\midrule
LCC (Exact) & 0 & 0 & 0 & 0 & 0 & 0 & 0 & 0 & 0 & 0 & 0 & 0 \\
Pairwise & 0 & 0 & 0 & 0 & 40 & 33 & 32 & 40 & 17 & 11 & 3 & 7 \\
$(k,r)$-core at $\tau$ & 15 & 17 & 4 & 13 & 41 & 44 & 91 & 97 & 34 & 40 & 69 & 85 \\
Size-matched SAC & 5 & 8 & 1 & 2 & 52 & 56 & 89 & 98 & 39 & 40 & 63 & 86 \\
\bottomrule
\end{tabular}

%% file: tables/tab_congress.tex
\begin{tabular}{lrrr}
\toprule
Method & (iii) holds & Members & Pairs \\
\midrule
LCC (Exact, Greedy, BallCore-F) & 100 & 0 & 0 \\
Pairwise & 0.7 & 0 & 0 \\
$(k,r)$-core at $\tau$ & 15.6 & 4.9 & 1.3 \\
Size-matched SAC & 2.2 & 8.2 & 1.5 \\
SAC & 87.1 & 14.3 & 14.2 \\
Global $(k,r)$-core & 1.7 & 56.7 & 19.1 \\
BallCore-R & 0.9 & 49.6 & 12.9 \\
Random, $k$-core, edge-accept & 0 & 100 & 71.6--77.4 \\
\bottomrule
\end{tabular}

%% file: sections/07_conclusion.tex
\section{Conclusion}\label{sec:conclusion}

Bounding group diameter by the smallest member tolerance connects community search to persistent acceptance: closed LCC groups preserve acceptance under bounded-confidence dynamics with any positive weights and avoid rejection under SJT dynamics, whereas the pairwise condition and other natural relaxations lose these guarantees. Finding a largest LCC is NP-hard even in the plane.
LatSweep reduces the search to subproblems with a common threshold, refines their bounds best-first, and returns certificates of optimality or absence, or a bound on the remaining gap. Its exact variant answers most evaluated queries faster than extracting $q$'s $k$-core component, handles graphs with 1.8 billion edges, and returns an LCC of at least 0.689 times a proven upper bound on the efficiency queries stopped early.
On real positions, groups found without tolerances often place a member's home beyond another member's travel radius, while enforcing every member's tolerance causes no measurable loss of co-visits relative to spatial communities of the same size.

The guarantee is for closed groups: with outside ties, acceptance can be lost early for many groups, and once tolerances span several communities, exact search exceeds our 60\,s budget on many queries (\cref{sec:exp:efficiency,sec:exp:dynamics}).
LCCs are candidate groups, not verdicts on individuals; tolerances estimated from check-ins or reviews are noisy and should not be used to label or exclude individuals.
Whether tolerances estimated from behaviour predict who interacts in opinion settings remains open; in two parliaments, median-based tolerances did not predict interaction beyond ideological distance~\trsec{sec:exp:congress}.
Other open questions include learning task-specific tolerances from longitudinal traces, maintaining LCCs as graphs evolve, and approximating \MaxLCC{} in fixed dimension.

%% file: sections/09_supplement.tex
\appendices

\section{Supplementary Results}\label{app:supp}

This appendix provides supplementary tables for \cref{sec:experiments}.
The tables retain the runs reported in the main text; any change in sample or sweep order is stated explicitly.

\subsection{Dynamics: All Variants and Methods}\label{app:dyn}

\begin{table*}[!tp]
\caption{Groups (\%) that lose an interaction accepted at the start in all 21 variants: methods that use tolerances.}\label{tab:tr:e1a}
\centering\scriptsize\setlength{\tabcolsep}{2.1pt}
\input{tables/tr_e1_aware}
\end{table*}

\begin{table*}[!tp]
\caption{As \cref{tab:tr:e1a}: methods that ignore tolerances.}\label{tab:tr:e1b}
\centering\scriptsize\setlength{\tabcolsep}{2.3pt}
\input{tables/tr_e1_agnostic}
\end{table*}

\Cref{tab:tr:e1a,tab:tr:e1b} extend \cref{tab:dyn} to all variants and methods, with the same abbreviations G, B, L and O; $<$1 marks shares above 0 and below 1\%, and -- marks runs not performed (edge-accept was not run on LiveJournal and Orkut).
A group is counted in the SJT rows when it shows a new rejection.
The last column of \cref{tab:tr:e1a} counts simulated LCC groups per variant, out of 1{,}000, 462, 95 and 115. Complete-graph and multi-run variants skip groups above a size cap (see Variants). Other methods have lower coverage because their larger groups are skipped more often (for gossip on Orkut, 37 of 115 $(k,r)$-cores at $\tau$).
For each method, the last row of both tables reports its groups under equal-weight HK on $G[H]$.
The groups are those of \cref{tab:dyn}: on Gowalla and Brightkite, 1{,}000 and 462 meetup queries with an LCC and 300 sampled queries without one, and on LiveJournal and Orkut, 200 default queries; the LCC, Greedy, $(k,r)$-cores at $\tau$, size-matched SAC and random groups exist only for queries with an LCC, whereas the other methods may also return groups for queries without an LCC.
BallCore-F returns subsets satisfying~(iii), and no BallCore-F group loses an interaction in a closed variant. We omit the $k$-core component because almost all its groups exceed 5{,}000 members, the largest group size we simulate.

\paragraph{Variants}
HK~\eqref{eq:hk} runs with equal weights, also for 3{,}000 steps with a stopping rule $10^6$ times stricter, or with weights drawn in each run from a Dirichlet distribution with $\alpha=1$ or $\alpha=0.1$.
In \emph{follow one neighbour}, each member moves halfway towards one accepted neighbour drawn at random in each step; in gossip~\cite{deffuant2000mixing}, a random edge interacts, and each end that accepts the other moves $0.3$ of their distance towards it, for 100 rounds of $|E(H)|$ such steps (50 rounds on complete graphs).
SJT~\eqref{eq:sjt} uses equal weights, $\tacc_v=0.5\,\theta_v$, $\trej_v=\theta_v$ and $\beta=0.05$; there we report new rejections, the event that \cref{thm:norejection} excludes.
Randomized variants run 10 times (gossip on complete graphs 5 times), and a group counts if any run loses an interaction.
On complete interaction graphs, every member hears every other; these variants run on groups of at most 1{,}000 members (equal weights, SJT, open), 300 (random weights, follow) or 100 (gossip), and the randomized variants on $G[H]$ on groups of at most 2{,}000 members.
The last three blocks test sensitivity to departures from the assumptions in \cref{sec:theory}.
In open groups, members also average with the outside neighbours they accept, which stay fixed, with the same weight as accepted members ($w=1$ in \cref{sec:exp:dynamics}; at most 200 outside neighbours per member, sampled at random); stubborn outsiders pull each member with weight $\lambda$ towards the fixed mean of all its outside neighbours, near or far; with tolerance noise, the dynamics use $\theta_v\exp(\nu Z_v)$ with standard normal $Z_v$, whereas the groups were found with $\theta_v$.

\paragraph{Observations}
In closed variants, no group returned by a method guaranteeing~(iii) (LCC, Greedy) loses an interaction; a tenfold longer run leaves every entry unchanged.
BallCore-R, whose sets need not satisfy~(iii), loses interactions under uneven influence in up to 76\% of its groups on $G[H]$.
On complete interaction graphs, pairwise groups also lose interactions under uneven influence (up to 30\%), and the $(k,r)$-cores at $\tau$ already under equal weights (18--54\%).
With stubborn outsiders, at least 77\% of the groups of every method lose an interaction on the meetup graphs, whose outside friends lie in all directions; on the synthetic graphs at $\lambda=0.01$, about 1\% of the LCC groups do, against 52--68\% of the $(k,r)$-cores at $\tau$.
Up to $\nu=0.2$, tolerance noise barely affects LCC and pairwise groups (below 1\%), while 16--39\% of $(k,r)$-cores at $\tau$ lose an interaction.

\begin{table*}[!tp]
\caption{LCC against pairwise groups: sizes, and pairwise groups (\%) that lose an accepted interaction on $G[H]$.}\label{tab:tr:pairwise}
\centering\scriptsize\setlength{\tabcolsep}{4pt}
\input{tables/tr_pairwise}
\end{table*}

\paragraph{LCC or pairwise}
\Cref{tab:tr:pairwise} provides the results for the paragraph ``LCC or pairwise'' of \cref{sec:exp:dynamics}.
It counts queries for which both LCC and pairwise groups exist and are proven optimal, the share where the pairwise group is strictly larger, and the median pairwise-to-LCC size ratio across these queries.
The other columns give the pairwise groups (\%) that lose an interaction accepted at the start, for the groups of \cref{tab:dyn}: in any of 10 runs, as in \cref{tab:dyn}, and on average per run; no LCC group ever does.
Pairwise groups are strictly larger than LCCs for most queries, and under uneven influence they lose an accepted interaction in a few of their runs: on average 5--13\% of the runs under concentrated weights, 1--5\% under gossip, and at most 2\% when each member follows one neighbour.
On Gowalla and Brightkite, the pairwise group is the LCC itself for 938 of the 978 and 177 of the 181 queries on which the two have the same size (the paired co-visit difference on these is $+0.14$\pp\ $[-0.03,0.36]$ and $0.00$\pp), and on the 1{,}491 and 281 queries on which it is larger (median 52 against 34 and 17 against 12 members), the LCC's pair rate is higher by $+1.49$\pp\ $[1.08,1.95]$ and $+1.77$\pp\ $[1.12,2.50]$.
The size-adjusted difference~(b) of \cref{sec:exp:meetup}, a linear regression of the difference on the log size ratio, gives intercepts of $-0.85$\pp\ $[-1.35,-0.37]$ and $-0.29$\pp\ $[-0.69,0.11]$.
The relation is far from linear: on Gowalla, the mean difference is $+0.14$\pp\ for log size ratios in $[0,0.25)$, $+1.2$ and $+0.8$\pp\ in $[0.25,0.5)$ and $[0.5,1)$, and $+7.6$ and $+27$\pp\ for the 67 and 32 queries with log ratios in $[1,2)$ and above 2, which set the slope; without the latter 99 queries, the intercept is $+0.10$\pp.
On Brightkite, the intercept stays negative without the queries of log ratio at least 1 ($-0.38$\pp).
The fitted line also misses the equal-size queries for size-matched SAC groups ($+0.01$ against $+0.19$\pp\ on Gowalla, $-0.05$ against $+0.30$\pp\ on Brightkite), so~(b) is a coarse adjustment for every baseline, and with pairwise groups, whose equal-size groups are almost all the LCC itself, it rests on extrapolation.
On Yelp, the pairwise group has the LCC's size for only 36 of the 4{,}951 queries where both searches return a group (median 443 against 365 members; about 10\% of the pairwise groups are not proven maximal within the limit or the cap on searched sets, so their sizes are lower bounds), so we compare them by the size-adjusted difference~(b): $-0.05$\pp\ $[-0.12,0.02]$ at the default setting, $-0.31$\pp\ at $k=4$ and $-0.13$\pp\ with tolerances at the 90th percentile (both CIs below 0), and CIs containing 0 at $k=2$, $k=5$ and the 70th percentile.

\subsection{Refusals in Pairwise Groups}\label{app:p4}

\begin{table}[!tp]
\caption{Refused interactions in pairwise groups under concentrated weights ($d=d(x_v,x_u)$; $v$ refuses $u$).}\label{tab:tr:p4}
\centering\scriptsize\setlength{\tabcolsep}{3pt}
\input{tables/tr_p4}
\end{table}

\begin{figure}[!tp]
\centering
\includegraphics[width=\columnwidth]{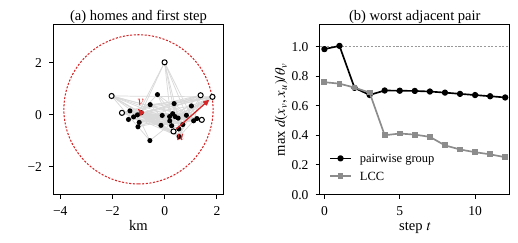}
\caption{A refusal in a Gowalla pairwise group: (a) homes and first moves; (b) worst adjacent pair per step.}\label{fig:tr:p4}
\end{figure}

\Cref{tab:tr:p4} examines the refusals behind the column of concentrated weights in \cref{tab:dyn} on Gowalla and Brightkite.
For the queries with an LCC, we replayed every run of the pairwise and LCC groups of that column with the same random streams and an instrumented copy of the simulator that records, for each directed interaction ($v$ accepts $u$) accepted at the start, the first step at which it is refused; the replay reproduces the stored results of every group and run exactly, and no LCC group ever refuses an interaction.
The table counts refused interactions over all runs; a refuser counts as the least tolerant member if its tolerance equals the group's minimum (ties included), and as one of the three least tolerant if it is the least tolerant or its average rank from the least tolerant is at most~3 (tied members share their average rank).

At the start, refused interactions are not close to tolerance (median $d/\theta_v$ of 0.55, and at least 0.9 for only 9\% and 12\%).
Refusal occurs mostly at the first step. During the run, the largest distance excess over $\theta_v$ has a median of 8\% and 10\%; by the end, all but about 1\% are accepted again.
The refuser is one of the three least tolerant members of the group in 73\% and 94\% of the refused interactions and is (or ties for) the least tolerant one in 45\% and 73\%; the least tolerant member is the one that bounds the diameter in~(iii).
In 84\% and 82\% of the refused interactions, the LCC of the same query excludes the refuser, the refused member or both, while the pairwise groups with a refusal have a median of 22.5 and 5 members outside the LCC.
These refusals use simulated influence weights. Closed Congress dynamics with weights proportional to observed interactions produce no loss in any pairwise group (\cref{sec:exp:congress}).

\Cref{fig:tr:p4} shows a median-sized case: among the 250 Gowalla groups of at most 80 members with a refusal, the group whose size is nearest the median of 33 (ties: lowest query), its first run with a refusal, and its earliest refused interaction.
Positions are shown relative to the group centroid.
Member $u$ ($\theta_u=15.3$\,km, outside the LCC) puts 99.8\% of its weight on one neighbour, also outside the LCC, and moves 2.0\,km in the first step; its distance to $v$ ($\theta_v=2.86$\,km, the second least tolerant of the 33 members, in the LCC of 27 members) grows from 0.50 to $1.004\,\theta_v$ and is back within $\theta_v$ from the second step on.
For the same query and run index, every adjacent pair in the LCC remains within 0.76 times its smaller tolerance.

\subsection{Heterogeneity of Tolerances}\label{app:sigma}

\begin{table*}[!tp]
\caption{Heterogeneity sweep on LiveJournal (L) and Orkut (O), count-order sweep.}\label{tab:tr:e2}
\centering\scriptsize\setlength{\tabcolsep}{3.2pt}
\input{tables/tr_e2}
\end{table*}

\Cref{tab:tr:e2} provides further results for the paragraph on heterogeneity in \cref{sec:exp:efficiency}.
It uses 200 default queries per setting and the count-order sweep (Exact 60\,s, the other searches 10\,s).
For each $\sigma$, it gives the queries (\%) with an LCC and those reaching the 60\,s limit, out of all; for the other methods, columns (iii) holds, Members and Size/LCC give the groups (\%) satisfying~(iii), the mean share (\%) of members who face another member beyond their own tolerance, and the median size relative to the LCC where both exist; Global $(k,r)$ is the heuristic $(k,r)$-core search with the median tolerance.
With equal tolerances, every $(k,r)$-core at $\tau$ and every pairwise group satisfies~(iii), and their median size equals the LCC's; Exact reaches the limit on 45\% of the Orkut queries there, the setting in which the LCC is a maximum $(k,r)$-core.
As $\sigma$ grows, fewer queries have an LCC, and the $(k,r)$-cores at $\tau$ grow to 1.25--4.21 times the LCC's size (median) while at most 9\% of them satisfy~(iii).
The reported time-limit shares use the count-order sweep; the best-first sweep is used in \cref{tab:eff}.

\subsection{Synthetic Robustness}\label{app:e4}

\begin{table*}[!tp]
\caption{Synthetic settings: scale, query sampling and community mixing.}\label{tab:tr:e4}
\centering\scriptsize\setlength{\tabcolsep}{4pt}
\input{tables/tr_e4}
\end{table*}

\Cref{tab:tr:e4} provides further results for the paragraph on the tolerance scale and the proven share in \cref{sec:exp:efficiency}, with 200 queries per setting and a 60\,s limit.
Column LCC counts the queries with an LCC; the next columns give shares (\%) of these queries: proven, reaching the limit, and stopped at the cap on searched sets; the last column gives, for the queries reaching the limit, the median ratio (\%) of the returned LCC to the proven upper bound.
Here the mean tolerance is $c$ times the median distance between friends (106.9 on LiveJournal and 126 on Orkut, mostly between communities), without calibration on Exact; $c=0.5$ already corresponds to scales $s$ of 9.4 and 11.1, against the calibrated 1.5 and 1.25.
We also draw queries uniformly from the whole $k$-core instead of from groups of at least 20 members, and move a share $\mu$ of the vertices to a random community.
Where tolerances reach the scale between communities ($c\ge0.75$ on Orkut, $c\ge1$ on LiveJournal), 35--99\% of the queries with an LCC reach the time limit and the bounds for these queries are loose; in the other settings, Exact proves every query.

\subsection{Caps and Time Limit at the Scale between Communities}\label{app:caps}

\begin{table*}[!tp]
\caption{Proven share (\%) of the queries with an LCC when one cap or the time limit changes (200 queries per setting).}\label{tab:tr:caps}
\centering\scriptsize\setlength{\tabcolsep}{3pt}
\input{tables/tr_caps}
\end{table*}

\Cref{tab:tr:caps} varies one parameter of \cref{sec:algorithm:exact} at a time on the three settings of \cref{tab:tr:e4} where Exact starts to time out, with the same instances and queries: the cap on searched sets (default 3{,}000), the cap on the root bound (default 20{,}000), the number of thresholds Greedy tries for the initial incumbent (default 20), and, for the queries left unproven at the default (C0), a time limit of 300\,s instead of 60\,s.
Raising a cap changes the proven share by at most $+1.1$\pp\ (two queries on Orkut at $c=0.75$); with a larger cap on searched sets, the queries that stopped at that cap under C0 (0.6--1.1\%) reach the time limit instead.
Letting Greedy try all thresholds lowers the share by 3.3 and 2.3\pp\ on LiveJournal at $c=1$ and Orkut at $c=0.75$, because the longer initial phase consumes part of the limit.
Under a 300\,s limit, Exact proves 5 of 55, 1 of 87 and 0 of 142 queries. Across configurations, the median returned-LCC-to-bound ratio remains 0.18--1.3\%.
Every configuration returns the same optimum wherever it proves one, in agreement with \cref{tab:tr:e4}; the queries with an LCC have a median of 1{,}218 (LiveJournal), 8{,}360 and 210{,}146 (Orkut, $c=0.75$ and 1) candidate vertices, and each engine process uses at most 2.7\,GB.

\subsection{Sweep against a Single Threshold}\label{app:single}

\begin{table*}[!tp]
\caption{The sweep and single-threshold $(k,r)$-core searches on the 200 default queries of \cref{tab:solvers} (60\,s limit).}\label{tab:tr:single}
\centering\scriptsize\setlength{\tabcolsep}{2.4pt}
\input{tables/tr_single}
\end{table*}

\Cref{tab:tr:single} runs our engine on one threshold only, as a maximum $(k,r)$-core search containing $q$ without personal tolerances, at the median tolerance of all vertices (the radius of the global baseline) and at the threshold $\tau^*$ of the optimal LCC, which the single-threshold search does not know in advance; the latter is run only on queries with an LCC.
Time ratios are computed per query with an LCC, flooring both times at 1\,ms; the last column reports the share of these queries for which the sweep is faster.
Query times exclude index construction, which the single-threshold search does not use. The index is built once per process; in this run it took 5.7\,s on LiveJournal and 14.3\,s on Orkut (a slower run than the 9.7\,s of \cref{sec:exp:efficiency}). On LiveJournal, where the sweep answers most queries in under 1\,ms, amortizing this cost over 200 queries would raise the median ratio to about 39.
The non-empty single-threshold answers satisfy~(iii) on 0\% of the queries at the median tolerance and on 4.2\% (LiveJournal) and 0\% (Orkut) at $\tau^*$.

\subsection{Unproven Efficiency Queries}\label{app:unproven}

\begin{table}[!tp]
\caption{The nine efficiency queries that Exact leaves unproven within 60\,s.}\label{tab:tr:unproven}
\centering\scriptsize\setlength{\tabcolsep}{2.4pt}
\input{tables/tr_unproven}
\end{table}

\Cref{tab:tr:unproven} lists the unproven queries of \cref{sec:exp:efficiency}, with the size of the returned LCC, the proven upper bound UB, their ratio (truncated), the candidate vertices (those with a non-empty survival interval), the thresholds evaluated out of all, and the branch-and-bound nodes.
Every candidate set $C(\tau)$ is a subset of the candidate vertices, of which these queries have at most 918 and no efficiency query has more than 1{,}502, so the size caps of \cref{sec:algorithm:exact} never apply; the search explores 60{,}421--551{,}931 nodes without settling all thresholds.
The bound pass after the limit, which query times exclude, takes at most 4\,\textmu{}s on these queries; on all other efficiency queries, it does not run.

\subsection{Meetups}\label{app:meetup}

\begin{table}[t]
\caption{The values of \cref{fig:meetup} (pp, 95\% bootstrap CIs).}\label{tab:meetup}
\centering\scriptsize\setlength{\tabcolsep}{2.2pt}
\input{tables/tab_meetup}
\end{table}

\Cref{tab:meetup} gives the values plotted in \cref{fig:meetup}: LCC minus size-matched SAC where both exist, in the share of member pairs that co-visit and of members who co-visit with anyone in the group: raw, on queries where SAC returns a group of the LCC's size with other members~(a), and as the intercept of a regression on the log size ratio~(b).

\subsection{Yelp on a Second Time Split}\label{app:yelpsplit}

\begin{table*}[!tp]
\caption{Yelp, LCC minus size-matched SAC (pp, 95\% bootstrap CIs) on the two time splits.}\label{tab:tr:yelpsplit}
\centering\scriptsize\setlength{\tabcolsep}{2.4pt}
\input{tables/tr_yelpsplit}
\end{table*}

The split of \cref{sec:exp:meetup} (training up to 2017, validation 2018, test 2019) was chosen from the co-visit rate of all friend pairs in the test period; the validation-period rate (6.82\%) and the number of validation-active users in the 3-core (12{,}351) lead to the same split.
As a robustness check, we rebuilt the dataset on a split that ends before 2019 (training up to 2016, validation 2017, test 2018): 52{,}527 users, 246{,}633 friendships, 9{,}343 queries, of which 4{,}416 have an LCC (median 323 members; $\theta_v$ median 4.07\,km).
\Cref{tab:tr:yelpsplit} repeats the comparison with size-matched SAC groups on both splits; the two splits have different users, homes and graphs, so they are not paired.
In pair rate, both splits give the same picture: the raw LCC advantage does not persist at equal size, as (a) and~(b) do not both exclude 0.
Against $(k,r)$-cores at $\tau$, the size-adjusted difference~(b) on the second split is $+0.16$\pp\ $[0.12,0.21]$ in pair rate and $+0.71$\pp\ $[0.42,1.01]$ in member rate.

\subsection{Queries without an LCC and Tolerance Scaling}\label{app:empty}

\begin{table*}[!tp]
\caption{Queries without an LCC ($k=3$) by cause, and with raised tolerances.}\label{tab:tr:p21}
\centering\scriptsize\setlength{\tabcolsep}{3.4pt}
\input{tables/tr_p21}
\end{table*}

\Cref{tab:tr:p21} provides the classification for the paragraph on empty LCCs in \cref{sec:exp:meetup}.
At the median tolerance and on all queries, a query without an LCC is in class A if the $k$-core of $q$'s component restricted to friendships whose ends accept each other is empty, in B if it is not but no pairwise group exists, in C if a pairwise group exists but no LCC, and in U if the LCC or pairwise search is not proven; columns A--U give shares (\%) of the queries without an LCC (Congress: counts).
The next two columns give the share of these queries that gain an LCC when every tolerance is the 70th or 90th percentile of the user's distances, and the last five the share of queries without an LCC at the 90th percentile and its classes; raised tolerances use the 2{,}000 sampled queries of the parameter study, and on Congress the two columns count all 18 queries without an LCC.
Raising tolerances to the 90th percentile gives an LCC to 605 of 1{,}698, 606 of 1{,}822 and 515 of 1{,}036 sampled queries without one. Queries of classes B and C almost all gain an LCC at the 90th percentile (B: 45 of 45, 50 of 51 and 67 of 68; C: 41 of 41, 8 of 8 and 47 of 48 on Gowalla, Brightkite and Yelp), but only 31--43\% of class A do (519 of 1{,}612, 548 of 1{,}763 and 390 of 907).
Newly obtained LCCs have median sizes of 21.5, 13 and 280 members at the 70th percentile, and 103, 42 and 522 at the 90th. On Yelp, 22 of 243 and 106 of 515 are not proven optimal, making their sizes lower bounds.

\begin{table*}[!tp]
\caption{Tolerance scaling: every tolerance multiplied by $s$.}\label{tab:tr:p24}
\centering\scriptsize\setlength{\tabcolsep}{3.4pt}
\input{tables/tr_p24}
\end{table*}

\Cref{tab:tr:p24} provides the full tolerance-scaling results for the same paragraph, on the same queries at every $s$ (Yelp: 2{,}000 sampled queries) with Exact limited to 60\,s.
It gives the queries (\%) with an LCC, the queries (\%) proven within the limit, the median LCC size, and the queries without an LCC by class (\cref{tab:tr:p21}).
For $s<1$, column Kept gives the share (\%) of the queries with an LCC at $s=1$ that keep one, with the median size ratio of the kept LCCs in parentheses; for $s>1$, column Gained gives the share of the queries without an LCC at $s=1$ that gain one.
The last two columns report the mean co-visiting share of member pairs in LCC groups and its difference from $s=1$, for queries with an LCC at both scales (pp, 95\% bootstrap CI).
Whenever both solutions are proven, increasing $s$ never shrinks an LCC (7 Yelp cases cannot be checked because a solution remains unproven).
Larger tolerances give more and larger LCCs whose pairs co-visit less, and smaller ones fewer and smaller LCCs whose pairs co-visit more; all paired differences exclude 0 except on Brightkite at $s=0.8$.
Class A dominates queries without an LCC at every $s$.
Shrinking tolerances also protects against noisy estimates: with tolerance noise $\nu=0.3$ in equal-weight HK on $G[H]$ (10 runs; a separate run on all LCC groups, with different random draws from those in \cref{tab:tr:e1a}), 1.17\% $[0.82,1.68]$ of the Gowalla and 3.68\% $[2.31,5.81]$ of the Brightkite LCC groups at $s=1$ lose an interaction accepted at the start, against 0\% $[0,0.19]$ and 1.08\% $[0.42,2.74]$ at $s=0.8$ (Wilson intervals; without noise, none does).

\subsection{Open Groups at Group Level}\label{app:open}

\begin{table*}[!tp]
\caption{Open LCC groups: losses and final opinion clusters.}\label{tab:tr:p22}
\centering\scriptsize\setlength{\tabcolsep}{2.2pt}
\input{tables/tr_p22}
\end{table*}

\Cref{tab:tr:p22} covers the LCC groups of \cref{tab:tr:e1a} on Gowalla and Brightkite and the 454 LCC groups on Congress, when each accepted outside friend has weight $w$ or, on Congress, a weight proportional to the observed interactions (obs.).
It gives the lost share of the interactions accepted at the start, in groups that lose one, and the largest final opinion cluster, by single linkage at $10^{-3}$ of the initial diameter after at most 300 steps (closed groups: 99.7--100\%); its last rows give LCC minus size-matched SAC in this cluster (closed groups: $+1.6$, $+1.7$ and 0\pp).
Interaction loss and group-wide convergence are separate outcomes: open groups lose only a small share of interactions, usually early.
Except on Congress at $w=0.01$, open groups also fail to reach a common opinion whether or not they lose an interaction: members stay spread between the outside friends they average with, so the largest final cluster holds a minority of the members of LCC groups that lose no interaction as well (22--35\% on Gowalla and Brightkite), and on Congress at $w=0.1$, where no LCC group loses an interaction, only 11.8\%.
Pairwise groups, size-matched SAC groups and $(k,r)$-cores at $\tau$ also have largest final clusters of 15--28\%. Paired differences relative to size-matched SAC vary in sign, so open LCC groups have no consistent group-level advantage.

%% file: tables/tr_e1_aware.tex
\begin{tabular}{lrrrrrrrrrrrrrrrrrrrrr}
\toprule
 & \multicolumn{4}{c}{LCC (Exact)} & \multicolumn{4}{c}{Greedy} & \multicolumn{4}{c}{BallCore-R} & \multicolumn{4}{c}{Pairwise} & \multicolumn{4}{c}{Edge-accept} & \multicolumn{1}{c}{LCC groups} \\
\cmidrule(lr){2-5}\cmidrule(lr){6-9}\cmidrule(lr){10-13}\cmidrule(lr){14-17}\cmidrule(lr){18-21}
Variant & G & B & L & O & G & B & L & O & G & B & L & O & G & B & L & O & G & B & L & O & G/B/L/O \\
\midrule
\multicolumn{22}{l}{\emph{Closed group, interactions on $G[H]$}} \\
\quad HK, equal weights & 0 & 0 & 0 & 0 & 0 & 0 & 0 & 0 & $<$1 & 0 & 0 & 0 & 0 & 0 & 0 & 0 & 67 & 63 & -- & -- & 1{,}000/462/95/115 \\
\quad HK, equal, 3{,}000 steps & 0 & 0 & 0 & 0 & 0 & 0 & 0 & 0 & $<$1 & 0 & 0 & 0 & 0 & 0 & 0 & 0 & 67 & 63 & -- & -- & 1{,}000/462/95/115 \\
\quad HK, random weights, $\alpha=1$ & 0 & 0 & 0 & 0 & 0 & 0 & 0 & 0 & 19 & 7 & 0 & $<$1 & 1 & $<$1 & 0 & 0 & 78 & 77 & -- & -- & 1{,}000/462/95/115 \\
\quad HK, random weights, $\alpha=0.1$ & 0 & 0 & 0 & 0 & 0 & 0 & 0 & 0 & 62 & 56 & 64 & 76 & 40 & 33 & 32 & 40 & 85 & 85 & -- & -- & 1{,}000/462/95/115 \\
\quad Follow one neighbour & 0 & 0 & 0 & 0 & 0 & 0 & 0 & 0 & 45 & 31 & 3 & 12 & 7 & 8 & 0 & $<$1 & 81 & 80 & -- & -- & 1{,}000/462/95/115 \\
\quad Gossip (DW) & 0 & 0 & 0 & 0 & 0 & 0 & 0 & 0 & 54 & 38 & 28 & 58 & 17 & 11 & 3 & 7 & 82 & 82 & -- & -- & 1{,}000/462/95/115 \\
\quad SJT, new rejection & 0 & 0 & 0 & 0 & 0 & 0 & 0 & 0 & 29 & 15 & 49 & 82 & $<$1 & 0 & 0 & $<$1 & 74 & 73 & -- & -- & 1{,}000/462/95/115 \\
\midrule
\multicolumn{22}{l}{\emph{Closed group, complete interaction graph}} \\
\quad HK, equal weights & 0 & 0 & 0 & 0 & 0 & 0 & 0 & 0 & $<$1 & 0 & 0 & 0 & 0 & 0 & 0 & 0 & 57 & 62 & -- & -- & 1{,}000/462/95/115 \\
\quad HK, random weights, $\alpha=1$ & 0 & 0 & 0 & 0 & 0 & 0 & 0 & 0 & 2 & 2 & 0 & 0 & 0 & 0 & 0 & 0 & 57 & 73 & -- & -- & 931/462/95/115 \\
\quad HK, random weights, $\alpha=0.1$ & 0 & 0 & 0 & 0 & 0 & 0 & 0 & 0 & 29 & 39 & 11 & 12 & 13 & 23 & 5 & 4 & 76 & 85 & -- & -- & 931/462/95/115 \\
\quad Follow one neighbour & 0 & 0 & 0 & 0 & 0 & 0 & 0 & 0 & 51 & 46 & 10 & 33 & 30 & 17 & 1 & $<$1 & 72 & 82 & -- & -- & 931/462/95/115 \\
\quad Gossip (DW) & 0 & 0 & 0 & 0 & 0 & 0 & 0 & 0 & 47 & 44 & 7 & 43 & 27 & 21 & 6 & 17 & 66 & 72 & -- & -- & 907/462/63/71 \\
\quad SJT, new rejection & 0 & 0 & 0 & 0 & 0 & 0 & 0 & 0 & 16 & 17 & 65 & 87 & $<$1 & 0 & 0 & 2 & 70 & 72 & -- & -- & 1{,}000/462/95/115 \\
\midrule
\multicolumn{22}{l}{\emph{Assumptions violated}} \\
\quad Open group, $G[H]$ & 65 & 62 & 0 & 0 & 63 & 63 & 0 & 0 & 74 & 70 & 0 & 0 & 72 & 67 & 0 & 0 & 83 & 85 & -- & -- & 1{,}000/462/95/115 \\
\quad Open group, complete & 61 & 57 & 0 & 0 & 57 & 57 & 0 & 0 & 67 & 62 & 0 & 0 & 65 & 60 & 0 & 0 & 78 & 86 & -- & -- & 1{,}000/462/95/115 \\
\quad Stubborn outsiders, $\lambda=0.01$ & 91 & 94 & 1 & $<$1 & 90 & 94 & 1 & 1 & 93 & 97 & 3 & 2 & 93 & 97 & 0 & 2 & 97 & 100 & -- & -- & 1{,}000/462/95/115 \\
\quad Stubborn outsiders, $\lambda=0.05$ & 98 & 99 & 49 & 33 & 98 & 99 & 39 & 26 & 98 & 100 & 73 & 46 & 98 & 100 & 65 & 38 & 99 & 100 & -- & -- & 1{,}000/462/95/115 \\
\quad Stubborn outsiders, $\lambda=0.1$ & 99 & 100 & 91 & 71 & 98 & 100 & 88 & 68 & 99 & 100 & 93 & 84 & 99 & 100 & 93 & 80 & 99 & 100 & -- & -- & 1{,}000/462/95/115 \\
\quad Tolerance noise, $\nu=0.1$ & 0 & 0 & 0 & 0 & 0 & 0 & 0 & 0 & 3 & 1 & 0 & 0 & 0 & 0 & 0 & 0 & 70 & 69 & -- & -- & 1{,}000/462/95/115 \\
\quad Tolerance noise, $\nu=0.2$ & 0 & $<$1 & 0 & 0 & $<$1 & 0 & 0 & 0 & 16 & 5 & $<$1 & 3 & 0 & $<$1 & 0 & 0 & 73 & 73 & -- & -- & 1{,}000/462/95/115 \\
\quad Tolerance noise, $\nu=0.3$ & 2 & 3 & 0 & 3 & 2 & 4 & 0 & 1 & 31 & 21 & 6 & 12 & 3 & 4 & $<$1 & 2 & 76 & 75 & -- & -- & 1{,}000/462/95/115 \\
\midrule
Groups (HK, equal, $G[H]$) & 1{,}000 & 462 & 95 & 115 & 887 & 447 & 83 & 93 & 1{,}004 & 467 & 107 & 130 & 1{,}003 & 464 & 107 & 129 & 1{,}011 & 471 & -- & -- & \\
\bottomrule
\end{tabular}

%% file: tables/tr_e1_agnostic.tex
\begin{tabular}{lrrrrrrrrrrrrrrrrrrrr}
\toprule
 & \multicolumn{4}{c}{$(k,r)$-core at $\tau$} & \multicolumn{4}{c}{Global $(k,r)$-core} & \multicolumn{4}{c}{SAC} & \multicolumn{4}{c}{Size-matched SAC} & \multicolumn{4}{c}{Random} \\
\cmidrule(lr){2-5}\cmidrule(lr){6-9}\cmidrule(lr){10-13}\cmidrule(lr){14-17}\cmidrule(lr){18-21}
Variant & G & B & L & O & G & B & L & O & G & B & L & O & G & B & L & O & G & B & L & O \\
\midrule
\multicolumn{21}{l}{\emph{Closed group, interactions on $G[H]$}} \\
\quad HK, equal weights & 15 & 17 & 4 & 13 & 60 & 33 & 7 & 6 & 18 & 32 & 1 & 3 & 5 & 8 & 1 & 2 & 37 & 27 & 0 & 0 \\
\quad HK, equal, 3{,}000 steps & 15 & 17 & 4 & 13 & 60 & 33 & 7 & 6 & 18 & 32 & 1 & 3 & 5 & 8 & 1 & 2 & 37 & 27 & 0 & 0 \\
\quad HK, random weights, $\alpha=1$ & 26 & 35 & 43 & 68 & 77 & 81 & 53 & 77 & 26 & 43 & 22 & 42 & 21 & 29 & 38 & 59 & 55 & 45 & 21 & 26 \\
\quad HK, random weights, $\alpha=0.1$ & 41 & 44 & 91 & 97 & 86 & 89 & 99 & 99 & 41 & 56 & 78 & 84 & 52 & 56 & 89 & 98 & 64 & 53 & 75 & 69 \\
\quad Follow one neighbour & 33 & 39 & 58 & 78 & 82 & 81 & 73 & 86 & 30 & 45 & 26 & 41 & 35 & 35 & 48 & 67 & 55 & 45 & 15 & 24 \\
\quad Gossip (DW) & 34 & 40 & 69 & 85 & 83 & 82 & 81 & 93 & 32 & 47 & 38 & 55 & 39 & 40 & 63 & 86 & 57 & 47 & 39 & 31 \\
\quad SJT, new rejection & 25 & 31 & 92 & 98 & 75 & 71 & 99 & 99 & 28 & 47 & 61 & 83 & 21 & 31 & 85 & 97 & 90 & 89 & 98 & 97 \\
\midrule
\multicolumn{21}{l}{\emph{Closed group, complete interaction graph}} \\
\quad HK, equal weights & 18 & 23 & 33 & 54 & 59 & 50 & 39 & 56 & 15 & 31 & 13 & 25 & 8 & 14 & 15 & 34 & 53 & 40 & 2 & 2 \\
\quad HK, random weights, $\alpha=1$ & 18 & 31 & 37 & 69 & 62 & 68 & 43 & 76 & 18 & 40 & 17 & 42 & 17 & 28 & 29 & 65 & 65 & 61 & 29 & 37 \\
\quad HK, random weights, $\alpha=0.1$ & 29 & 45 & 89 & 95 & 79 & 87 & 97 & 99 & 31 & 51 & 74 & 83 & 37 & 47 & 88 & 98 & 74 & 68 & 81 & 84 \\
\quad Follow one neighbour & 36 & 43 & 70 & 89 & 77 & 86 & 78 & 96 & 30 & 44 & 40 & 58 & 45 & 48 & 63 & 88 & 67 & 61 & 39 & 39 \\
\quad Gossip (DW) & 27 & 42 & 54 & 78 & 73 & 81 & 74 & 84 & 28 & 43 & 34 & 53 & 41 & 46 & 50 & 80 & 66 & 61 & 33 & 30 \\
\quad SJT, new rejection & 24 & 29 & 92 & 100 & 71 & 75 & 99 & 99 & 23 & 44 & 67 & 85 & 19 & 30 & 93 & 97 & 78 & 78 & 93 & 97 \\
\midrule
\multicolumn{21}{l}{\emph{Assumptions violated}} \\
\quad Open group, $G[H]$ & 75 & 80 & 4 & 12 & 84 & 84 & 5 & 9 & 61 & 74 & 4 & 11 & 73 & 79 & 5 & 14 & 49 & 37 & 0 & $<$1 \\
\quad Open group, complete & 73 & 81 & 33 & 55 & 80 & 87 & 39 & 57 & 58 & 72 & 14 & 31 & 70 & 80 & 15 & 44 & 60 & 46 & 2 & 3 \\
\quad Stubborn outsiders, $\lambda=0.01$ & 96 & 99 & 52 & 68 & 97 & 99 & 64 & 76 & 92 & 97 & 24 & 29 & 95 & 99 & 38 & 45 & 80 & 77 & 12 & 9 \\
\quad Stubborn outsiders, $\lambda=0.05$ & 98 & 100 & 83 & 90 & 99 & 100 & 97 & 97 & 97 & 99 & 59 & 60 & 98 & 100 & 79 & 82 & 89 & 86 & 48 & 35 \\
\quad Stubborn outsiders, $\lambda=0.1$ & 99 & 100 & 96 & 97 & 99 & 100 & 99 & 100 & 98 & 99 & 80 & 84 & 99 & 100 & 95 & 97 & 91 & 88 & 80 & 70 \\
\quad Tolerance noise, $\nu=0.1$ & 17 & 22 & 16 & 32 & 65 & 58 & 15 & 20 & 17 & 33 & 4 & 10 & 9 & 13 & 6 & 10 & 44 & 32 & 0 & $<$1 \\
\quad Tolerance noise, $\nu=0.2$ & 23 & 27 & 21 & 39 & 70 & 65 & 24 & 34 & 20 & 37 & 8 & 12 & 15 & 19 & 9 & 14 & 48 & 37 & 0 & 3 \\
\quad Tolerance noise, $\nu=0.3$ & 28 & 35 & 25 & 38 & 74 & 73 & 31 & 40 & 26 & 39 & 10 & 16 & 23 & 29 & 15 & 23 & 51 & 42 & 1 & 2 \\
\midrule
Groups (HK, equal, $G[H]$) & 1{,}000 & 462 & 95 & 115 & 994 & 428 & 150 & 162 & 1{,}243 & 745 & 153 & 178 & 1{,}000 & 462 & 95 & 115 & 1{,}000 & 462 & 95 & 115 \\
\bottomrule
\end{tabular}

%% file: tables/tr_pairwise.tex
\begin{tabular}{lrrrrrrrrr}
\toprule
 & & \multicolumn{2}{c}{Pairwise vs.\ LCC} & \multicolumn{2}{c}{Concentrated} & \multicolumn{2}{c}{Gossip} & \multicolumn{2}{c}{Follow one} \\
\cmidrule(lr){3-4}\cmidrule(lr){5-6}\cmidrule(lr){7-8}\cmidrule(lr){9-10}
Graph & Queries & Larger (\%) & Size ratio & Any & Per run & Any & Per run & Any & Per run \\
\midrule
Gowalla & 2{,}469 & 60.4 & 1.08 & 40.4 & 13.2 & 16.6 & 5.1 & 6.9 & 1.9 \\
Brightkite & 462 & 60.8 & 1.09 & 33.4 & 8.1 & 10.6 & 2.9 & 7.8 & 1.5 \\
LiveJournal & 87 & 93.1 & 1.33 & 31.8 & 4.8 & 2.8 & 0.8 & 0.0 & 0.0 \\
Orkut & 102 & 97.1 & 1.52 & 39.5 & 7.8 & 7.0 & 1.7 & 0.8 & 0.2 \\
\bottomrule
\end{tabular}

%% file: tables/tr_p4.tex
\begin{tabular}{lrr}
\toprule
 & Gowalla & Brightkite \\
\midrule
Pairwise groups with a refusal & 402 of 1{,}000 & 153 of 462 \\
Refused interactions, all runs & 2{,}200 & 592 \\
Initial $d/\theta_v$, median [quartiles] & 0.55 [0.36, 0.73] & 0.55 [0.28, 0.77] \\
Initial $d/\theta_v\ge0.9$ (\%) & 8.9 & 11.8 \\
Step of refusal, median & 1 & 1 \\
Largest $d/\theta_v$, median & 1.08 & 1.10 \\
Still refused at the end (\%) & 1.2 & 1.0 \\
Refuser is the least tolerant member (\%) & 45.3 & 73.3 \\
Refuser among the three least tolerant (\%) & 72.9 & 93.9 \\
Refuser outside the LCC (\%) & 69.1 & 71.1 \\
Refuser or refused outside the LCC (\%) & 83.7 & 82.4 \\
Pairwise members outside the LCC, median & 22.5 & 5 \\
\bottomrule
\end{tabular}

%% file: tables/tr_e2.tex
\begin{tabular}{lrrrrrrrrrrrrrr}
\toprule
 & \multicolumn{2}{c}{With LCC} & \multicolumn{2}{c}{Exact limit} & \multicolumn{6}{c}{$(k,r)$-core at $\tau$} & \multicolumn{2}{c}{Pairwise} & \multicolumn{2}{c}{Global $(k,r)$} \\
\cmidrule(lr){2-3}\cmidrule(lr){4-5}\cmidrule(lr){6-11}\cmidrule(lr){12-13}\cmidrule(lr){14-15}
 & & & & & \multicolumn{2}{c}{(iii) holds} & \multicolumn{2}{c}{Members} & \multicolumn{2}{c}{Size/LCC} & \multicolumn{2}{c}{(iii) holds} & \multicolumn{2}{c}{Members} \\
\cmidrule(lr){6-7}\cmidrule(lr){8-9}\cmidrule(lr){10-11}\cmidrule(lr){12-13}\cmidrule(lr){14-15}
$\sigma$ & L & O & L & O & L & O & L & O & L & O & L & O & L & O \\
\midrule
0.0 & 76.5 & 85.5 & 0.5 & 45.0 & 100.0 & 100.0 & 0 & 0 & 1.00 & 1.00 & 100.0 & 100.0 & 0 & 0 \\
0.2 & 70.0 & 74.5 & 14.0 & 26.5 & 8.6 & 4.7 & 12 & 23 & 1.25 & 1.50 & 4.1 & 0.6 & 27 & 38 \\
0.4 (default) & 47.5 & 57.5 & 4.0 & 7.0 & 4.2 & 0.0 & 31 & 43 & 1.80 & 2.19 & 5.6 & 2.3 & 40 & 44 \\
0.6 & 39.0 & 44.0 & 0.5 & 3.0 & 2.6 & 1.1 & 43 & 52 & 2.54 & 2.93 & 7.1 & 3.1 & 45 & 47 \\
0.8 & 30.5 & 33.0 & 0.0 & 0.0 & 1.6 & 0.0 & 49 & 61 & 2.72 & 3.87 & 9.0 & 2.7 & 47 & 49 \\
1.0 & 21.0 & 27.5 & 0.0 & 0.0 & 2.4 & 1.8 & 55 & 65 & 3.50 & 4.21 & 11.1 & 6.9 & 49 & 48 \\
\bottomrule
\end{tabular}

%% file: tables/tr_e4.tex
\begin{tabular}{llrrrrr}
\toprule
 & & & Proven & Limit & Size cap & Solution/ \\
Graph & Setting & LCC & (\%) & (\%) & (\%) & bound (\%) \\
\midrule
LiveJournal & $c=0.5$ & 153 & 100.0 & 0.0 & 0.0 & -- \\
LiveJournal & $c=0.75$ & 153 & 100.0 & 0.0 & 0.0 & -- \\
LiveJournal & $c=1$ & 153 & 64.1 & 35.3 & 0.7 & 0.33 \\
LiveJournal & $c=1.5$ & 150 & 15.3 & 84.0 & 0.7 & 0.04 \\
LiveJournal & $c=2$ & 119 & 3.4 & 95.0 & 1.7 & 0.03 \\
Orkut & $c=0.5$ & 177 & 100.0 & 0.0 & 0.0 & -- \\
Orkut & $c=0.75$ & 177 & 50.8 & 48.0 & 1.1 & 0.91 \\
Orkut & $c=1$ & 177 & 19.8 & 79.7 & 0.6 & 0.18 \\
Orkut & $c=1.5$ & 161 & 3.1 & 96.3 & 0.6 & 0.40 \\
Orkut & $c=2$ & 177 & 0.6 & 99.4 & 0.0 & 7.6 \\
\midrule
LiveJournal & $q$ from the $k$-core & 91 & 100.0 & 0.0 & 0.0 & -- \\
Orkut & $q$ from the $k$-core & 92 & 100.0 & 0.0 & 0.0 & -- \\
\midrule
LiveJournal & mixing $\mu=0.1$ & 80 & 100.0 & 0.0 & 0.0 & -- \\
LiveJournal & mixing $\mu=0.2$ & 68 & 100.0 & 0.0 & 0.0 & -- \\
LiveJournal & mixing $\mu=0.4$ & 42 & 100.0 & 0.0 & 0.0 & -- \\
Orkut & mixing $\mu=0.1$ & 103 & 100.0 & 0.0 & 0.0 & -- \\
Orkut & mixing $\mu=0.2$ & 83 & 100.0 & 0.0 & 0.0 & -- \\
Orkut & mixing $\mu=0.4$ & 51 & 100.0 & 0.0 & 0.0 & -- \\
\bottomrule
\end{tabular}

%% file: tables/tr_caps.tex
\begin{tabular}{lrrrrrr}
\toprule
 & & \multicolumn{4}{c}{Proven (\%) with} & 300\,s: proven / \\
\cmidrule(lr){3-6}
Setting (LCC) & C0 & cap 20{,}000 & cap 50{,}000 & root 100{,}000 & Greedy all $\tau$ & unproven under C0 \\
\midrule
LiveJournal $c=1$ (153) & 64.1 & 64.1 & 64.1 & 64.1 & 60.8 & 5 / 55 \\
Orkut $c=0.75$ (177) & 50.8 & 51.4 & 52.0 & 50.8 & 48.6 & 1 / 87 \\
Orkut $c=1$ (177) & 19.8 & 20.3 & 20.3 & 19.8 & 19.8 & 0 / 142 \\
\bottomrule
\end{tabular}

%% file: tables/tr_single.tex
\begin{tabular}{llrrrrrr}
\toprule
 & & \multicolumn{3}{c}{Queries (of 200)} & Median & \multicolumn{2}{c}{Time ratio, sweep / single} \\
\cmidrule(lr){3-5}\cmidrule(lr){7-8}
Graph & Search & Proven non-empty & Proven empty & Limit & time (ms) & Median & Sweep faster (\%) \\
\midrule
LiveJournal & Sweep (Exact) & 95 & 105 & 0 & 0.13 & -- & -- \\
LiveJournal & $(k,r)$-core at median $\theta$ & 131 & 50 & 19 & 2.1 & 0.89 & 55 \\
LiveJournal & $(k,r)$-core at $\tau^*$ (95 queries) & 79 & -- & 16 & 6.2 & 0.86 & 56 \\
Orkut & Sweep (Exact) & 115 & 85 & 0 & 3.2 & -- & -- \\
Orkut & $(k,r)$-core at median $\theta$ & 96 & 33 & 71 & 959 & 0.065 & 92 \\
Orkut & $(k,r)$-core at $\tau^*$ (115 queries) & 78 & -- & 37 & 2{,}586 & 0.135 & 94 \\
\bottomrule
\end{tabular}

%% file: tables/tr_unproven.tex
\begin{tabular}{llrrrrrr}
\toprule
Graph & Setting & LCC & UB & LCC/UB & Candidates & Thresholds & Nodes \\
\midrule
Friendster & default & 147 & 212 & 0.693 & 587 & 22/25 & 519{,}058 \\
Friendster & default & 305 & 336 & 0.907 & 560 & 83/84 & 263{,}320 \\
Friendster & default & 374 & 398 & 0.939 & 738 & 69/78 & 102{,}156 \\
LiveJournal & $2\times$ scale & 552 & 578 & 0.955 & 803 & 34/35 & 60{,}421 \\
Orkut & $2\times$ scale & 719 & 732 & 0.982 & 918 & 154/851 & 70{,}370 \\
Pokec & default & 169 & 245 & 0.689 & 622 & 77/83 & 539{,}455 \\
Pokec & default & 179 & 246 & 0.727 & 645 & 73/79 & 417{,}817 \\
Pokec & default & 189 & 253 & 0.747 & 578 & 323/484 & 551{,}931 \\
Pokec & default & 369 & 394 & 0.936 & 787 & 379/539 & 199{,}822 \\
\bottomrule
\end{tabular}

%% file: tables/tab_meetup.tex
\begin{tabular}{lrrr}
\toprule
 & Gowalla & Brightkite & Yelp \\
\midrule
Pairs, raw & $+0.75$\,{\scriptsize[0.50,0.99]} & $+0.56$\,{\scriptsize[0.15,1.01]} & $+0.14$\,{\scriptsize[0.11,0.18]} \\
\quad (a) same size & $+0.44$\,{\scriptsize[$-$0.15,1.02]} & $+0.74$\,{\scriptsize[$-$0.28,1.74]} & $+0.03$\,{\scriptsize[$-$0.01,0.08]} \\
\quad (b) adjusted & $+0.01$\,{\scriptsize[$-$0.24,0.24]} & $-0.05$\,{\scriptsize[$-$0.49,0.47]} & $+0.04$\,{\scriptsize[0.00,0.08]} \\
Members, raw & $-1.35$\,{\scriptsize[$-$1.83,$-$0.87]} & $+0.12$\,{\scriptsize[$-$0.84,1.08]} & $+1.03$\,{\scriptsize[0.84,1.22]} \\
\quad (a) same size & $-0.30$\,{\scriptsize[$-$1.52,0.99]} & $+2.86$\,{\scriptsize[0.96,4.79]} & $+0.84$\,{\scriptsize[0.61,1.06]} \\
\midrule
Queries, all / (a) & 2{,}469 / 610 & 462 / 108 & 5{,}010 / 2{,}387 \\
\bottomrule
\end{tabular}

%% file: tables/tr_yelpsplit.tex
\begin{tabular}{llrrr}
\toprule
Split & Measure & Raw & (a) Same size & (b) Intercept \\
\midrule
precovid (test 2019) & member pairs & $+0.14$ $[0.11,0.18]$ & $+0.03$ $[-0.01,0.08]$ & $+0.04$ $[0.00,0.08]$ \\
precovid (test 2019) & members & $+1.03$ $[0.84,1.22]$ & $+0.84$ $[0.61,1.06]$ & $+0.91$ $[0.73,1.10]$ \\
shift2018 (test 2018) & member pairs & $+0.19$ $[0.15,0.24]$ & $-0.01$ $[-0.06,0.04]$ & $+0.04$ $[-0.01,0.09]$ \\
shift2018 (test 2018) & members & $+1.55$ $[1.37,1.74]$ & $+0.98$ $[0.73,1.22]$ & $+1.27$ $[1.07,1.48]$ \\
\bottomrule
\end{tabular}

%% file: tables/tr_p21.tex
\begin{tabular}{lrrrrrrrrrrrrr}
\toprule
 & & & \multicolumn{4}{c}{Without LCC, median tolerance (\%)} & \multicolumn{2}{c}{Solved when raised (\%)} & \multicolumn{5}{c}{90th percentile, sample (\%)} \\
\cmidrule(lr){4-7}\cmidrule(lr){8-9}\cmidrule(lr){10-14}
Data & Queries & No LCC & A & B & C & U & 70th & 90th & No LCC & A & B & C & U \\
\midrule
Gowalla & 15{,}829 & 13{,}360 (84.4\%) & 95.0 & 3.2 & 1.8 & 0 & 15.0 & 35.6 & 54.7 & 85.1 & 11.8 & 3.1 & 0 \\
Brightkite & 5{,}260 & 4{,}798 (91.2\%) & 96.7 & 2.6 & 0.6 & 0 & 10.0 & 33.3 & 60.8 & 81.7 & 14.3 & 3.9 & 0 \\
Yelp & 10{,}363 & 5{,}353 (51.7\%) & 86.3 & 7.8 & 4.4 & 1.5 & 23.5 & 49.7 & 26.1 & 67.4 & 15.9 & 9.6 & 7.1 \\
Congress & 472 & 18 (3.8\%) & 14 & 2 & 2 & 0 & 12/18 & 16/18 & -- & -- & -- & -- & -- \\
\bottomrule
\end{tabular}

%% file: tables/tr_p24.tex
\begin{tabular}{lrrrrrrrrrrr}
\toprule
 & & & & \multicolumn{4}{c}{Queries without LCC} & & & \multicolumn{2}{c}{Pairs that co-visit} \\
\cmidrule(lr){5-8}\cmidrule(lr){11-12}
$s$ & LCC (\%) & Proven (\%) & Size & A & B & C & U & Kept (\%) & Gained (\%) & Share (\%) & vs.\ $s=1$ (pp) \\
\midrule
\multicolumn{12}{l}{\emph{Gowalla (15{,}829 queries)}} \\
0.5 & 7.9 & 100.0 & 9 & 14{,}204 & 244 & 129 & 0 & 50.7 (0.58) & -- & 11.12 & $+$3.73 [3.15,4.40] \\
0.67 & 10.5 & 100.0 & 10 & 13{,}617 & 392 & 159 & 0 & 67.3 (0.67) & -- & 9.44 & $+$1.86 [1.52,2.26] \\
0.8 & 12.5 & 100.0 & 11 & 13{,}237 & 424 & 194 & 0 & 80.0 (0.82) & -- & 8.84 & $+$1.30 [1.03,1.59] \\
1 & 15.6 & 100.0 & 16 & 12{,}691 & 423 & 246 & 0 & -- & -- & 8.42 & -- \\
1.25 & 19.3 & 100.0 & 22 & 11{,}942 & 557 & 270 & 0 & -- & 4.4 & 7.33 & $-$1.29 [$-$1.57,$-$1.03] \\
1.5 & 22.7 & 100.0 & 27 & 11{,}431 & 552 & 252 & 0 & -- & 8.4 & 6.67 & $-$2.18 [$-$2.51,$-$1.87] \\
2 & 27.3 & 100.0 & 40 & 10{,}657 & 583 & 274 & 0 & -- & 13.8 & 5.93 & $-$3.09 [$-$3.50,$-$2.69] \\
3 & 33.9 & 100.0 & 53 & 9{,}683 & 563 & 224 & 0 & -- & 21.6 & 5.13 & $-$3.84 [$-$4.28,$-$3.41] \\
\midrule
\multicolumn{12}{l}{\emph{Brightkite (5{,}260 queries)}} \\
0.5 & 4.5 & 100.0 & 8 & 4{,}970 & 48 & 6 & 0 & 51.1 (0.78) & -- & 8.58 & $+$1.18 [0.34,2.13] \\
0.67 & 5.9 & 100.0 & 8 & 4{,}849 & 77 & 22 & 0 & 67.5 (0.80) & -- & 7.56 & $+$0.68 [0.18,1.21] \\
0.8 & 7.1 & 100.0 & 8 & 4{,}762 & 106 & 21 & 0 & 80.3 (0.89) & -- & 6.44 & $+$0.24 [$-$0.08,0.59] \\
1 & 8.8 & 100.0 & 10 & 4{,}640 & 127 & 31 & 0 & -- & -- & 5.67 & -- \\
1.25 & 11.0 & 100.0 & 12 & 4{,}473 & 146 & 62 & 0 & -- & 2.4 & 4.32 & $-$1.35 [$-$1.91,$-$0.85] \\
1.5 & 12.9 & 100.0 & 13 & 4{,}326 & 172 & 85 & 0 & -- & 4.5 & 3.60 & $-$2.11 [$-$2.74,$-$1.52] \\
2 & 17.1 & 100.0 & 16 & 4{,}088 & 175 & 95 & 0 & -- & 9.2 & 3.15 & $-$3.19 [$-$4.18,$-$2.32] \\
3 & 23.5 & 100.0 & 24 & 3{,}746 & 186 & 92 & 0 & -- & 16.1 & 2.52 & $-$3.75 [$-$4.80,$-$2.79] \\
\midrule
\multicolumn{12}{l}{\emph{Yelp (2{,}000 queries)}} \\
0.5 & 29.7 & 99.9 & 144.5 & 1{,}282 & 79 & 41 & 4 & 61.6 (0.32) & -- & 2.53 & $+$1.07 [0.90,1.26] \\
0.67 & 36.7 & 99.7 & 217 & 1{,}117 & 96 & 38 & 15 & 76.1 (0.52) & -- & 2.26 & $+$0.71 [0.59,0.84] \\
0.8 & 42.2 & 99.5 & 271 & 1{,}020 & 74 & 51 & 11 & 87.4 (0.69) & -- & 1.97 & $+$0.39 [0.28,0.49] \\
1 & 48.2 & 95.7 & 379 & 907 & 68 & 48 & 13 & -- & -- & 1.70 & -- \\
1.25 & 53.8 & 88.9 & 477 & 792 & 78 & 34 & 21 & -- & 10.7 & 1.59 & $-$0.25 [$-$0.32,$-$0.18] \\
1.5 & 58.1 & 87.1 & 549 & 692 & 88 & 31 & 27 & -- & 19.2 & 1.44 & $-$0.43 [$-$0.52,$-$0.35] \\
2 & 65.0 & 84.9 & 679 & 572 & 71 & 34 & 24 & -- & 32.5 & 1.19 & $-$0.70 [$-$0.82,$-$0.60] \\
3 & 74.2 & 84.0 & 824 & 402 & 67 & 33 & 15 & -- & 50.2 & 0.98 & $-$0.88 [$-$1.01,$-$0.76] \\
\bottomrule
\end{tabular}

%% file: tables/tr_p22.tex
\begin{tabular}{lrrrrrrrrrr}
\toprule
 & \multicolumn{3}{c}{Gowalla} & \multicolumn{3}{c}{Brightkite} & \multicolumn{4}{c}{Congress} \\
\cmidrule(lr){2-4}\cmidrule(lr){5-7}\cmidrule(lr){8-11}
$w$ & 0.01 & 0.1 & 1 & 0.01 & 0.1 & 1 & 0.01 & 0.1 & 1 & obs. \\
\midrule
Groups that lose an interaction (\%) & 34.1 & 48.4 & 65.2 & 22.5 & 37.7 & 62.1 & 0 & 0 & 80.6 & 98.2 \\
\quad lost share (median, \%) & 4.0 & 8.5 & 11.0 & 0.4 & 11.8 & 10.8 & -- & -- & 2.1 & 5.3 \\
\quad first loss (median step) & 1 & 1 & 1 & 1 & 1 & 1 & -- & -- & 2 & 1 \\
Largest final cluster (mean, \%) & 23.5 & 17.7 & 16.8 & 24.0 & 18.0 & 18.8 & 71.6 & 11.8 & 5.2 & 3.9 \\
\midrule
Largest cluster, LCC $-$ size-matched SAC (pp) & $-$1.0 & $+$0.8 & $+$0.9 & $-$1.4 & $+$0.4 & $+$0.3 & $-$4.7 & $-$1.5 & $-$0.3 & $-$0.4 \\
\quad 95\% CI & [$-$1.7,$-$0.3] & [0.4,1.2] & [0.5,1.2] & [$-$2.4,$-$0.3] & [$-$0.0,0.9] & [$-$0.1,0.7] & [$-$6.4,$-$3.1] & [$-$2.3,$-$0.8] & [$-$0.4,$-$0.1] & [$-$0.5,$-$0.3] \\
\bottomrule
\end{tabular}

%% file: sections/08_statements.tex
\ifdefined\NOREFS\else
\iffull\else\clearpage\fi
\section*{AI-Generated Content Acknowledgement}
We used Anthropic Claude Opus 5.5 to debug and test code, and OpenAI GPT-6 Astra to polish the manuscript.
\fi

%% file: main.bbl
\begin{thebibliography}{10}
\providecommand{\url}[1]{#1}
\csname url@samestyle\endcsname
\providecommand{\newblock}{\relax}
\providecommand{\bibinfo}[2]{#2}
\providecommand{\BIBentrySTDinterwordspacing}{\spaceskip=0pt\relax}
\providecommand{\BIBentryALTinterwordstretchfactor}{4}
\providecommand{\BIBentryALTinterwordspacing}{\spaceskip=\fontdimen2\font plus
\BIBentryALTinterwordstretchfactor\fontdimen3\font minus
  \fontdimen4\font\relax}
\providecommand{\BIBforeignlanguage}[2]{{%
\expandafter\ifx\csname l@#1\endcsname\relax
\typeout{** WARNING: IEEEtran.bst: No hyphenation pattern has been}%
\typeout{** loaded for the language `#1'. Using the pattern for}%
\typeout{** the default language instead.}%
\else
\language=\csname l@#1\endcsname
\fi
#2}}
\providecommand{\BIBdecl}{\relax}
\BIBdecl

\bibitem{sozio2010community}
M.~Sozio and A.~Gionis, ``The community-search problem and how to plan a
  successful cocktail party,'' in \emph{KDD}, 2010, pp. 939--948.

\bibitem{fang2020survey}
Y.~Fang, X.~Huang, L.~Qin, Y.~Zhang, W.~Zhang, R.~Cheng, and X.~Lin, ``A survey
  of community search over big graphs,'' \emph{The VLDB Journal}, vol.~29,
  no.~1, pp. 353--392, 2020.

\bibitem{zhang2017engagement}
F.~Zhang, Y.~Zhang, L.~Qin, W.~Zhang, and X.~Lin, ``When engagement meets
  similarity: Efficient $(k,r)$-core computation on social networks,''
  \emph{PVLDB}, vol.~10, no.~10, pp. 998--1009, 2017.

\bibitem{wang2018radius}
K.~Wang, X.~Cao, X.~Lin, W.~Zhang, and L.~Qin, ``Efficient computing of
  radius-bounded $k$-cores,'' in \emph{ICDE}, 2018, pp. 233--244.

\bibitem{chen2018colocated}
L.~Chen, C.~Liu, R.~Zhou, J.~Li, X.~Yang, and B.~Wang, ``Maximum co-located
  community search in large scale social networks,'' \emph{PVLDB}, vol.~11,
  no.~10, pp. 1233--1246, 2018.

\bibitem{fang2017spatial}
Y.~Fang, R.~Cheng, X.~Li, S.~Luo, and J.~Hu, ``Effective community search over
  large spatial graphs,'' \emph{PVLDB}, vol.~10, no.~6, pp. 709--720, 2017.

\bibitem{liu2020vac}
Q.~Liu, Y.~Zhu, M.~Zhao, X.~Huang, J.~Xu, and Y.~Gao, ``{VAC}: Vertex-centric
  attributed community search,'' in \emph{ICDE}, 2020, pp. 937--948.

\bibitem{sherif1961social}
M.~Sherif and C.~I. Hovland, \emph{Social Judgment: Assimilation and Contrast
  Effects in Communication and Attitude Change}.\hskip 1em plus 0.5em minus
  0.4em\relax Yale University Press, 1961.

\bibitem{hegselmann2002opinion}
R.~Hegselmann and U.~Krause, ``Opinion dynamics and bounded confidence: Models,
  analysis and simulation,'' \emph{J.\ Artif.\ Soc.\ Soc.\ Simul.}, vol.~5,
  no.~3, 2002.

\bibitem{lorenz2010heterogeneous}
J.~Lorenz, ``Heterogeneous bounds of confidence: Meet, discuss and find
  consensus!'' \emph{Complexity}, vol.~15, no.~4, pp. 43--52, 2010.

\bibitem{cui2014local}
W.~Cui, Y.~Xiao, H.~Wang, and W.~Wang, ``Local search of communities in large
  graphs,'' in \emph{SIGMOD}, 2014, pp. 991--1002.

\bibitem{huang2014querying}
X.~Huang, H.~Cheng, L.~Qin, W.~Tian, and J.~X. Yu, ``Querying $k$-truss
  community in large and dynamic graphs,'' in \emph{SIGMOD}, 2014, pp.
  1311--1322.

\bibitem{fang2016effective}
Y.~Fang, R.~Cheng, S.~Luo, and J.~Hu, ``Effective community search for large
  attributed graphs,'' \emph{PVLDB}, vol.~9, no.~12, pp. 1233--1244, 2016.

\bibitem{guo2026skyline}
F.~Guo, X.~Luo, S.~Xu, H.~Gao, Y.~Liu, H.~Shen, and X.~Cheng, ``Skyline
  community search over edge-attributed bipartite graphs,'' in \emph{DASFAA},
  ser. LNCS, vol. 16536.\hskip 1em plus 0.5em minus 0.4em\relax Springer, 2026,
  pp. 367--383.

\bibitem{fang2020heterogeneous}
Y.~Fang, Y.~Yang, W.~Zhang, X.~Lin, and X.~Cao, ``Effective and efficient
  community search over large heterogeneous information networks,''
  \emph{PVLDB}, vol.~13, no.~6, pp. 854--867, 2020.

\bibitem{liu2024sach}
Y.~Liu, F.~Guo, B.~Xu, P.~Bao, H.~Shen, and X.~Cheng, ``{SACH}:
  Significant-attributed community search in heterogeneous information
  networks,'' in \emph{ICDE}, 2024, pp. 3283--3296.

\bibitem{chen2024fcs}
G.~Chen, F.~Guo, Y.~Wang, Y.~Liu, P.~Yu, H.~Shen, and X.~Cheng, ``{FCS-HGNN}:
  Flexible multi-type community search in heterogeneous information networks,''
  in \emph{CIKM}, 2024, pp. 207--217.

\bibitem{shen2023density}
C.-Y. Shen, S.-H. Ko, G.-S. Lee, W.-C. Lee, and D.-N. Yang, ``Density
  personalized group query,'' \emph{PVLDB}, vol.~16, no.~4, pp. 615--628, 2022.

\bibitem{bonchi2019discovering}
F.~Bonchi, E.~Galimberti, A.~Gionis, B.~Ordozgoiti, and G.~Ruffo, ``Discovering
  polarized communities in signed networks,'' in \emph{CIKM}, 2019, pp.
  961--970.

\bibitem{ordozgoiti2020finding}
B.~Ordozgoiti, A.~Matakos, and A.~Gionis, ``Finding large balanced subgraphs in
  signed networks,'' in \emph{WWW}, 2020, pp. 1378--1388.

\bibitem{liu2026rethinking}
Y.~Liu, F.~Guo, H.~Shen, and X.~Cheng, ``Rethinking community consistency in
  heterogeneous networks,'' in \emph{CIKM}, 2026, pp. 3293--3304.

\bibitem{yang2012socio}
D.-N. Yang, C.-Y. Shen, W.-C. Lee, and M.-S. Chen, ``On socio-spatial group
  query for location-based social networks,'' in \emph{KDD}, 2012, pp.
  949--957.

\bibitem{zhu2017geosocial}
Q.~Zhu, H.~Hu, C.~Xu, J.~Xu, and W.-C. Lee, ``Geo-social group queries with
  minimum acquaintance constraints,'' \emph{The VLDB Journal}, vol.~26, no.~5,
  pp. 709--727, 2017.

\bibitem{li2020skyline}
Q.~Li, Y.~Zhu, and J.~X. Yu, ``Skyline cohesive group queries in large
  road-social networks,'' in \emph{ICDE}, 2020, pp. 397--408.

\bibitem{guo2019cohesive}
F.~Guo, Y.~Yuan, G.~Wang, L.~Chen, X.~Lian, and Z.~Wang, ``Cohesive group
  nearest neighbor queries over road-social networks,'' in \emph{ICDE}, 2019,
  pp. 434--445.

\bibitem{guo2021cohesive}
------, ``Cohesive group nearest neighbor queries on road-social networks under
  multi-criteria,'' \emph{IEEE Trans. Knowl. Data Eng.}, vol.~33, no.~11, pp.
  3520--3536, 2021.

\bibitem{guo2021multi}
F.~Guo, Y.~Yuan, G.~Wang, X.~Zhao, and H.~Sun, ``Multi-attributed community
  search in road-social networks,'' in \emph{ICDE}, 2021, pp. 109--120.

\bibitem{ameryahia2009group}
S.~Amer-Yahia, S.~Basu~Roy, A.~Chawla, G.~Das, and C.~Yu, ``Group
  recommendation: Semantics and efficiency,'' \emph{PVLDB}, vol.~2, no.~1, pp.
  754--765, 2009.

\bibitem{basuroy2015group}
S.~Basu~Roy, L.~V.~S. Lakshmanan, and R.~Liu, ``From group recommendations to
  group formation,'' in \emph{SIGMOD}, 2015, pp. 1603--1616.

\bibitem{degroot1974reaching}
M.~H. DeGroot, ``Reaching a consensus,'' \emph{J.\ Am.\ Stat.\ Assoc.},
  vol.~69, no. 345, pp. 118--121, 1974.

\bibitem{friedkin1990social}
N.~E. Friedkin and E.~C. Johnsen, ``Social influence and opinions,'' \emph{J.\
  Math.\ Sociol.}, vol.~15, no. 3--4, pp. 193--206, 1990.

\bibitem{deffuant2000mixing}
G.~Deffuant, D.~Neau, F.~Amblard, and G.~Weisbuch, ``Mixing beliefs among
  interacting agents,'' \emph{Advances in Complex Systems}, vol.~3, no. 1--4,
  pp. 87--98, 2000.

\bibitem{mirtabatabaei2012opinion}
A.~Mirtabatabaei and F.~Bullo, ``Opinion dynamics in heterogeneous networks:
  Convergence conjectures and theorems,'' \emph{SIAM J.\ Control Optim.},
  vol.~50, no.~5, pp. 2763--2785, 2012.

\bibitem{su2017partial}
W.~Su, Y.~Gu, S.~Wang, and Y.~Yu, ``Partial convergence of heterogeneous
  {Hegselmann-Krause} opinion dynamics,'' \emph{Sci.\ China Technol.\ Sci.},
  vol.~60, no.~9, pp. 1433--1438, 2017.

\bibitem{hegselmann2015opinion}
R.~Hegselmann and U.~Krause, ``Opinion dynamics under the influence of radical
  groups, charismatic leaders, and other constant signals: A simple unifying
  model,'' \emph{Netw.\ Heterog.\ Media}, vol.~10, no.~3, pp. 477--509, 2015.

\bibitem{jager2005uniformity}
W.~Jager and F.~Amblard, ``Uniformity, bipolarization and pluriformity captured
  as generic stylized behavior with an agent-based simulation model of attitude
  change,'' \emph{Comput.\ Math.\ Organ.\ Theory}, vol.~10, no.~4, pp.
  295--303, 2005.

\bibitem{monti2020learning}
C.~Monti, G.~De~Francisci~Morales, and F.~Bonchi, ``Learning opinion dynamics
  from social traces,'' in \emph{KDD}, 2020, pp. 764--773.

\bibitem{luan2026homophily}
Y.~Luan, C.~Ancona, C.~Bernardo, V.~Pansanella, F.~Lo~Iudice, G.~Rossetti,
  F.~Vasca, X.~Wu, and C.~Altafini, ``Quantifying opinion homophily in online
  social networks: A bounded confidence perspective,'' arXiv:2604.23157, 2026.

\bibitem{lappas2009finding}
T.~Lappas, K.~Liu, and E.~Terzi, ``Finding a team of experts in social
  networks,'' in \emph{KDD}, 2009, pp. 467--476.

\bibitem{seidman1983network}
S.~B. Seidman, ``Network structure and minimum degree,'' \emph{Social
  Networks}, vol.~5, no.~3, pp. 269--287, 1983.

\bibitem{batagelj2003cores}
V.~Batagelj and M.~Zaver{\v{s}}nik, ``An {$O(m)$} algorithm for cores
  decomposition of networks,'' arXiv:cs/0310049, 2003.

\bibitem{moreau2005stability}
L.~Moreau, ``Stability of multiagent systems with time-dependent communication
  links,'' \emph{IEEE Trans.\ Autom.\ Control}, vol.~50, no.~2, pp. 169--182,
  2005.

\bibitem{karp1972reducibility}
R.~M. Karp, ``Reducibility among combinatorial problems,'' in \emph{Complexity
  of Computer Computations}.\hskip 1em plus 0.5em minus 0.4em\relax Plenum
  Press, 1972, pp. 85--103.

\bibitem{zuckerman2007linear}
D.~Zuckerman, ``Linear degree extractors and the inapproximability of max
  clique and chromatic number,'' \emph{Theory of Computing}, vol.~3, pp.
  103--128, 2007.

\bibitem{maehara1984space}
H.~Maehara, ``Space graphs and sphericity,'' \emph{Discrete Applied
  Mathematics}, vol.~7, no.~1, pp. 55--64, 1984.

\bibitem{clark1990unit}
B.~N. Clark, C.~J. Colbourn, and D.~S. Johnson, ``Unit disk graphs,''
  \emph{Discrete Mathematics}, vol.~86, no. 1--3, pp. 165--177, 1990.

\bibitem{bonamy2021eptas}
M.~Bonamy, {\'E}.~Bonnet, N.~Bousquet, P.~Charbit, P.~Giannopoulos, E.~J. Kim,
  P.~Rz{\k{a}}{\.z}ewski, F.~Sikora, and S.~Thomass{\'e}, ``{EPTAS} and
  subexponential algorithm for maximum clique on disk and unit ball graphs,''
  \emph{J.\ ACM}, vol.~68, no.~2, 2021, article 9.

\bibitem{tomita2003efficient}
E.~Tomita and T.~Seki, ``An efficient branch-and-bound algorithm for finding a
  maximum clique,'' in \emph{DMTCS}, ser. LNCS, vol. 2731.\hskip 1em plus 0.5em
  minus 0.4em\relax Springer, 2003, pp. 278--289.

\bibitem{cho2011friendship}
E.~Cho, S.~A. Myers, and J.~Leskovec, ``Friendship and mobility: User movement
  in location-based social networks,'' in \emph{KDD}, 2011, pp. 1082--1090.

\bibitem{yang2012defining}
J.~Yang and J.~Leskovec, ``Defining and evaluating network communities based on
  ground-truth,'' in \emph{ICDM}, 2012, pp. 745--754.

\bibitem{takac2012pokec}
L.~Takac and M.~Zabovsky, ``Data analysis in public social networks,'' in
  \emph{Proc.\ Int.\ Scientific Conf.\ \& Int.\ Workshop Present Day Trends of
  Innovations}, {\L}om{\.z}a, Poland, May 2012.

\bibitem{leskovec2014snap}
J.~Leskovec and A.~Krevl, ``{SNAP Datasets}: {Stanford} large network dataset
  collection,'' \url{http://snap.stanford.edu/data}, Jun. 2014.

\bibitem{raghavan2007near}
U.~N. Raghavan, R.~Albert, and S.~Kumara, ``Near linear time algorithm to
  detect community structures in large-scale networks,'' \emph{Physical Review
  E}, vol.~76, no.~3, p. 036106, 2007.

\bibitem{huangfu2018parallelizing}
Q.~Huangfu and J.~A.~J. Hall, ``Parallelizing the dual revised simplex
  method,'' \emph{Math.\ Program.\ Comput.}, vol.~10, no.~1, pp. 119--142,
  2018.

\bibitem{ortools}
{Google}, ``{OR-Tools}, version 9.15, with the {CP-SAT} solver,''
  \url{https://developers.google.com/optimization/}, 2026.

\bibitem{yelp2022dataset}
{Yelp}, ``{Yelp Open Dataset},'' \url{https://www.yelp.com/dataset}, 2022,
  accessed 2026.

\bibitem{poole2007ideology}
K.~T. Poole and H.~Rosenthal, \emph{Ideology and Congress}, 2nd~ed.\hskip 1em
  plus 0.5em minus 0.4em\relax New Brunswick, NJ: Transaction Publishers, 2007.

\bibitem{lewis2023voteview}
J.~B. Lewis, K.~Poole, H.~Rosenthal, A.~Boche, A.~Rudkin, and L.~Sonnet,
  ``{Voteview: Congressional Roll-Call Votes Database},''
  \url{https://voteview.com/}, 2026.

\bibitem{govinfo_billstatus}
{U.S. Government Publishing Office}, ``Bill status ({BILLSTATUS}) bulk data,
  117th {Congress},'' \url{https://www.govinfo.gov/bulkdata/BILLSTATUS}, 2026,
  accessed Oct. 2026.

\bibitem{fink2023congress}
C.~G. Fink, N.~Omodt, S.~Zinnecker, and G.~Sprint, ``A {Congressional}
  {Twitter} network dataset quantifying pairwise probability of influence,''
  \emph{Data Brief}, vol.~50, 2023, {Art.\ no.}\ 109521.

\bibitem{fink2023centrality}
C.~G. Fink, K.~Fullin, G.~Gutierrez, N.~Omodt, S.~Zinnecker, G.~Sprint, and
  S.~McCulloch, ``A centrality measure for quantifying spread on weighted,
  directed networks,'' \emph{Physica A}, vol. 626, 2023, {Art.\ no.}\ 129083.

\bibitem{cherepnalkoski2016mep}
D.~Cherepnalkoski, A.~Karpf, I.~Mozeti{\v{c}}, and M.~Gr{\v{c}}ar, ``Cohesion
  and coalition formation in the {European Parliament}: Roll-call votes and
  {Twitter} activities,'' \emph{PLoS ONE}, vol.~11, no.~11, 2016, {Art.\ no.}\
  e0166586; data: CLARIN.SI 11356/1071.

\end{thebibliography}
